\documentclass[12pt,preprint]{aastex}
\usepackage{epsfig}
\usepackage{amsthm}
\usepackage{hyperref}
\usepackage{amsmath}
\newtheorem{prob}{Problem}[section]
\numberwithin{equation}{section}
\newcommand{\en}{\mathcal{E}}
\begin{document}

\begin{figure}[!htb]
\centering
\includegraphics[scale=0.8]{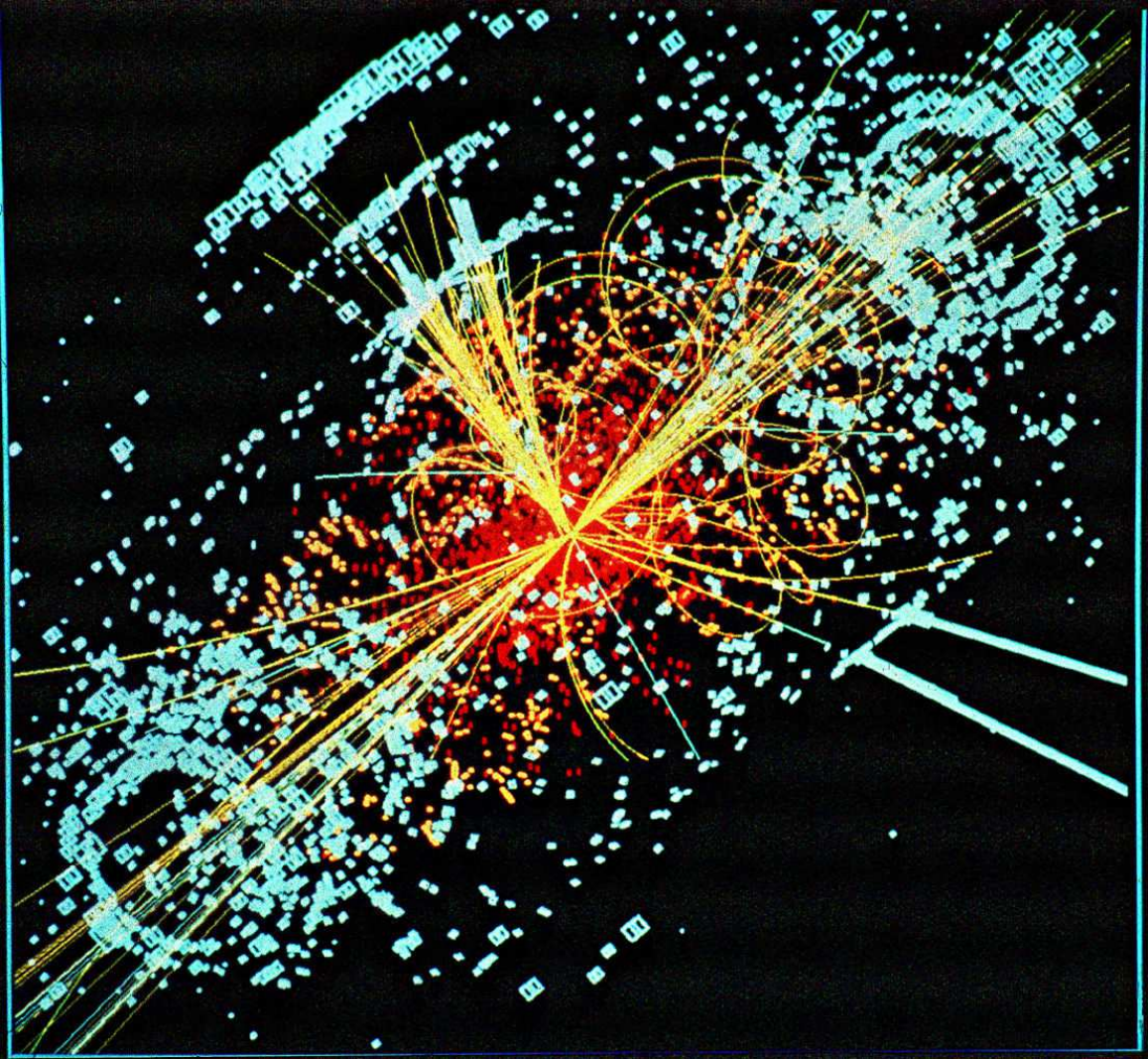}
\caption{Large Hadron Collider (LHC) Compact Muon Solenoid (CMS) event simulation. A Higgs boson decays into two jets of hadrons and two electrons. The lines represent the possible paths of particles produced by the proton-proton collision in the detector while the energy these particles deposit is shown in blue. Image by Lucas Taylor. From \url{http://commons.wikimedia.org/wiki/File:CMS_Higgs-event.jpg}.}
\end{figure}

\clearpage
\title{Theory of Special Relativity}

\author{Nadia L. Zakamska\altaffilmark{*}}

\altaffiltext{*}{Department of Physics \& Astronomy, Johns Hopkins University, 3400 N. Charles St., Baltimore, MD 21218, USA}

\begin{abstract}
Special Relativity is taught to physics sophomores at Johns Hopkins University in a series of eight lectures. Lecture 1 covers the principle of relativity and the derivation of the Lorentz transform. Lecture 2 covers length contraction and time dilation. Lecture 3 covers Minkowski diagrams, simultaneous events and causally connected events, as well as velocity transforms. Lecture 4 covers energy and momentum of particles and introduces 4-vectors. Lecture 5 covers energy and momentum of photons and collision problems. Lecture 6 covers Doppler effect and aberration. Lecture 7 covers relativistic dynamics. Optional Lecture 8 covers field transforms. The main purpose of these notes is to introduce 4-vectors and the matrix notation and to demonstrate their use in solving standard problems in Special Relativity. The pre-requisites for the class are calculus-based Classical Mechanics and Electricity \& Magnetism, and Linear Algebra is highly recommended. {\bf These are lecture notes, not original research work by the author. Use at your own risk; please report typos and other problems to the author. This is version 3.1 (September 2018).}\footnote{Changes from v.1.0 to v.2.0: 4-vector notation changed from $\underline{P}\circ\underline{P}$ to $P^{\mu}P_{\mu}$; small editorial changes. Changes from v.2.0 to v.3.0: added section 7.4; small editorial changes. Changes from v.3.0 to v.3.1: fixed typos; added clarifications; small editorial changes.}
\end{abstract}

\tableofcontents

\clearpage
\begin{figure}[!htb]
\centering
\includegraphics[scale=0.8]{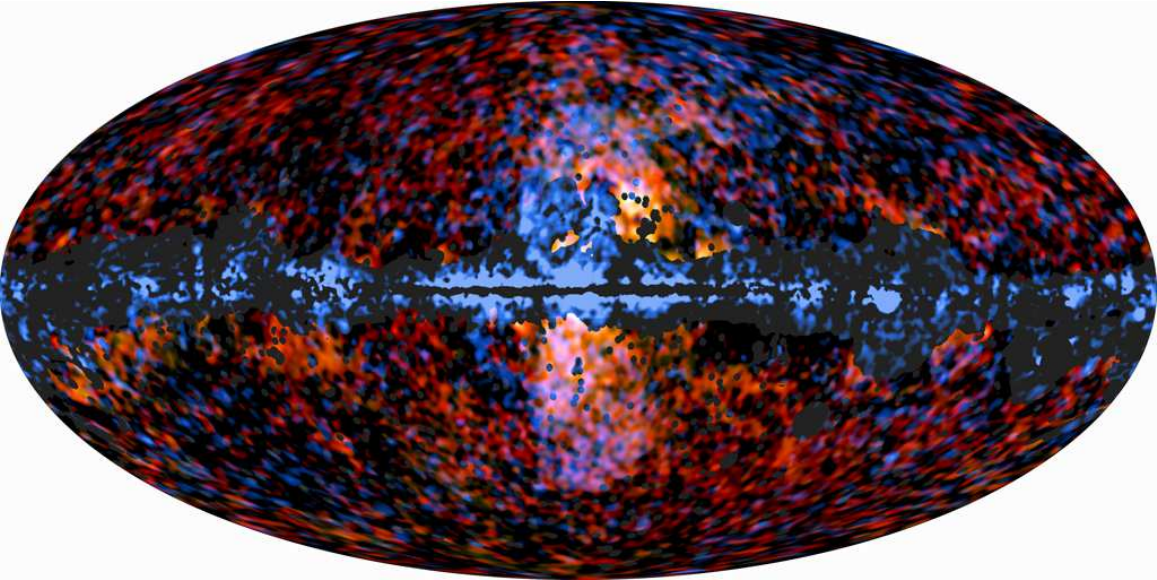}
\caption{Superposition of the maps of the microwave emission (from Planck telescope) and the gamma-ray emission (from Fermi telescope) in Galactic coordinates, revealing the mysterious ``bubbles'' above and below the Galactic plane. Both types of emission are likely produced by relativistic particles accelerated by shocks due to a powerful explosion that happened in the Galactic nucleus several million years ago \citep{su10}. Image by ESA/Planck, NASA/DoE/Fermi and Douglas Finkbeiner, from \url{http://planck.caltech.edu/news20120213.html}.}
\end{figure}

\clearpage
\section{Principle of Relativity}

{\bf Q.} Before we start, what is the difference between Special and General Relativity? -- {\bf A.} Special Relativity addresses the geometry of space-time in empty space, whereas General Relativity addresses a much more complicated issue of the space-time in the presence of gravitating (massive) bodies. Thus in Special Relativity we consider motions of particles and lightwaves in empty space. 

\subsection{Galilean transform in classical mechanics}

In order to formulate the principle of relativity, we need some apparatus, much of it familiar from classical mechanics. {\bf A reference frame} is necessary to describe the motion of bodies. It includes a coordinate system to describe their positions and a clock. There exist many reference frames called {\bf inertial frames} where the free particles move with constant velocities on straight lines. If we have one inertial frame, then another frame moving at constant velocity $\vec{v}$ relative to the first one is also inertial. 

The {\bf principle of relativity} states simply that all laws of nature are the same in all inertial frames. Let us see how the principle of relativity applies to classical mechanics, but before doing that we need to properly set up the reference frames.  

\begin{figure}[htb]
\centering
\includegraphics[scale=1.0, clip=true, trim=1cm 0cm 6cm 19cm]{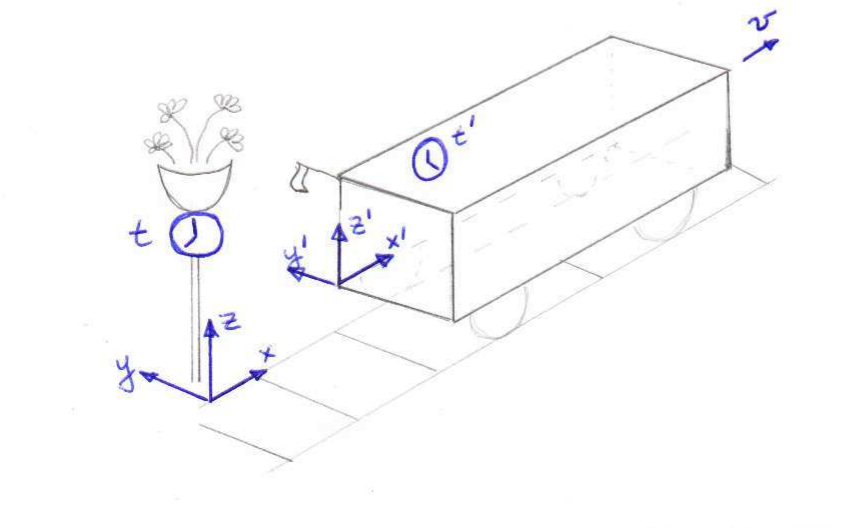}
\caption{A sketch introducing the train and the tracks reference systems.}
\label{pic_frames}
\end{figure}

Our main example of reference frames will consist of railroad tracks and a train moving along the tracks (Figure \ref{pic_frames}). The tracks will be called reference frame $S$ and the train will be called reference frame $S'$. In the tracks frame $S$, there is a ruler lying on the ground along the tracks, starting at the station which we place at $x=0, y=0, z=0$. This is the $x$-axis in the $S$ system. The $z$ axis points vertically toward the sky, and the $y$ axis is perpendicular to tracks along the ground. Everywhere along the tracks we have clocks that are synchronized amongst themselves. We will discuss a bit later how to do this in Special Relativity, but in classical mechanics we can just use several phones to call up the stations along the tracks and ask them to set their timer clocks to zero on our command.

We also have a train moving with constant velocity $v$ along the tracks in the positive $x$ direction, as seen in the frame $S$. We get on the train, i.e., into the frame $S'$. Inside the train, there is a ruler starting at the back of the train; it starts at $x'=0$ at the back of the last car, with $x'$ increasing toward the head of the train. 

How do we synchronize the clocks between the station and the train? We will do this in a way that can be applied to both classical mechanics and Special Relativity. Last car of the train has a flag sticking out the window at $x'=0$, while the station has a flower pot at $x=0$. When the last car passes by the station, the flag breaks against the flower pot. When this happens, the observer at the station sets the clock to $t=0$ and the observer on the train sets the clock to $t'=0$. We know what happens next in classical mechanics: if the train were to stop at one of the next stations, the clocks on the train and at the station would be showing the same time. The flow of time is the same in all inertial frames in classical mechanics. 

The breaking of the flag against the pot is {\bf a physical event}. An event is something that happens independently of the reference frame. It happens at a point in space and at an instant in time. This particular event has {\bf space-time coordinates} $(0,0,0,0)$ in both frames $S$ and $S'$: $t=0, x=0, y=0, z=0$ and $t'=0, x'=0, y'=0, z'=0$. Events are easier to visualize if something irreversible happens, e.g., when something breaks, so a few things will be broken before these lectures are over. 

There are also some kids playing soccer near the tracks (while maintaining the highest safety standards!). Right as the flag is breaking against the flower pot, one of the kids kicks the ball away from the pot. Ball being kicked is also a physical event, which also happens at space-time coordinates (0,0,0,0) in both frames. At some later time, the ball lands on a sharp spike of a fence and pops. It happens at space-time coordinates $(t,x,y,z)$ in the $S$ system and at space-time coordinates $(t',x',y',z')$ in the $S'$ system. 

How are the space-time coordinates of the ball-popping related in classical mechanics? This is known as {\bf Galilean coordinate transform} and is familiar from classical mechanics:
\begin{eqnarray}
t'=t,\\
x'=x-vt,\\
y'=y,\\
z'=z.
\end{eqnarray}
We have written down this coordinate transformation for just one event -- ball being popped in the fence -- but we can pop the ball at any point in its trajectory. Therefore, if we know the trajectory of the ball in one frame, we can derive it in another frame. 

Now that we have set up the coordinate systems and remembered how to transform between them in classical mechanics, we will check how the relativity principle applies to classical mechanics. The relativity principle states that all laws of nature, including the laws describing the motion of the ball, should be the same in both frames $S$ and $S'$. The laws of classical mechanics are Newton laws, so we need to see the principle of relativity applies to each law. 

Law number 1 states that an object moving with $\vec{u}=$const (including the case of an object at rest, const$=0$) will continue to move with this velocity unless a force is applied to it. According to the Galilean transform, this object is moving with $\vec{u'}=\vec{u}-\vec{v}$ in the frame $S'$ moving with $\vec{v}$. Thus, if $\vec{v}$ is constant (i.e., $S'$ is also an inertial frame), $\vec{u'}$ is also constant, so Newton's first law is satisfied in $S'$. 

Law number 2 states that an object affected by net force $\vec{F}$ is moving with acceleration $\vec{a}$ such that $\vec{F}=m\vec{a}$. In classical mechanics, neither $\vec{F}$ nor $\vec{a}$ change when we move from one inertial frame $S$ to another $S'$, thus the law is satisfied in all inertial frames. Law number 3 is about opposing forces, which again do not change from one frame to the next, so this law is also satisfied in all inertial frames. 

We have just seen that some quantities changed because we moved from one system to another (they {\bf transformed}), whereas others did not (they remained {\bf invariant}). In classical mechanics, the values that remain {\bf invariant under Galilean transform} include: acceleration $\vec{a}$, time $t$, size of the ball, for example the distance between its front and back $x_2-x_1$. The values that are not invariant include: velocity of the ball $\vec{u}$, its coordinates $(x,y,z)$. 

\subsection{Galilean transform and electromagnetism}
\label{sec_maxwell}

Now that we have considered classical mechanics, we consider electromagnetism. We have the same situation: train and tracks, but instead of a ball game we are observing light waves, which are electromagnetic radiation. One of the kids now flashes a flashlight, pointing it parallel to the tracks. If the light from the flashlight moves with velocity $c$ in the tracks frame, then according to the Galilean transformation it should move with velocity $c-v$ in the train frame, because in classical mechanics this is exactly analogous to the situation with the ball. Thus, Galilean transform implies that observers in different inertial frames should measure different speed of light. 

Is this compatible with the laws of electromagnetism? The laws of electromagnetism are mathematically described by Maxwell's equations. Making a brief foray into the previous semester, we remember how one derives electromagnetic radiation from Maxwell's equations:
\begin{eqnarray}
{\rm div} \vec{E}\equiv \vec{\nabla}\cdot\vec{E}=4\pi\rho\mbox{ in cgs units or }\frac{\rho}{\epsilon_0}\mbox{ in MKS units}\\
{\rm div} \vec{B}\equiv \vec{\nabla}\cdot\vec{B}=0\mbox{ in either system of units}\\
{\rm curl} \vec{E}\equiv \vec{\nabla}\times\vec{E}=-\frac{1}{c}\frac{\partial \vec{B}}{\partial t}\mbox{ in cgs units or }-\frac{\partial \vec{B}}{\partial t}\mbox{ in MKS units}\\
{\rm curl} \vec{B}\equiv \vec{\nabla}\times\vec{B}=\frac{1}{c}\frac{\partial \vec{E}}{\partial t}+\frac{4\pi}{c}\vec{j}\mbox{ in cgs units or }\mu_0\epsilon_0\frac{\partial \vec{E}}{\partial t}+\mu_0\vec{j}\mbox{ in MKS units }.
\end{eqnarray}
To get vacuum electromagnetic radiation from this, we set $\rho=0$ and $\vec{j}=0$ because there are no charges and no currents: in the light wave, the changing $\vec{E}$ field gives rise to changing $\vec{B}$ field and vice versa. We do this in the cgs units, but the same idea works in MKS as well: 
\begin{eqnarray}
\vec{\nabla}\cdot\vec{E}=0\\
\vec{\nabla}\cdot\vec{B}=0\\
\vec{\nabla}\times\vec{E}=-\frac{1}{c}\frac{\partial \vec{B}}{\partial t}\\
\vec{\nabla}\times\vec{B}=\frac{1}{c}\frac{\partial \vec{E}}{\partial t}.
\end{eqnarray}
We take curl of the 3rd equation, change the order of derivatives and plug in the curl from the 4th equation:
\begin{equation}
\vec{\nabla}\times(\vec{\nabla}\times\vec{E})=-\frac{1}{c}\vec{\nabla}\times\frac{\partial\vec{B}}{\partial t}=-\frac{1}{c}\frac{\partial}{\partial t}(\vec{\nabla}\times\vec{B})=-\frac{1}{c^2}\frac{\partial^2}{\partial t^2}\vec{E}.
\end{equation}
In the introductory electromagnetism classes, one would usually consider a one-dimensional case, for example a plane electromagnetic wave propagating along the $x$ axis with electric field directed along the $z$ axis, in which case this simplifies to
\begin{equation}
\frac{\partial^2{E_z}}{\partial x^2}=\frac{1}{c^2}\frac{\partial^2 E_z}{\partial t^2}\mbox{ in cgs units or }=\mu_0\epsilon_0\frac{\partial^2E_z}{\partial t^2}\mbox{ in MKS units}. 
\end{equation}
Because the fundamental constants $\mu_0$ and $\epsilon_0$ are defined in the MKS system to satisfy the relation $c^2=\frac{1}{\mu_0\epsilon_0}$, this equation is actually the same in both systems. 

In the electromagnetism classes, you probably considered sinusoidal waves, but in principle any function of the variable $(x\pm ct)$ is the solution to this equation, from where we get that the constant $c$ used in the Maxwell's equations is not just an arbitrary constant, but the speed of propagation of electromagnetic waves. 

As written, Maxwell's equations seem to imply that $c$ is a fundamental constant, because it enters explicitly into the equations. But we just saw that the physical meaning of this constant is that it is the speed of the propagation of electromagnetic radiation. If the Galilean transform is applicable, we would measure different velocity of light in different frames; thus, it would seem that different Maxwell's equations should apply in different frames, so perhaps Maxwell's equations are not fundamental laws of electromagnetism after all? We can already see that there is some tension between Maxwell's theory and Galilean transforms: one or the other must be wrong.  

We take one further step. Our experience tells us that information cannot propagate instantaneously. If something happens in one point in space, all our experimental evidence suggests that we need to wait for a while before this can affect bodies very far away. Therefore, there ought to be a maximal value at which the signals propagate. This seems like a good candidate for a fundamental limit, a fundamental law of nature, and thus according to the relativity principle this value should be the same in all inertial frames. We can make this intuitive requirement compatible with Maxwell's equations by saying that this maximum propagation speed is the same as the speed of electromagnetic waves in vacuum. 

The arguments above are somewhat vague, but they sum up the growing realization which occurred in the beginning of the 20th century that Galilean transformations were compatible with Newton's mechanics but not with Maxwell's electromagnetism. The resolution of this controversy did not happen overnight, and several options were seriously considered by the leading experimentalists and theoreticians:

$\bullet$ The first possibility is that the relativity principle is valid for mechanics, but not valid for the electromagnetism. The relativity principle states that all inertial frames must be equivalent for the laws of nature. If it does not apply to the electromagnetism, this means that there is a special inertial frame where the Maxwell's equations are completely correct and where the electromagnetic radiation propagates with $c$. Such frame is called {\bf the ether, or aether}, and much experimental effort was dedicated to the search of the velocity of our planet relative to this frame. 

$\bullet$ Another option is that the relativity principle is valid for both mechanics and electromagnetism, but Maxwell's laws are wrong. 

$\bullet$ Finally, it is also possible that the relativity principle is valid for both mechanics and electromagnetism, but Newton's laws are wrong. 

Faced with these options, physicists conducted new experiments and developed new theories. A review of some of these attempts is presented in \citet{resn68} and other places. The key experiment demonstrated the constancy of the speed of light in all inertial frames and thus rejected the ether theories. It was performed by Michelson and Morley, and Michelson became the first American recipient of a Nobel Prize for this work. While the Michelson-Morley experiment ruled out the existence of an ether frame, other experiments confirmed the validity of Maxwell's equations, while all theoretical attempts to reconcile Maxwell's equations with Galilean transformations eventually failed. 

Thus, the third option -- rather astonishingly -- proved to be the one worth pursuing. The Newton's laws, the foundation that begins the study of physics, are not in fact correct. Rather, they are an approximation of the true laws of physics in the limit when things are moving much slower than the speed of light. In these lectures, we will set out to amend the laws of classical mechanics to make them compatible with {\bf Einstein's relativity principle} which now consists of two parts:

$\bullet$ Physical laws are the same in all inertial frames; 

$\bullet$ There is a maximum speed of propagation of signals which is constant in all frames.

The speed of light, being the fundamental constant of nature, is now defined to be exactly 299,792,458 m/sec, and the standard of length is derived from this value and the time standard based on atomic clocks. We will use $3\times 10^8$ m/sec and $3\times 10^{10}$ cm/sec in this class. 

We are developing a theory that is more general than the classical mechanics. We should be able to recover the classical mechanics from our new theory by setting all velocities to be $\ll c$, or alternatively by setting $c\rightarrow\infty$. 

\clearpage
\begin{figure}[!htb]
\centering
\includegraphics[scale=0.8]{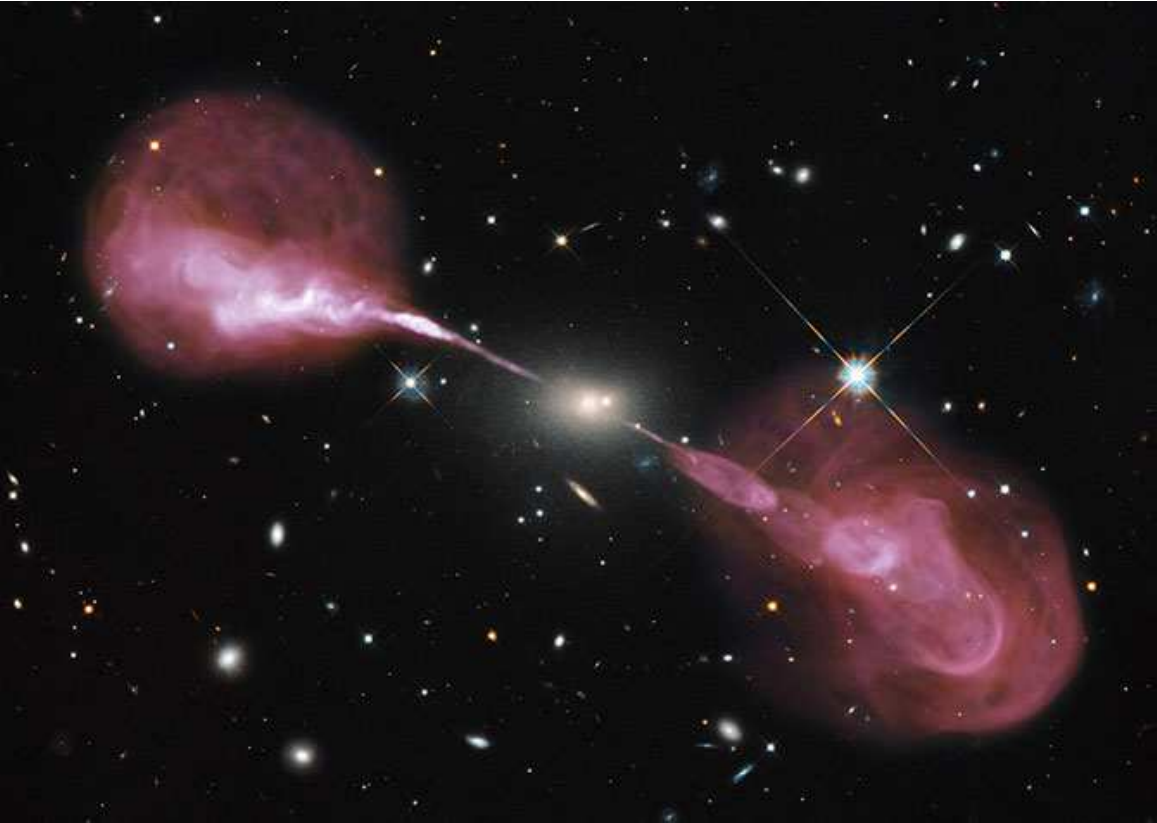}
\caption{A color-composite image of the radio galaxy Hercules A. White, yellow and blue show the main galaxy and surrounding galaxies as seen in the visible (optical) light, whereas radio synchrotron emission is shown in red. The supermassive black hole in the center of Hercules A launches powerful collimated outflows (``jets'') which move with velocity close to the speed of light and propagate far into the intergalactic space, before being slowed down by the interactions with surrounding diffuse gas. Image by NASA, ESA, STScI, S. Baum, C. O'Dea, R. Perley, and W. Cotton, from \url{http://www.nasa.gov/mission_pages/hubble/science/hercules-a.html}.}
\end{figure}

\clearpage
\section{Lorentz transform}
\label{sec_lorentz}

This derivation closely follows the one by \citet{resn68}. Our task is now to amend the Galilean transform to satisfy both parts of the Einstein's principle of relativity. 

We consider the same situation with the track frame $S$, the train frame $S'$ and kids playing soccer. We have properly started both the tracks and the train clocks using the broken flag method. The task is to write the transform of the coordinates of the ball while accommodating Einstein's principle of relativity. We will look for this transform in the form:
\begin{eqnarray}
ct'=a_{11}ct+a_{12}x+a_{13}y+a_{14}z\\
x'=a_{21}ct+a_{22}x+a_{23}y+a_{24}z\\
y'=a_{31}ct+a_{32}x+a_{33}y+a_{34}z\\
z'=a_{41}ct+a_{42}x+a_{43}y+a_{44}z,\label{eq_lin_transform}
\end{eqnarray}
where we have now multiplied $t$ and $t'$ by $c$ in both frames so that all space-time coordinates have the same units of length. Therefore, all unknown values $a_{ij}$ are dimensionless numbers which will depend on the velocity between the frames (i.e., the velocity of the train). In matrix form, 
\begin{equation}
\left(\begin{array}{c} ct' \\ x' \\ y' \\ z' \end{array}\right)=\left(\begin{array}{cccc} a_{11} & a_{12} & a_{13} & a_{14} \\ a_{21} & a_{22} & a_{23} & a_{24} \\  a_{31} & a_{32} & a_{33} & a_{34} \\  a_{41} & a_{42} & a_{43} & a_{44} \end{array}\right)\left(\begin{array}{c} ct \\ x \\ y \\ z \end{array}\right).
\end{equation}

If the train is not moving, we know that both the train system and the tracks system are the same, so for $v=0$ the transform matrix is a unit matrix, or identity matrix. For $v\ll c$, we should get the Galilean transform:
\begin{equation}
\left(\begin{array}{c} ct' \\ x' \\ y' \\ z' \end{array}\right)=\left(\begin{array}{cccc} 1 & 0 & 0 & 0 \\ -v/c & 1 & 0 & 0 \\ 0 & 0 & 1 & 0 \\ 0 & 0 & 0 & 1 \end{array}\right)\left(\begin{array}{c} ct \\ x \\ y \\ z \end{array}\right). \label{eq_galilean}
\end{equation}

In writing down equations (\ref{eq_lin_transform}), we have already made a big assumption: that of linearity of space. This assumption follows from the assumed {\bf homogeneity} of space: the choice of the origin of the frame should not matter. What would happen if we had a non-linear term in this transform, for example $x'=bx^2$? Suppose we have a ruler in the frame $S$ which has the coordinates of the ends $x_1=0$ and $x_2=1$. In the moving frame, we will measure $x_2'-x_1'=b$. Now we displace the ruler slightly, to a location with $x_1=1$ and $x_2=2$. It is still the same object, but now in the moving frame, we will measure $x_2'-x_1'=3b$, so the measured length of the ruler now depends on its location. 

This is exactly what will happen in General Relativity: properties of space-time will depend critically on how far the location is from a gravitating object. Therefore, in General Relativity the assumption of linearity will no longer be valid, and the fabric of space-time is no longer homogeneous. For Special Relativity, the assumption of linearity will greatly simplify our lives, but it must be understood that this is just an assumption, which happens to be true in empty space but not around massive objects. 

We will now simplify our task even further by using some of the properties of the coordinate systems that we have set up and some other properties of space. Because the $x$ axis is aligned with $x'$ and the tracks have $y'=0, z'=0$ for all $x', ct'$ and the train has $y=0, z=0$ for all $x, ct$, we can set $a_{31}=0, a_{32}=0, a_{41}=0$ and $a_{42}=0$. This would no longer be true if the direction of motion was not along the $x$-axis; we will have examples of such motion in Section \ref{sec_notx}. 

Now imagine a ruler sticking out of the train parallel to the grass, i.e., in the $y'$ direction, with $z'=0$. As the train is moving and the ruler is touching the tips of the grass, we would like for this to be true in the track frame as well. Therefore, the ruler will not twist upwards and it will remain with $z=0$ in the tracks frame. Thus we can set $a_{34}=0$. A similar argument can be made for $a_{43}=0$. 

We now have relationships for coordinates perpendicular to the tracks: $y'=a_{33}y$ and $z'=a_{44}z$. We consider a ruler with length $L$ in the tracks frame $S$ lying on the grass perpendicular to the tracks. According to our transform, we will measure $a_{33}L$ in the train frame. Now we put the same ruler on the train; according to our transforms, we will measure $L/a_{33}$ in the tracks frame. But consider the situation from the ruler's point of view: as far as it is concerned, the only difference between these two cases is that the frames in which it is being measured are moving in opposite directions. Thus, if we would like our space to be {\bf isotropic} we have to postulate that $a_{33}=1/a_{33}$ and thus $a_{33}=1$ (and similarly $a_{44}=1$). 

We deal with the first equation: $ct'=a_{11}ct+a_{12}x+a_{13}y+a_{14}z$. Suppose we have clocks in the tracks frame $S$ which are on either side of the tracks: one clock is at $(x,y,z)$ and another one is at $(x,-y,z)$. If the space is {\bf symmetric} relative to the tracks, both those clocks should be showing the same time in the train frame. Thus, $a_{13}=0$ (and similarly $a_{14}=0$). 

As for the second equation, $x'= a_{21}ct+a_{22}x+a_{23}y+a_{24}z$. For any $y$, $z$ and $t$, a point moving with $x=vt$ in the $S$ frame should be at rest ($x'=0$) in the $S'$ frame (this describes the movement of frame $S'$ relative to frame $S$). Therefore, our equations (\ref{eq_lin_transform}) reduce to
\begin{eqnarray}
x'=a_{22}(x-vt)\\
ct'=a_{11}ct+a_{12}x\\
y'=y\\
z'=z.\label{eq_lin_simple}
\end{eqnarray}
So far we have only used the properties of space and the definitions of the coordinate systems. In classical mechanics, the space is homogeneous and isotropic and symmetric, so Galilean transform satisfies these equations with $a_{22}=1$, $a_{11}=1$ and $a_{12}=0$. But it is now time to use the postulate that the speed of light is a fundamental constant equal to $c$ in all frames. 

As the flag breaks against the flower pot, it makes a spark which sends out a spherical light wavefront in all directions. Because its propagation speed is the same in all frames and equal to $c$, we can write the equations for the wavefront in both frames:
\begin{eqnarray}
x^2+y^2+z^2=c^2t^2\mbox{ in frame $S$}\label{eq_lin_wave1}\\
x'^2+y'^2+z'^2=c^2t'^2\mbox{ in frame $S'$}.\label{eq_lin_wave2}
\end{eqnarray}
The idea is to get rid of all $S'$ values in the latter by using the transform equations (\ref{eq_lin_simple}) and compare to the former. Plugging transform equations into (\ref{eq_lin_wave2}), we find
\begin{equation}
a_{22}^2(x-vt)^2+y^2+z^2=(a_{11}ct+a_{12}x)^2.
\end{equation}
We open parentheses and rearrange the terms to make this more directly comparable to (\ref{eq_lin_wave1}): 
\begin{equation}
x^2(a_{22}^2-a_{12}^2)+y^2+z^2+2xt(-va_{22}^2-ca_{11}a_{12})=(a_{11}^2c^2-a_{22}^2v^2)t^2.
\end{equation}
But the wavefront should satisfy equation (\ref{eq_lin_wave1}); therefore
\begin{eqnarray}
a_{22}^2-a_{12}^2=1\\
-va_{22}^2-ca_{11}a_{12}=0\\
a_{11}^2c^2-a_{22}^2v^2=c^2.
\end{eqnarray}
There are three equations for three unknown coefficients. This system is solvable after some algebra; for example, we can express $a_{12}$ as a function of $a_{22}$ from the first equation and $a_{11}$ as a function of $a_{22}$ from the third equation and plug them both into the second equation to find $a_{11}=a_{22}=1/\sqrt{1-v^2/c^2}$ and $a_{12}=-\frac{v/c}{\sqrt{1-v^2/c^2}}$.

We define {\bf the Lorentz factor}
\begin{equation}
\gamma=\frac{1}{\sqrt{1-v^2/c^2}}
\end{equation}
and dimensionless velocity measured in units of the speed of light:
\begin{equation}
\beta=\frac{v}{c}. 
\end{equation}
Using these definitions, we arrive at the {\bf Lorentz transform equations} in their most popular form:
\begin{equation}
\left(\begin{array}{c} ct' \\ x' \\ y' \\ z' \end{array}\right)=\left(\begin{array}{cccc} \gamma & -\gamma\beta & 0 & 0 \\ -\gamma\beta & \gamma & 0 & 0 \\ 0 & 0 & 1 & 0 \\ 0 & 0 & 0 & 1 \end{array}\right)\left(\begin{array}{c} ct \\ x \\ y \\ z \end{array}\right).\label{eq_lorentz1}
\end{equation}

\begin{figure}[htb]
\centering
\includegraphics[scale=0.8, clip=true, trim=0cm 10cm 0cm 0cm]{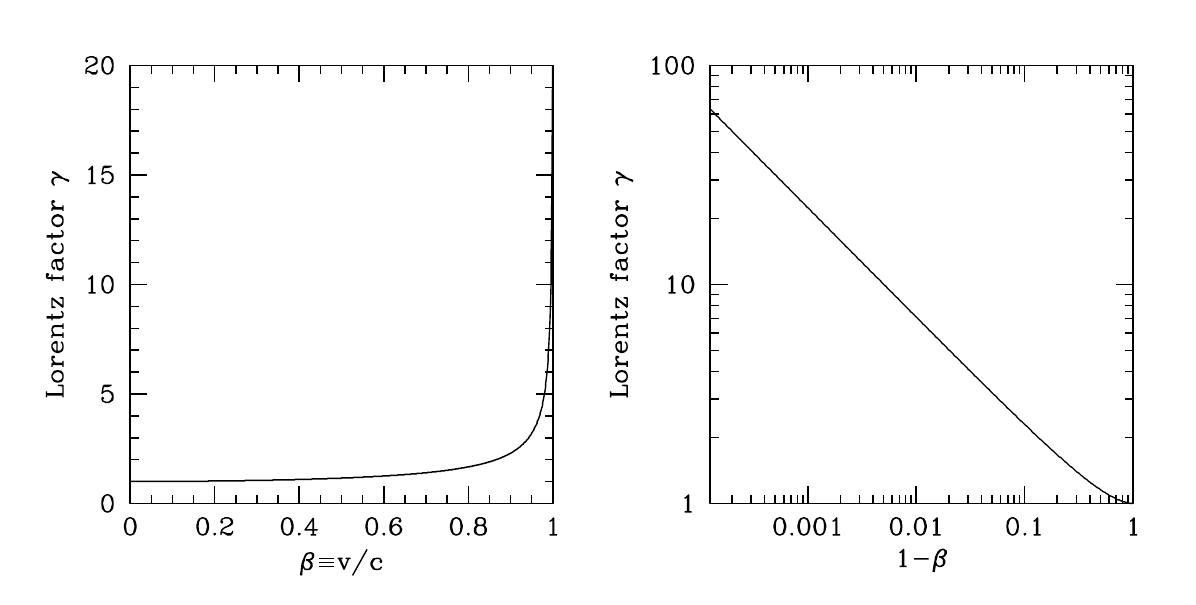}
\caption{Lorentz factor in linear and logarithmic coordinates plotted against the velocity and the difference between the speed of light and the velocity.}
\label{pic_gamma}
\end{figure}

In Figure \ref{pic_gamma} we show the behavior of the Lorentz factor as a function of the velocity. It stays close to unity (non-relativistic case) for a wide range of velocities; for example, for $\beta=0.5$ we find $\gamma=1.15$, only a 15\% change from the non-relativistic value, even though the velocity is half the speed of light. However, the closer we get to the speed of light, the stronger the relativistic effects, and $\gamma$ increases very rapidly. For $\beta=0.999$, we find $\gamma=22$. For {\bf ultra-relativistic velocities}, when $1-\beta\ll1$, we can approximate
\begin{equation}
\gamma=\frac{1}{\sqrt{(1+\beta)(1-\beta)}}\simeq\frac{1}{\sqrt{2(1-\beta)}}.\label{eq_gamma_ultrarel} 
\end{equation}
Thus for ultra-relativistic velocities, $\gamma$ as a function of $1-\beta$ is a power-law which appears as a straight line on the log-log plot of the right panel of Figure \ref{pic_gamma}. For these particles, we can approximate $1-\beta=1/(2\gamma^2)$ using equation (\ref{eq_gamma_ultrarel}), and therefore for large values of $\gamma$ it is inconvenient to talk about their velocities measured in m/sec -- it makes more sense to think in terms of how close their velocity is to the speed of light. 

At the Large Hadron Collider, protons are accelerated to Lorentz factors $\gamma=7000$, and therefore $1-\beta=10^{-8}$; but those are not by any stretch of the imagination the fastest moving particles we know of: ultra-relativistic cosmic rays, whose origin remains poorly understood, arrive at Earth with Lorentz factors $\gamma>5\times 10^{10}$.

\clearpage
\begin{figure}[!htb]
\centering
\includegraphics[scale=0.8]{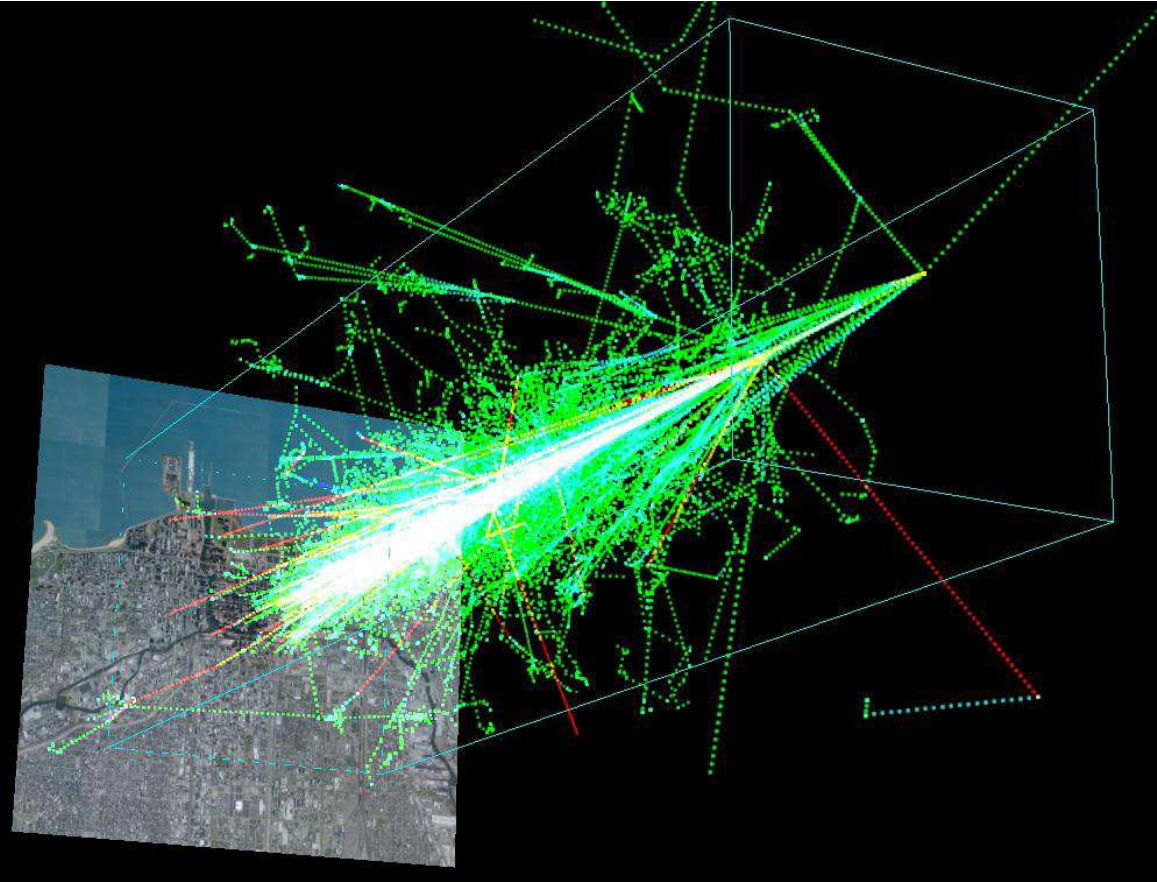}
\caption{A simulation of a cosmic ray shower produced when a proton with energy 1 TeV hits the atmosphere about 20 km above the ground. The shower is in a 20 km $\times$ 5 km $\times$ 5 km box superimposed on a scale map of Chicago's lakefront. Image by Maximo Ave et al., from \url{http://astro.uchicago.edu/cosmus/projects/aires/}.}
\label{pic_crsim}
\end{figure}

\clearpage
\section{Using the Lorentz transform: length contraction, time dilation}

\subsection{Some introductory comments}

We heroically derived the Lorentz transform, and for the next couple of lectures we will try to understand what it means and how to use it to calculate things. We now have the essential language of Special Relativity defined, and all of these terms will now enter our active vocabulary:
\begin{itemize}
\item Reference frame;
\item Inertial frames;
\item Principle of relativity;
\item Galilean transform;
\item Maximal signal propagation speed;
\item Einstein's principle of relativity;
\item Physical event;
\item Invariant;
\item Space-time coordinates of an event;
\item Lorentz transform;
\item Lorentz factor and dimensionless velocity.
\end{itemize}

Here are some typical questions that are asked by students in the aftermath of the first lecture. 

{\bf Q.} Why do we care? In what area of physics do we really have to deal with trains and soccer balls moving with velocities close to the speed of light? -- {\bf A.} We care for two reasons. One is that in astrophysics and particle physics we routinely encounter motions that are close to the speed of light; thus even the contrived mechanical problems we are considering here actually have direct applicability in these areas. Instead of trains and soccer balls, we will start to increasingly consider problems of more direct relevance to these fields. In everyday life, modern global positioning system technology in your phone needs to incorporate relativistic corrections in order to accurately show you on the map. But the more important reason is that, as we saw, Special Relativity is the foundation of electromagnetism and many other swaths of physics. For example, we cannot understand why the sky is blue without Special Relativity (the sky is blue because the particles scatter sunlight much more efficiently at shorter wavelengths, but an understanding of the scattering process requires Special Relativity). 

{\bf Q.} In MKS, Maxwell's equations contain $\epsilon_0$ and $\mu_0$ instead of $c$. How do we know which one of these three values is the ``fundamental constant of nature''? Why is $c$ any better than $\epsilon_0$ and $\mu_0$ for this purpose? -- {\bf A.} Following the steps we used to derive electromagnetic waves from the Maxwell's equations, we find that the speed with which they propagate is equal to $1/\sqrt{\mu_0\epsilon_0}$, which we denote $c$. There are indeed three constants in MKS, but they are related through one constraint, thus there are only two independently defined values, for example $c$ and $\mu_0$, with the third one defined via the first two, $\epsilon_0=1/(\mu_0 c^2$). 

Still, it is clear that electromagnetism in MKS has one extra constant compared to electromagnetism in cgs which has only one constant ($c$) in its equations. To understand what happened here, we go back to Coulomb law for two identical charges expressed in cgs units $F=q^2/r^2$ and in MKS units $F=q^2/(4 \pi \epsilon_0 r^2)$. In cgs, two identical charged particles placed 1 cm apart exerting force of 1 dyne on each other must have charge equal to 1 cgs charge unit. The cgs unit of charge is measured in $\sqrt{{\rm gram}\times{\rm cm}^3/{\rm sec}^2}$, which involves a square root of other units, which students find weird. But, really, why is this any weirder than the units of velocity (cm/sec), momentum (g$\times$cm/sec) or kinetic energy (g$\times$cm$^2$/sec$^2$)? In these units, an electron has charge $4.8\times 10^{-10}$ cgs units. In contrast, in MKS charge has its own unit (Coulomb) which is defined independently of meters, kilograms and seconds for historical reasons. Thus an extra constant is necessary to relate the Coulomb to the previously defined units of force and distance: two charges one Coulomb each placed 1 m apart exert a force of $8.98755...\times 10^{9}$ Newtons. Therefore, $\epsilon_0$ is simply a constant that converts one set of units of charge to another -- it is analogous to the number 100 used to convert centimeters (units of length in cgs) to meters (units of length in MKS), even though it is not a dimensionless number. 

{\bf Q.} Because the maximal signal speed is finite, do events actually happen before we observe them? -- {\bf A.} The short answer is yes. But that really does make things more complicated, doesn't it. Suppose the kids are playing soccer some distance away, then how do we make a measurement of the trajectory of the ball, if the light takes time to propagate to us and therefore is delayed by different amounts depending on where the ball is? All of a sudden the task of measuring the trajectory of the ball in one system (forget any moving ones!) became very complicated. 

One way to do this is to turn this problem into events. I like to break things, because then you know it really did happen. So we put down a coordinate grid and synchronized clocks everywhere on the soccer field. A girl kicks the ball, the ball rolls on the ground knocking down clocks, which break and stop. Then we can walk around the soccer field collecting the broken clocks and recording the time that they are showing and the coordinates where they were found. Now we know where the ball was: $x(t), y(t), z(t)$. Doing the same thing in the frame of the train requires more imagination, because now the coordinate grid and the clocks are moving together with the train, but nonetheless the same approach will allow us to measure $x'(t'), y'(t'), z'(t')$. 

\begin{figure}[htb]
\centering
\includegraphics[scale=0.8, clip=true, trim=3cm 12cm 2cm 11.5cm]{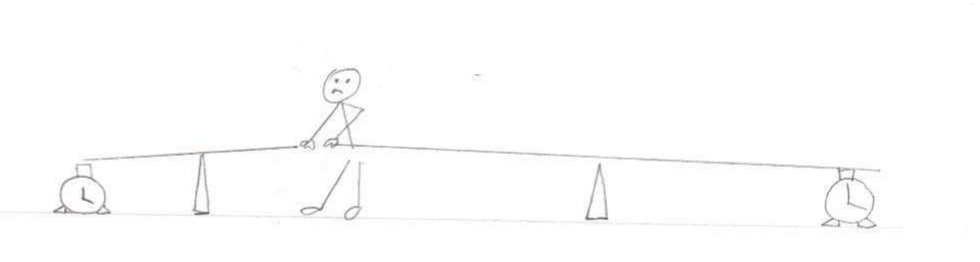}
\caption{An unsuccessful attempt to synchronize clocks. The student is trying to use the contraption to push the start buttons ``at the same time'' but did not take into account the speed of the propagation of signals along the seesaws.}
\label{pic_seesaw}
\end{figure}

But now we have hit another impediment: we have all along been talking about {\bf synchronized clocks} in frame $S$, but how exactly do we accomplish that? Enterprising students at this point typically come up with a clever contraption of long seesaws (Figure \ref{pic_seesaw}), with one end of each near the student and the other end hovering above the start button of a clock on the coordinate grid. The near ends of the seesaws are all tied together with a rope, so at some point the student pulls on the rope, lifting the near ends of the seesaws, lowering the far ends which whack the start buttons on the clocks. 

Are the clocks properly synchronized? No. Remember that there exists a maximal speed of signal propagation. So no matter how neatly you tie the seesaws together, the far ends of them cannot know instantaneously that the near ends have been pulled upwards. In fact, the signals in mechanical systems propagate much slower than the speed of light; the typical speeds involved are on the order of the speed of sound in the solids, which are only a few km/sec. 

The best way to synchronize clocks is to use the fundamental maximal signal propagation speed. At time $t=0$, we send out a spherical electromagnetic wavefront from the station $(x=0,y=0,z=0)$. When it reaches an observer at $(x,y,z)$, the observer sets the clock to $\sqrt{x^2+y^2+z^2}/c$. This clock is now synchronized with the clock at the station in system $S$. If you are standing midway between two clocks (to cancel out light propagation effects) and take a picture of each of the two clocks at the same time, the pictures will show the same thing. 

{\bf Q.} Do we really need linear algebra for this? -- {\bf A.} No. Resnick's book will teach you to do all we need to do in this class without the linear algebra. However, Special Relativity calculations are more transparent and elegant when done in the matrix notation, and as the complexity of the problems increases it becomes more and more difficult to do them without linear algebra. The specific terms for typical matrices in Special Relativity are 4-vectors and tensors; we will have some examples later of calculations done with and without the 4-vector apparatus. 

Some of the properties of the matrix multiplication that we will need in this course:
\begin{itemize}
\item Two matrices can be multiplied if and only if the number of columns of the first matrix is equal to the number of rows of the second one. If the first matrix has dimensions $l\times m$ and the second one $m\times n$, the product has dimensions $l\times n$.
\item In the matrix product $C=AB$, the $ij$ element of the product is $C_{ij}=\sum_{k}A_{ik}B_{kj}$. 
\item Matrix multiplication is associative: $A(BC)=(AB)C$. 
\item Matrix multiplication is distributive: $(A+B)C=AC+BC$.
\item Matrix multiplication is in general not commutative: $AB\ne BA$. 
\end{itemize}

\begin{prob}
Using the definition of matrix multiplication, prove that the matrix multiplication is associative. 
\end{prob}

\begin{prob} 
The frame $S$ and and the frame $S'$ are not moving relative to one another, and the clocks are synchronized. Suppose we have an event with space-time coordinates $(ct',x',y',z')$. Write down in matrix form the transformation from the frame $S'$ into the frame $S$ for the following cases:
\begin{itemize}
\item The origins of the frames $S'$ and $S$ coincide, the $x'$ axis is aligned with the $y$ axis, the $y'$ axis is aligned with the $z$ axis and the $z'$ axis is aligned with the $x$ axis. 
\item The origins of frames $S'$ and $S$ coincide, their $z$ axes are aligned with one another, but the $x'-y'$ plane of $S'$ is rotated by an angle $\phi$ counter-clockwise relative to the the $x-y$ plane of $S$.
\item The axes are aligned, but the origin of the coordinate system $S'$ is located at $(x_0,y_0,z_0)$ as measured in $S$.
\end{itemize} 
\end{prob}

\begin{prob} 
The frame $S'$ moves with velocity $v$ relative to the frame $S$ along the $x$ axis, and the origins of the two frames are synchronized. What is the velocity of the frame $S$ relative to the frame $S'$? Suppose we have an event with space-time coordinates $(ct',x',y',z')$ in the frame $S'$. Use appropriate transformations to shift from $S'$ to $S$ (i.e., find the space-time coordinates of this event in $S$) and back to $S'$. Show explicitly that you recover the space-time coordinates that you started with. -- {\bf Solution.} The frame $S$ moves with velocity $-v$ relative to the frame $S'$. We have derived the transform from $S$ to $S'$; in order to derive the transform from $S'$ to $S$, we can find the inverse of the transform matrix in equation (\ref{eq_lorentz1}) -- or simply flip the sign of velocity. Indeed, the relativity principle tells us that the frames $S$ and $S'$ are equivalent, thus the transform from one to the other should depend only on the relative velocity between the two. Thus, we find another popular form of the Lorentz transform:
\begin{equation}
\left(\begin{array}{c} ct \\ x \\ y \\ z \end{array}\right)=\left(\begin{array}{cccc} \gamma & \gamma\beta & 0 & 0 \\ \gamma\beta & \gamma & 0 & 0 \\ 0 & 0 & 1 & 0 \\ 0 & 0 & 0 & 1 \end{array}\right)\left(\begin{array}{c} ct' \\ x' \\ y' \\ z' \end{array}\right).\label{eq_lorentz2}
\end{equation}
Now we verify that transforming from $S'$ to $S$ and back to $S'$ will let us recover the same coordinates with started with. To calculate the final space-time coordinates in $S'$ we need to calculate a matrix product
\begin{equation}
\left(\begin{array}{cccc} \gamma & -\gamma\beta & 0 & 0 \\ -\gamma\beta & \gamma & 0 & 0 \\ 0 & 0 & 1 & 0 \\ 0 & 0 & 0 & 1 \end{array}\right) \left[\left(\begin{array}{cccc} \gamma & \gamma\beta & 0 & 0 \\ \gamma\beta & \gamma & 0 & 0 \\ 0 & 0 & 1 & 0 \\ 0 & 0 & 0 & 1 \end{array}\right)\left(\begin{array}{c} ct' \\ x' \\ y' \\ z' \end{array}\right)\right].
\end{equation}
Using the associative property of matrix multiplication, we know that we can multiply the 4x4 matrices first. Their product is
\begin{equation}
\left(\begin{array}{cccc} \gamma^2-\beta^2\gamma^2 & \beta\gamma^2-\beta\gamma^2 & 0 & 0 \\ -\beta\gamma^2+\beta\gamma^2 & -\beta^2\gamma^2 +\gamma^2& 0 & 0 \\ 0 & 0 & 1 & 0 \\ 0 & 0 & 0 & 1 \end{array}\right).
\end{equation}
Remembering now the relationship between $\gamma$ and $\beta$ we find that this is a unit matrix. Thus doing Lorentz transform twice with opposite velocities is equivalent to no transform at all, and we recover the initial space-time coordinates.
\label{prob_inverse}
\end{prob}

It will become progressively more inconvenient to only work in perfectly synchronized systems. Without any loss of generality and because of the linearity of the Lorentz transform, we may place the station at non-zero coordinates, or set the train clock during flag breaking to a non-zero time. Then everything we have said above would apply to the difference between the space-time coordinates of the event in question and of the flag-breaking event. In fact, there is nothing special about the flag-breaking event, either. Thus, the differences between space-time coordinates of any two physical events should transform according to the Lorentz transform:
\begin{eqnarray}
\left(\begin{array}{c} c\Delta t' \\ \Delta x' \\ \Delta y' \\ \Delta z' \end{array}\right)=\left(\begin{array}{cccc} \gamma & -\gamma\beta & 0 & 0 \\ -\gamma\beta & \gamma & 0 & 0 \\ 0 & 0 & 1 & 0 \\ 0 & 0 & 0 & 1 \end{array}\right)\left(\begin{array}{c} c\Delta t \\ \Delta x \\ \Delta y \\ \Delta z \end{array}\right)\mbox{ and }\label{eq_lorentz_final1}\\
\left(\begin{array}{c} c\Delta t \\ \Delta x \\ \Delta y \\ \Delta z \end{array}\right)=\left(\begin{array}{cccc} \gamma & \gamma\beta & 0 & 0 \\ \gamma\beta & \gamma & 0 & 0 \\ 0 & 0 & 1 & 0 \\ 0 & 0 & 0 & 1 \end{array}\right)\left(\begin{array}{c} c\Delta t' \\ \Delta x' \\ \Delta y' \\ \Delta z' \end{array}\right).\label{eq_lorentz_final2}
\end{eqnarray}

By comparing the matrix forms for the Galilean transform (\ref{eq_galilean}) and for the Lorentz transform (\ref{eq_lorentz_final1}-\ref{eq_lorentz_final2}) we notice that the Lorentz matrix is symmetric, whereas the Galilean one is not. This tells us something profound about the nature of time and space in Special Relativity. In classical mechanics, time is absolute. In Special Relativity, time is intertwined with space and no longer flows the same way in all reference frames. Events that are simultaneous in one frame are not necessarily simultaneous in the other frame. Events that happen in one order can be happening in another order in a different frame. How do we sort out this mess? Because our intuition is so profoundly based on the everyday notion of absolute time, we must be careful, and defining events (i.e., breaking things) will be enormously helpful. 

\subsection{Length contraction}

We are on the train, in the frame $S'$. We can lay down the ruler and measure the length of the train $L$. This is called {\bf proper or co-moving length}, since the rulers are moving together with the train. The task is now to measure the train in the track frame. Wait, but what does it mean to measure the length of a moving object? We never had to worry about this in classical mechanics (where as we saw previously the sizes were invariant under Galilean transformation). 

One possibility is to stand far away from the train and to take a photo. If we are equidistant from the front and the back of the train, then the light propagation effects will cancel out. We arrive at the conclusion that to measure the length of a moving object, we want to get the position of the front of the object and the back of the object at the same time. 

To make this clearer, we re-cast the ``measurement'' in terms of events. We set up an axe at the front of the train and one at the back of the train. The observer sees both axes falling down at the same time, and they conveniently make dents on the tracks as they do so. The train passes, the observer comes with the ruler and measures the distance between the dents. This is the observed length of the train in the tracks frame; we calculate it in terms of the proper length $L$ and the train speed $v$. 

The two events are the two axes dropping down at the front and the back of the train. What do we know about them? We know that the observer saw them happening simultaneously, with $\Delta t=0$. We know that in the co-moving frame the length of the train is $\Delta x'=L$. We will use the Lorentz transform in the form
\begin{equation}
\left(\begin{array}{c} c\Delta t' \\ \Delta x' \\ \Delta y' \\ \Delta z' \end{array}\right)=\left(\begin{array}{cccc} \gamma & -\gamma\beta & 0 & 0 \\ -\gamma\beta & \gamma & 0 & 0 \\ 0 & 0 & 1 & 0 \\ 0 & 0 & 0 & 1 \end{array}\right)\left(\begin{array}{c} c\Delta t \\ \Delta x \\ \Delta y \\ \Delta z \end{array}\right).\label{eq_prob_length}
\end{equation} 
In particular, the second equation reads $\Delta x'=-\beta\gamma c\Delta t + \gamma \Delta x$. Since $\Delta t=0$, we get $\Delta x'=\gamma \Delta x$. The observed length of the train 
\begin{equation}
\Delta x=\frac{L}{\gamma}
\end{equation}
is always smaller than the proper length because $\gamma>1$. This phenomenon is called {\bf length contraction}. 

Here are some of the questions that come up in the aftermath of this calculation:

{\bf Q.} This seems like a trick: how did you know to use the Lorentz transform this way and not the other way? -- {\bf A.} Let us try to use the other one. 
\begin{equation}
\left(\begin{array}{c} c\Delta t \\ \Delta x \\ \Delta y \\ \Delta z \end{array}\right)=\left(\begin{array}{cccc} \gamma & \gamma\beta & 0 & 0 \\ \gamma\beta & \gamma & 0 & 0 \\ 0 & 0 & 1 & 0 \\ 0 & 0 & 0 & 1 \end{array}\right)\left(\begin{array}{c} c\Delta t' \\ \Delta x' \\ \Delta y' \\ \Delta z' \end{array}\right).
\end{equation}
We still know the same two things, $\Delta t=0$ and $\Delta x'=L$ and we still need to find $\Delta x$. The first two equations can then be written as 
\begin{eqnarray}
0=\gamma c \Delta t'+\beta \gamma L\\
\Delta x = \beta \gamma c \Delta t' + \gamma L.
\end{eqnarray}
There are two equations for two unknowns ($\Delta x$ and $\Delta t'$) which can be solved to find $\Delta x=L/\gamma$ and $c\Delta t'=-\beta L$. So we recover the same result for the length and we also find that the two events that were simultaneous in the station frame ($\Delta t=0$) are no longer simultaneous in the train frame, which can also be found from the first equation of (\ref{eq_prob_length}). 

We have two events: number 1 is the axe falls off the back of the train; number 2 is the axe falls off the front of the train. In the tracks frame, the two events happen simultaneously ($\Delta t=0$) and they are separated by length $\Delta x=L/\gamma$ between the nicks on the tracks. In the train frame, $c \Delta t'=-\beta L$ and $\Delta x'=L$. Thus event number 2 happens before event number 1. So the axe falls off the front of the train first, then the last car drives some more and the axe falls of the back of the train. So the people on the train are saying to each other: ``Well, of course those silly people on the platform will measure a shorter length! They waited a while for the back of the train to catch up to the measurement of the front of the train!'' But for the people on the platform doing things this way is the only way that makes sense to conduct the measurement of length. 

{\bf Q.} Is this real? Or are we somehow imagining that the train is shorter, but it is actually the same length? -- {\bf A.} If we take a picture of the train (when its ends are equidistant from us, so as to cancel out propagation events), on the picture it will appear with the length $L/\gamma$. If we stand closer to the train and take pictures of people on the train, the people will appear squashed in the direction parallel to the direction of motion. If there are two sign-posts $L/\gamma$ apart next to the tracks, then there will be a moment in time when the entire train is observed to fit between the sign-posts. Thus, this is as real as it gets. 

There is a big difference between the length parallel to the direction of motion and other directions. If the ruler on the train is parallel to the train and we take its picture, it will appear shorter on the photos. If however the ruler is sticking vertically up from the roof in the positive $z$ direction and we take a picture, the ruler will appear with length $L$. 

{\bf Q.} How do the people on the train know when to pull the axes down? This seems artificial. -- {\bf A.} Because in the frame of the train the axes do not fall down at the same time, we need to make arrangements with the people on the train, so that they can time the axes exactly right so that we will see the two events simultaneously in the tracks frame. If we are unhappy making the arrangements with the people on the train, we can push our thought experiment a bit further. In the station frame, we can set up a flash of light (from a source that is equidistant from the front and the back of the train, to cancel out propagation effects, or a plane-parallel light wave propagating along the y axis and hitting all points on the train at the same time). The light reaches the front and the back of the train, activates a photon-sensitive switch, and the axes fall simultaneously in the station frame. There is no need to talk to those squashed people on the train after all. 

{\bf Q.} In classical mechanics, I can time how long it takes for an object to pass by some fixed point, then multiply by $v$ to get the length. Does this work here? -- {\bf A.} Event number 1 is the front of the train is passing by the station. Event number 2 is the back of the train is passing by the station. (If you are not convinced that these are events, you can always break more flags.) In the $S'$ frame, the distance between these two events is the proper length of the train: $\Delta x'=-L$ (we use a negative sign because of our coordinate convention: the front of the train is $x'_1=L$, the back of the train is $x'_2=0$, thus $\Delta x'=x'_2-x'_1=-L$). We also know that the observer does not move: $\Delta x=0$. To make the suggested measurement, we first need to find the elapsed time $\Delta t$. We use the second equation of the Lorentz transform from $S$ to $S'$ (\ref{eq_lorentz_final1}): $\Delta x'=-\gamma\beta c\Delta t +\gamma \Delta x$. Plugging in $\Delta x'=-L$ and $\Delta x=0$, we find $\Delta t=L/(c\gamma\beta)$. The new method of length measurement calls for multiplying this by $v=\beta c$ to get the length; doing this results in a length measurement$=L/\gamma$. Thus, this method (timing the front and the back of the train) gives us the same measurement of length as the first method (simultaneous measurements of the positions of the front and the back). 

\subsection{Time dilation}
\label{sec_timedil}

A student is standing in the last car of the train with a vase in his hands. We synchronize the train and the tracks systems using the broken flag method, and thus we set all clocks to show 0 at that moment. The student waits an amount of time $\tau$ after the synchronization event and smashes the vase. Pieces of the vase fly out the window and break some clocks in the tracks frame. What are these clocks showing? 

We have two events, event 1 is the flag breaking and event 2 is the vase breaking. We know that $\Delta t'=\tau$ (this is called {\bf the proper, or co-moving, life-time} of the vase). The student did not move, so $\Delta x'=0$. Then we can find the elapsed observer time from the first equation of the $S \rightarrow S'$ Lorentz transform in the form (\ref{eq_lorentz_final2}): $c \Delta t=\gamma c \Delta t'+\beta \gamma \Delta x'$, giving us
\begin{equation}
\Delta t=\gamma \tau. 
\end{equation}
Thus the elapsed time is longer than the proper time; the vase is observed to have survived longer in the non-moving frame. 

\begin{prob}
A muon ($\mu^-$) is created in a high-energy collision of a primary cosmic ray in the Earth’s atmosphere 10 km above sea level. The muon descends vertically with velocity 0.999c and then decays. In its proper frame, its lifetime is 2.2 microseconds. At what altitude above sea level does it decay? -- {\bf Solution.} Let us first talk about about the concept of the ``lifetime'' of an elementary particle. Elementary particles obey laws of quantum electrodynamics; as you will learn in your future classes, many quantum mechanical problems have a probabilistic character to them. When we say that the proper lifetime of an elementary particle is $\tau$, we do not mean that all particles live exactly $\tau$ when at rest in the lab frame and then decay. Typically, we start off with some (large) number of particles, and as time goes on some of them decay and fewer and fewer remain, with the number of remaining particles following an exponential function appropriate for this probabilistic process. Thus by ``lifetime'' one typically means either the mean lifetime of the population of these particles, or the half lifetime after which half of all particles have decayed. 

Having said that, we assume that this particular muon really lived in its proper frame for $\tau=$2.2 microseconds. In the observed frame it has lived $\gamma \tau$ and it has traveled $\gamma \tau v$. Plugging in all the values, we find $\gamma=22.4$ and the distance traveled is 14.8 km. Not only would this particle reach the surface, but it would penetrate into the ground! Without time dilation, it would have decayed after traveling 0.7 km -- still in the upper atmosphere.

Taking into account the distribution of proper lifetimes and Lorentz factors, we might expect that if there were no time dilation, few muons should reach the ground, whereas if time dilation is present we might expect many. In the 1940s, in the Rossi-Hall experiment, counting muons far from their source confirmed the presence of time dilation in agreement with Special Relativity and enabled the measurement of the muon's proper lifetime. 
\end{prob}

{\bf Q.} Waaait a second! Now we have a frame where time moves faster and another frame where time moves slower. Surely this cannot square with the relativity principle! -- {\bf A.} To demonstrate that the train and the tracks frames are equivalent from the standpoint of the relativity principle, we need to set up the same experiment in either system. In experiment number 1, we are standing at the station and taking pictures of the moving train, $T_0$ minutes apart. We see some squashed clocks in our pictures; what are they showing? Let us find out. We know $\Delta t=T_0$ between the pictures and the observer did not move, so $\Delta x=0$. From the first equation of the Lorentz transform (\ref{eq_lorentz_final1}), we find $c \Delta t'=cT_0\gamma$. Thus if I take one picture at 1 minute after frame synchronization, the squashed train clock in my picture will show time $\gamma\times$1 minute. 

Experiment number 2 is set up in a very similar way, but we switch the frames. We are on the train, in the last car, with fields and stations flying by in the windows. We are taking pictures $\tau$ minutes apart. The stations and people and clocks appear squashed in our pictures. What are the clocks showing? We have $\Delta t'=\tau$ between the pictures, while $\Delta x'=0$ (we are standing in the last car of the train). From the first equation of the Lorentz transform (\ref{eq_lorentz_final2}) we find $\Delta t=\gamma \tau$. Thus if we take a picture 1 minute after synchronization, the clocks outside the train will show $\gamma\times$1 minute (this experiment is the same as the vase smashing problem). 

Therefore, when we are at the station taking pictures of the train the clocks appear squashed and show greater elapsed time. Similarly, when we are on the train taking pictures of the outside stations, the station clocks appear squashed and show greater elapsed time. The same experiment leads to the same result regardless of which frame we are in, and therefore the principle of relativity is satisfied. 

The problems we have had in this section all use the notion of the {\bf thought experiment}, which is a problem setup that does not contradict the postulates of our theory, but is often carried out to an extreme degree and thus impractical. We can make people moving with very high speeds ($v=0.9c$, for example) part of our thought experiment. It is not practical, but can be done in principle. However, we cannot make reference frames move relative to one another with a speed greater than the speed of light. This would contradict the postulates of our theory (that there exists maximal propagation speed which is the same in all frames) and would be unphysical.  

\subsection{Motions not along the $x$ axis}
\label{sec_notx}

All equations so far have been derived under the assumption that the frame $S'$ is moving in the positive $x$ direction as seen in the $S$ frame. While a generalized Lorentz transform can be written for frames moving at an arbitrary direction as seen in the $S$ frame (see for example the Wikipedia page\footnote{\url{http://en.wikipedia.org/wiki/Lorentz_transformation}}), the simplest way to deal with this issue is to introduce a new system of coordinates in the $S$ frame such that the $S'$ frame is moving along the positive $x$ axis. In particular, often the problems are given without coordinates explicitly introduced; the first step in such problems (e.g., Problem \ref{prob_lorentz}) is to identify the $S$ and the $S'$ frames and to set up the $x$ and the $x'$ axes parallel to the relative motion of the two frames. 

On a rare occasion something moves along the $y$ or the $z$ axis, and then we can easily anticipate what the Lorentz transforms should look like. For example for the motion along the positive $y$ axis:
\begin{equation}
\left(\begin{array}{c} c\Delta t \\ \Delta x \\ \Delta y \\ \Delta z \end{array}\right)=\left(\begin{array}{cccc} \gamma & 0 & \beta\gamma & 0 \\ 0 & 1 & 0 & 0 \\ \beta\gamma & 0 & \gamma & 0 \\ 0 & 0 & 0 & 1 \end{array}\right)\left(\begin{array}{c} c\Delta t' \\ \Delta x' \\ \Delta y' \\ \Delta z' \end{array}\right).
\end{equation}

Finally, if you absolutely need transforms for the case of arbitrarily directed velocities, then here are the steps. Suppose $S'$ is moving with velocity $\vec{v}=(v_x,v_y,v_z)$ as measured in the $S$ frame, and frame $S$ is moving with velocity $-\vec{v}=(-v_x,-v_y,-v_z)$ as measured in the $S'$ frame. The trick is to introduce ``temporary'' frames $T$ and $T'$ which are stationary rotations of $S$ and $S'$, such that $T$ and $T'$ are moving relative to one another along the $x_T$ and $x_T'$ axes and the regular Lorentz transform applies. 

In step 1, we transform from $S$ to $T$, which is not moving with respect to $S$, but whose $x_T$ axis is directed along $\vec{v}$; it is a stationary rotation. In step 2, we transform from $T$ to $T'$ using equation (\ref{eq_lorentz_final1}). In step 3, we rotate from $T'$ (in which $-x_T'$ axis is directed along $-\vec{v}$) to $S'$. Then you need to multiply the three matrices (rotation -- Lorentz transform -- rotation) to obtain the final equations on Wikipedia.

\clearpage
\begin{figure}[!htb]
\centering
\includegraphics[scale=0.8]{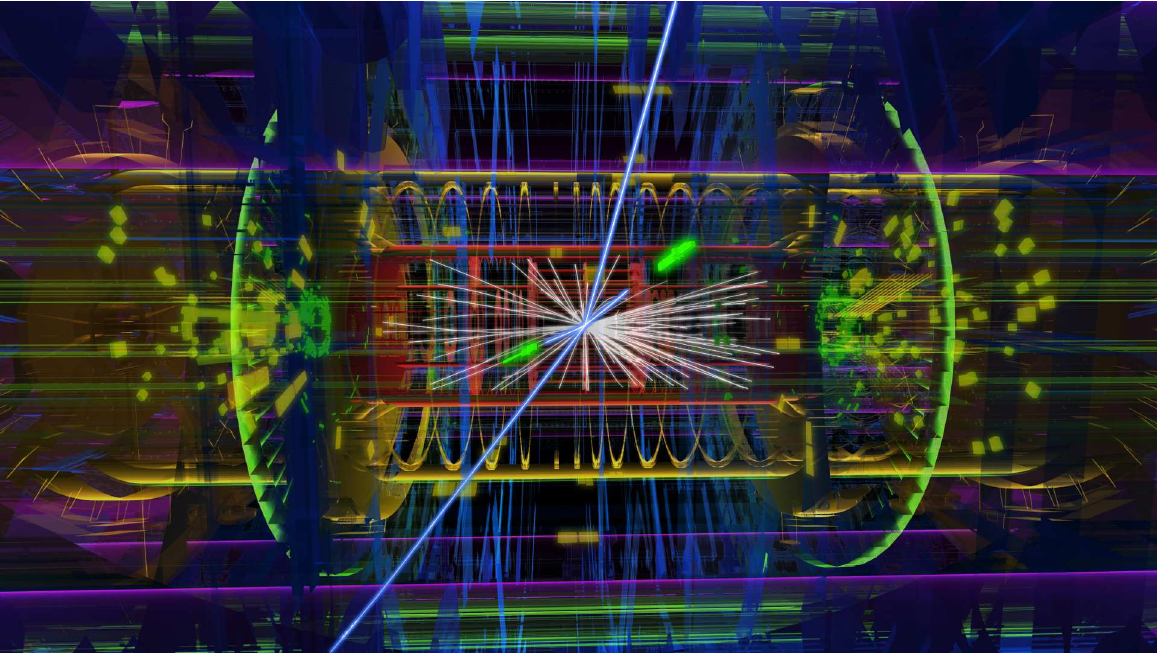}
\caption{An LHC ATLAS (``A Toroidal LHC Apparatus'') event with two muons and two electrons - a candidate for a Higgs-like decay. The two muons are picked out as long blue tracks, the two electrons as short blue tracks matching green clusters of energy in the calorimeters. By ATLAS Experiment, from \url{http://www.quantumdiaries.org/tag/computing/}.}
\end{figure}

\clearpage
\section{Simultaneous events, causally connected events}
\label{sec_time}

\subsection{Minkowski diagram}

\begin{figure}[htb]
\centering
\includegraphics[scale=0.8, clip=true, trim=3cm 18cm 1cm 1cm]{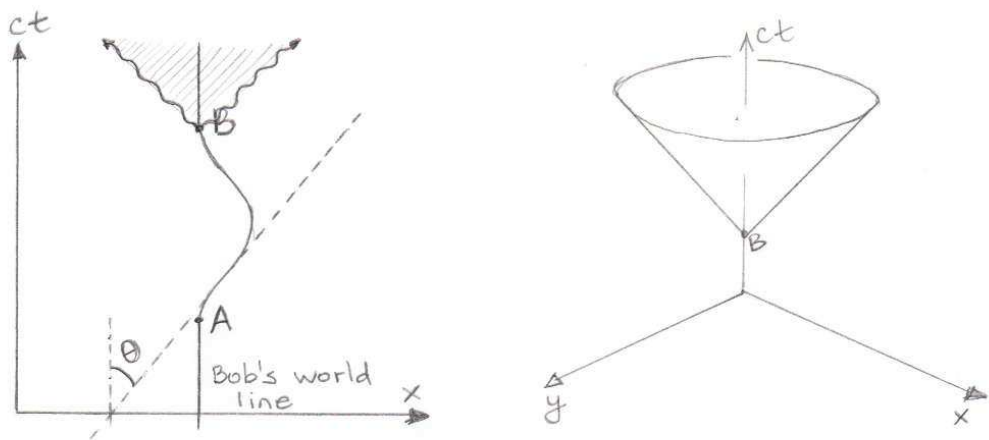}
\caption{Minkowski diagrams for 2-dimensional and 3-dimensional space-time.}
\label{pic_mink}
\end{figure}

{\bf Minkowski diagram} is the visualization of the space-time coordinates of objects and events in the $x$-$ct$ plane. Figure \ref{pic_mink}, left shows the space-time coordinates of a guy named Bob who was sitting on a couch, then got up to get a drink from the fridge (event A), then returned to the couch and sat back down (event B). The locus of space-time coordinates of Bob at all times is called Bob's {\bf world line}. We denote $\theta$ the angle between the $ct$ axis and the tangent line to Bob's world line. Because Bob cannot move faster than the speed of light, the angle $\theta$ can never be greater than 45$^{\rm o}$. 

At any point on his world line, Bob cannot cause something to happen infinitely far away, because signals from Bob cannot propagate faster than the speed of light. Thus for any event in Bob's life (B: Bob sat on the couch) only events within the shaded triangle can be affected by the event. Of course events can happen outside this triangle, but they cannot be caused by the event B. The triangle is bracketed by the world line of photons that Bob emitted when he sat on the couch. They are represented by wavy lines as per the usual convention to distinguish them from massive particles, e.g., Bob. This convention is used in many areas of physics, e.g., in quantum field theory for Feynman diagrams, but the actual world lines of photons are straight in $x-ct$ coordinates.

This triangle is called {\bf the light cone} of the event B. Why a cone? Because we have only considered one spatial dimension ($x$, the coordinate running between the couch and the fridge), but if we do this with the more appropriate 2D coordinates ($xy$ plane -- the plane of Bob's apartment floor), then event B can only cause events within the 3D cone in the $ct, x, y$ space (Figure \ref{pic_mink}, right). Ideally we really need to consider all three spatial dimensions, but this would be rather hard to visualize. 

\begin{figure}[htb]
\centering
\includegraphics[scale=0.7, clip=true, trim=0cm 9.5cm 10cm 0cm]{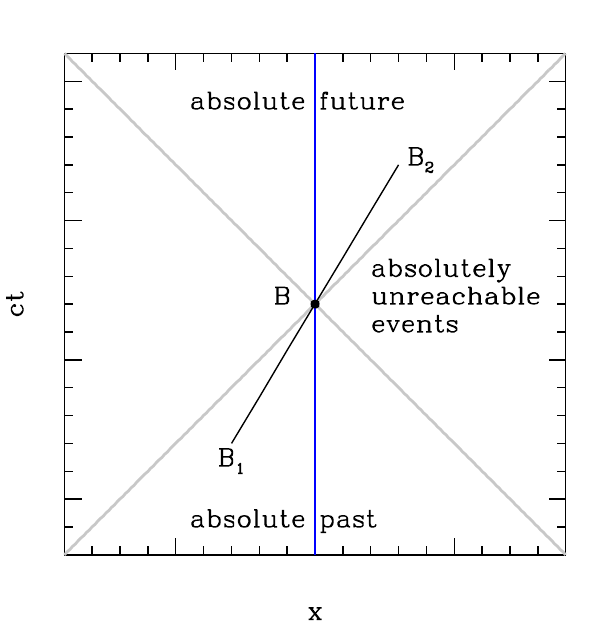}%
\includegraphics[scale=0.9, clip=true, trim=2cm 10cm 10cm 9.5cm]{hand_drawn.pdf}\\
\caption{Left: the separation of Minkowski space into absolute past, absolute future and absolutely unreachable events. Right: Minkowski diagram for the lighthouse paradox.}
\label{pic_mink2}
\end{figure}

Any event divides the 4-dimensional space-time into the following different parts (Figure \ref{pic_mink2}, left): absolute future, absolute past and absolutely unreachable events. The 4D space with these properties is called {\bf Minkowski space}. Any event (for example B$_2$) which causally follows from event B must lie within the future light cone of the event B. Any event which caused B (for example, B$_1$) must lie within the past light cone of the event B. There could be many events within each other's light cones which are not causally connected, for example a midterm on Tuesday and rain in the same location on Wednesday. But these events are not obligated to lie within each other's light cones, whereas the causally connected ones are. 

{\bf Lighthouse paradox} is a thought experiment in which we take a flashlight and rotate it with angular rotation speed $\Omega$ and record the movement of the light spot from the flashlight on a cylindrical wall surrounding us with radius $D$. The light spot moves with velocity $\Omega D$ on the wall. If $D$ is sufficiently large, the speed of the light spot can be greater than $c$. Does this contradict Special Relativity? No physical object is moving with this speed, and the subsequent spots are not causally connected to the previous ones; thus there is no contradiction with Special Relativity.

We examine this problem with the Minkowski diagram (Figure \ref{pic_mink2}, right). The world line of the flashlight is depicted with the solid line and the world lines of the walls are depicted with the long-dashed lines. During the event A the flashlight is pointed to Wall 1 and emits a photon in that direction; it produces a light spot during event W$_1$. Shortly thereafter (at B) we point the flashlight in the opposite direction, and another light spot is produced during event W$_2$. Both W$_1$ and W$_2$ are causally connected to the flashlight via the world lines of the corresponding photons (wavy lines). But W$_1$ and W$_2$ lie outside each other's light cones. It is perfectly fine for events to lie outside each other's light cones (as I am typing this, a supernova explodes in a distant galaxy), as long as these events are not causally connected.

We can use Minkowski diagrams to examine the space-time coordinates of events in different frames. When we change frames of reference, the world lines tilt around, but the separation of space into absolute past, absolute future and absolutely unreachable events remains. This means that if two events are causally related in one frame (and therefore one is within the light cone of the other), they will remain so in other frames. We cannot make a Lorentz transformation to make these events simultaneous or make the later event precede the earlier one; thus, there is no time-travel in Special Relativity. We will demonstrate these statements quantitatively in subsequent sections.

\begin{figure}[htb]
\centering
\includegraphics[scale=0.8, clip=true, trim=0cm 9.5cm 10cm 0cm]{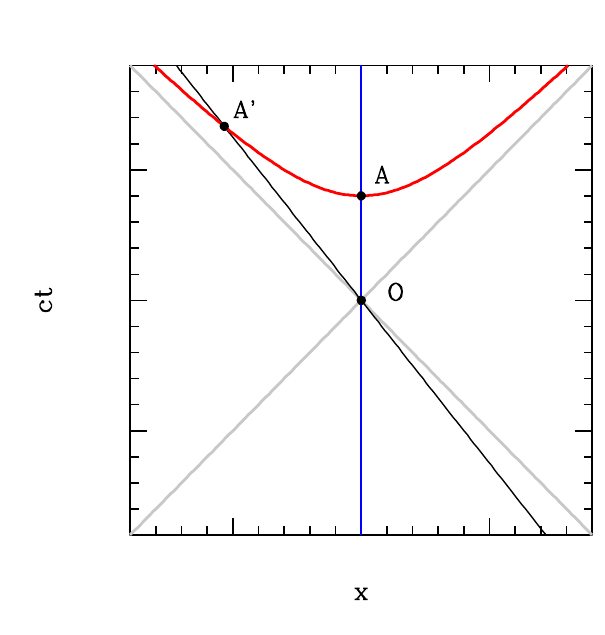}
\caption{Minkowski diagram showing the locus of the same event A as a result of Lorentz transforms with different velocities. The blue line shows the world line of the petal in the station frame, the black line shows the world line of the petal as seen in one of the train frames, and the grey lines show the light cone of the event O.}
\label{pic_mink3}
\end{figure}

In Figure \ref{pic_mink3}, we illustrate frame transformations on a Minkowski diagram. When the train $S'$ passes the station $S$, the flag breaks against the flower pot (event O) and knocks down a petal which falls on the platform right underneath the pot some time $t$ after (event A). The world line of the petal in the tracks frame, whose $x$ coordinate does not change with respect to the tracks, is shown with the blue line. The space-time coordinates of event A are $(ct, 0, 0, 0)$. 

In the train frame, the petal is moving backwards along the black world line and hits the platform as event A$'$. The space-time coordinates of A and A$'$ are related via Lorent transform: $x'=-\beta\gamma ct$ and $c t'=\gamma c t$, where we have taken into account that $x=0$. Then we discover that $(ct')^2-(x')^2=(ct)^2(\gamma^2-\beta^2\gamma^2)=(ct)^2$, which is a value that is independent of the velocity of the train. Thus if we were to observe event A from many different trains moving with different velocities, the locus of events A$'$ on the Minkowski diagram would obey the equation $(ct')^2-(x')^2=(ct)^2$. This is the equation for a hyperbola (shown in red in Figure \ref{pic_mink3}), which is why it is sometimes said that Special Relativity is described by {\bf hyperbolic geometry}. 

Many problems in Special Relativity can be solved geometrically using Minkowski diagrams, though this is not the method we are following in this course; if you are interested in this approach, use the Appendix of \citet{resn68}. 

\subsection{Experimental status of Special Relativity}

There is a lot of confusion surrounding Special Relativity, and there are many confident-sounding internet users who claim to have ``disproved'' it. Therefore, be very cautious when exploring various topics in Special Relativity online. Some of the phenomena in Special Relativity seem exotic, especially those that involve the non-absolute nature of time, which contradicts our everyday experiences with mechanical systems. Nonetheless, Special Relativity is not a niche speculative area of physics; it has vast body of experimental confirmation and is at the foundation of all modern physics. 
\begin{itemize}
\item As was already hinted in Section \ref{sec_maxwell}, Special Relativity is at the core of all electrodynamics. This means that Special Relativity is required for explaining a vast array of everyday phenomena. For example, the sky is blue because air particles scatter shorter wavelength radiation more efficiently than the longer wavelength radiation. The derivation of this simple statement involves using `delayed potentials' which are products of Special Relativity theory. 
\item Special Relativity is exquisitely confirmed by many astronomical measurements; furthermore, even General Relativity (which is vastly more complex than Special Relativity, but has the latter as its empty space limit) has been confirmed to better than 1\% using a double-neutron star system which is losing energy to gravitational waves \citep{tayl92}, a discovery that earned R. Hulse and J. Taylor a Nobel Prize in physics in 1993. Gravitational waves were discovered by direct observations in September 2015 \citep{abbo16} by the Laser Interferometer Gravitational-Wave Observatory, resulting in the 2017 Physics Nobel Prize to K. Thorne, R. Weiss, and B. Barish. 
\item Quantum electrodynamics -- the relativistic version of quantum mechanics -- is tested in accelerators around the world every day. As energetic particles collide in an accelerator, new particles are produced and fly out. Predicting which particles might result from these collisions is the subject of quantum electrodynamics, but we can use Special Relativity to calculate the trajectories and energies of resulting particles (Section \ref{sec_momenergy}). One of the older insights from Special Relativity is the understanding that muons produced in the upper atmosphere can reach the surface of the Earth despite their very short laboratory lifetime because of time dilation. 
\item In everyday applications, relativistic corrections are necessary for precise measurements of time and location, e.g., by the global positioning systems.
\end{itemize}

Whenever an argument or an experiment is put forward as disproving or undermining Special Relativity, it usually falls into one of two categories: (1) The experiment turns out to be  wrong; (2) There is confusion regarding which velocity can and cannot be faster than the speed of light. 

A recent example of the former is the potential discovery of superluminal neutrinos in 2011 at the European Center for Nuclear Research (CERN)\footnote{\url{http://en.wikipedia.org/wiki/Faster-than-light_neutrino_anomaly}}. Early reports from the OPERA experiment suggested that neutrinos traveled between Switzerland and Italy with velocity exceeding $c$ ($v=c(1+\delta)$, where $\delta=1/40,000$), leading to a lot of excitement in the physics community. 

Many astronomers were immediately suspicious of the claimed detection: in 1987, observers on Earth discovered a supernova in a nearby galaxy\footnote{\url{http://en.wikipedia.org/wiki/SN_1987A}}, with both light and neutrinos arriving on the same day (this was the first ever detection of neutrinos from an astronomical source other than the Sun). The distance between Earth and SN 1987A is $D=168,000$ light years; thus if neutrinos had been faster than light by $c\delta$, we would have seen neutrinos arriving earlier by $D \delta/c=168,000$ years$\times \delta=4.2$ years.

These arguments were rejected by superluminal neutrino enthusiasts on the grounds that the neutrinos from SN 1987A were of a different type and different energies than those in the OPERA experiment, and many research papers were written in the span of a few weeks following the announcement. Eventually, it was found that a faulty cable and a fast clock were responsible for the erroneous measurement. With fixed experimental apparatus, the experiments were repeated and it was found that the speed of neutrinos was consistent with the speed of light. 

Another source of confusion is the issue of which events can and cannot be connected by apparent speeds faster than the speed of light. We just saw in the previous chapter that the spot from the flashlight in the lighthouse experiment can move from one wall to the next with an apparent velocity $\gg c$. This thought experiment does not seem like much of a paradox, but analogous situations can get rather complicated. A slightly more complex problem is illustrated by Problem \ref{prob_rod}. 

So what can and what cannot move faster than the speed of light? Identifying massive particles (which cannot move faster than the speed of light) is usually fairly straightforward (though Problem \ref{prob_rod} attempts to obfuscate the issue). But other examples can get so subtle that they continue to be subject of ongoing research. It is usually said that anything that transports energy and / or information has to move at velocity less than $c$, but it is sometimes non-trivial to identify which sequences of events do and do not lead to transport of energy and / or information. 

Examples include the apparent superluminal motions of astronomical sources, which can be easily understood as a projection effect (Problem \ref{prob_superlum}). However, there are continued arguments as to whether this is all there is to the story\footnote{\url{http://en.wikipedia.org/wiki/Superluminal_motion}}. Alternative explanations involve differences between group velocity of perturbations (which is normally smaller than the speed of light) and phase velocity (which can be higher). The group velocity in turn is also not free from speed of light controversy\footnote{\url{http://en.wikipedia.org/wiki/Group_velocity}}. In another example, understanding of quantum entanglement involves lengthy discussions of what does and does not constitute information transport in quantum mechanics (e.g., \citealt{barr14}).

Paradoxes are especially confusing, and Special Relativity is famous for them. Many paradoxes are resolved in the context of the theory itself\footnote{\url{http://en.wikipedia.org/wiki/Ladder_paradox}}. But some are not so easy and have on occasion resulted in new insights and in developments in physics and mathematics\footnote{\url{http://en.wikipedia.org/wiki/Ehrenfest_paradox}}.

\begin{figure}[htb]
\centering
\includegraphics[scale=0.4]{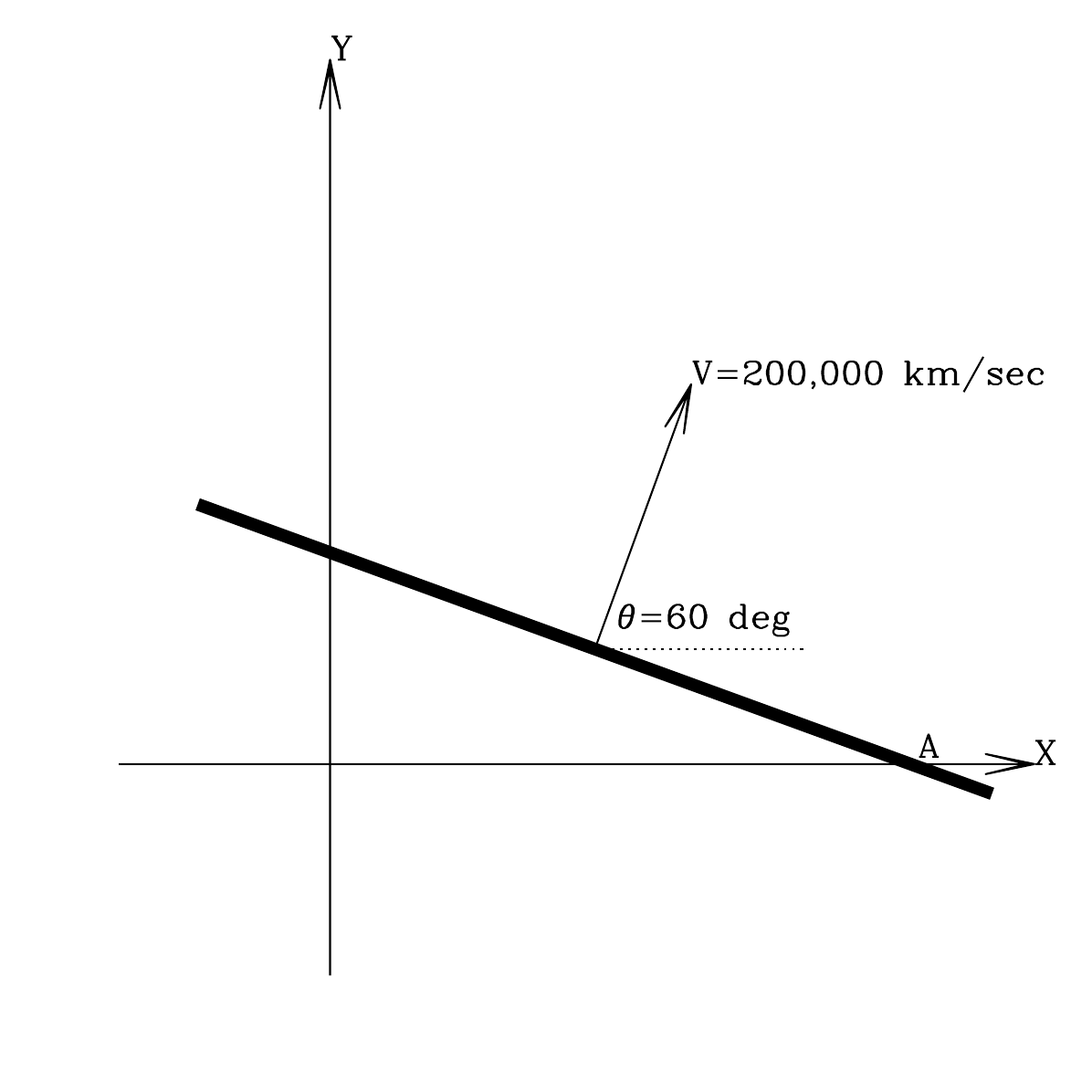}
\caption{The rod from problem \ref{prob_rod}.}
\label{pic_prob_rod}
\end{figure}

\begin{prob}
A thin rod is moving with velocity $v$ which is directed perpendicular to the rod and at an angle $\theta$ to the $x$-axis of the observer (Figure \ref{pic_prob_rod}). What is the speed of $A$, the point of intersection between the rod and the $x$-axis? Evaluate for $v=200,000$ km/sec and $\theta=60$ degrees in units of the speed of light (i.e., your answer should be in the form $v_A=?\times c$). -- {\bf Solution.} We set the time so that the rod passes through the origin at $t=0$. In that case the length of the perpendicular dropped from the origin to the rod at any instant $t$ is $p(t)=vt$. From the geometry of the setup we see that the coordinate of the point A at the same instant is $x_{A}(t)=p(t)/\cos(\theta)=vt/\cos(\theta)$. Hence $v_{A}=dx_{A}/dt=v/\cos(\theta)\simeq 1.33c$, greater than the speed of light. This does not contradict the second postulate of Special Relativity because the trajectory of A is just the locus of a moving geometric point which coincides with different material points on the rod at different instants of time. So, this superluminal velocity does not correspond to the motion of any physical particle.
\label{prob_rod}
\end{prob}

\subsection{Interval and proper time}

{\bf Relativistic invariants} are functions of physical values which remain invariant under Lorentz transformation. We define an {\bf interval} between two events, for example the ball being kicked and the ball being caught:
\begin{equation}
\Delta s^2=c^2 \Delta t^2 - \Delta x^2 - \Delta y^2 - \Delta z^2.
\end{equation}
In the train frame $S'$, the interval is $\Delta s'^2=c^2 \Delta t'^2 - \Delta x'^2 - \Delta y'^2 - \Delta z'^2$. Even though individual terms in this expression do change when we move from one reference frame to another, their combination -- the interval -- is invariant under Lorentz transformation. To verify this, we express values in the $S$ frame in terms of values from the $S'$ frame using the Lorentz transform:
\begin{eqnarray}
\Delta s^2=(\gamma c \Delta t'+\beta \gamma \Delta x')^2 - (\beta \gamma c \Delta t'+\gamma \Delta x')^2 - \Delta y'^2-\Delta z'^2=\nonumber\\
(\gamma^2-\beta^2\gamma^2)c^2\Delta t'^2-(\gamma^2-\beta^2\gamma^2)\Delta x'^2-\Delta y'^2-\Delta z'^2.
\end{eqnarray}
But $\gamma^2-\beta^2\gamma^2\equiv 1$, and therefore $\Delta s^2\equiv \Delta s'^2$. Thus, an interval between two events is a relativistic invariant.

If two events are causally connected, then the interval between them is non-negative. Indeed, in the definition of the interval $\Delta s^2=c^2 \Delta t^2 - \Delta \vec{r}^2$ for causally connected events $\Delta \vec{r}=\vec{v_s}\Delta t$, where $v_s\le c$ is the signal propagation speed. Intervals such that $\Delta s^2>0$ are called {\bf time-like intervals}. Because intervals are relativistic invariants, if an interval is time-like in one frame it will remain time-like in all other frames. Conversely, intervals with $\Delta s^2<0$ are called {\bf space-like intervals} and they remain such in all frames. {\bf Light-like intervals} are those that are $\Delta s^2=0$ in all frames.

\begin{prob} 
If we have two events that are not simultaneous in a given frame, can we get on a train and make them simultaneous in the train frame? -- {\bf Solution.} If we find such a frame $S'$, then in this frame the interval between the events is $\Delta s'^2=c^2\Delta t'^2 - \Delta \vec{r'}^2<0$ since $\Delta t'=0$. Interval is relativistically invariant; therefore this is only possible if in the original frame the events are separated by a space-like interval. The corollary of this argument is that causally connected events (which are separated by a time-like interval) cannot be made simultaneous in any frame. 
\end{prob}

\begin{prob} 
Bob, who is in college in Maryland, picks up the phone and calls Mary in Connecticut 250 miles away. Mary answers one minute later. How fast do we have to fly on a plane along the Maryland - Connecticut line in order for these two events to happen simultaneously? -- {\bf Solution.} Since Bob's dialing the phone and Mary's picking up the phone are causally connected events, there is no such frame. If we try to find such a frame, we would use the Lorentz transform to write $c\Delta t'=0=\gamma c\Delta t - \beta \gamma \Delta x$, and therefore $\beta=c\Delta t/\Delta x$. But $\Delta t=1$ min and $\Delta x=250$ miles, so we would get $\beta=4.5\times 10^4$ for the velocity of the plane, which is not possible. 
\end{prob}

Another relativistic invariant is {\bf the proper time of a particle}. Suppose the ball is traveling in frame $S$ with velocity $\vec{u}$ for a time span $\Delta t$. In frame $S'$, we will record some other velocity $\vec{u'}$ and some other time span $\Delta t'$. Both $t$ and $\vec{u}$ transform from one frame to another; we do not yet know how to transform $\vec{u}$ between two different frames -- it is in fact a pretty complicated transformation. But it turns out that the following combination of these measurements will be the same in both frames:
\begin{equation}
\Delta \tau=\Delta t \sqrt{1-u^2/c^2}=\Delta t' \sqrt{1-u'^2/c^2}.
\end{equation}
To prove this, we remember the definition of velocity. As the ball travels through space, it breaks some clocks which can be used to define two events separated by $\Delta t$ and $\Delta \vec{r}$, with $\vec{u}\equiv \Delta \vec{r}/\Delta t$. Therefore,
\begin{equation}
\Delta \tau=\sqrt{\Delta t^2-u^2 \Delta t^2/c^2}=\sqrt{\Delta t^2-\Delta \vec{r}^2/c^2}=\sqrt{\Delta s^2/c^2}.
\end{equation}
Thus, proper time is related to the interval between the two clock-breaking events which is in turn a relativistic invariant as we showed previously. 

The value $\Delta \tau$ is called proper time because in the frame co-moving with the particle (the ball in our case), $\vec{u}=0$ and thus $\Delta \tau$ is simply the elapsed time in the co-moving (proper) frame. 

{\bf Twin paradox} is one of the most famous paradoxes of Special Relativity. Bob and Mary, who are the same age, split up and Mary goes traveling on fast rockets in the outer space, while Bob continues to sit on the couch. Space travel gets lonely after a while, so Mary changes her mind and returns to Bob. From the standpoint of Bob, Mary was the one traveling and thus because of the time dilation she should have aged less than he has by the time of her return. However, from the standpoint of Mary, Bob is the one who was traveling, thus she would conclude that he should have aged less. 

The resolution of the paradox is that only Bob's frame is inertial. In order for Mary to go away from him and to come back to him and compare clocks, she must have moved with some acceleration at least at some points during her trip (e.g., at take-off, at turn-around and at landing). Thus the situation is in fact not symmetric. Special Relativity theory can predict the behavior of accelerated objects as long as in the formulation of physical laws we take the view of the inertial observer. 

\begin{figure}[htb]
\centering
\includegraphics[scale=0.8, clip=true, trim=0cm 9.5cm 0cm 0cm]{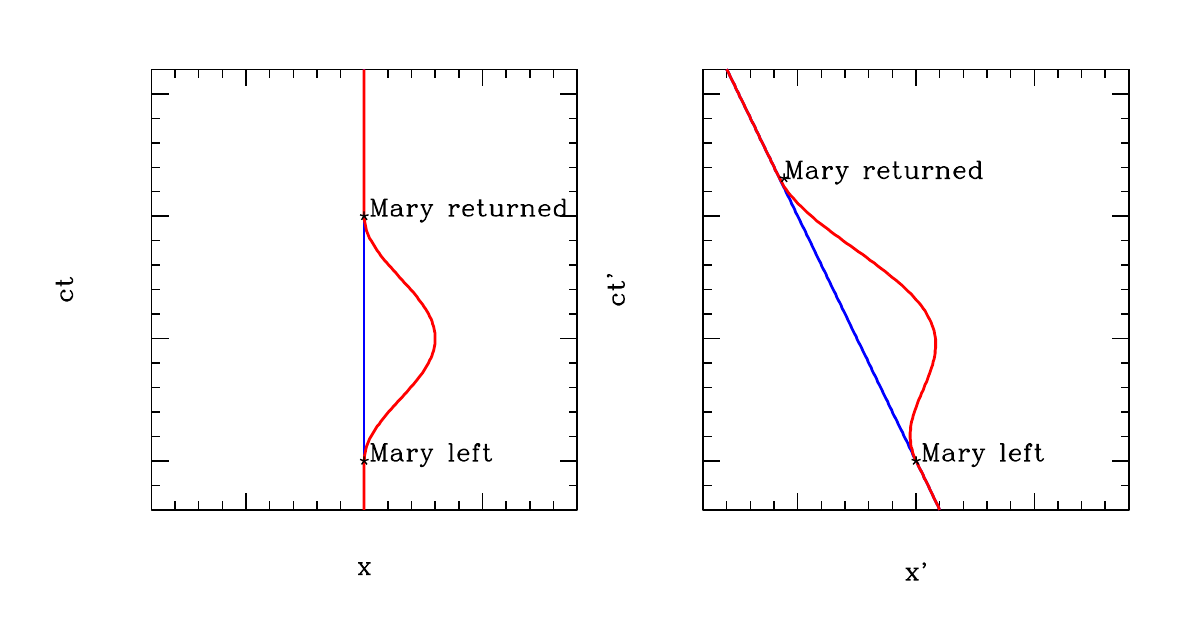}
\caption{Twin paradox in Minkowski diagrams for two different inertial frames: Bob on the couch (left; Bob's blue world line is vertical in this frame) and Lady Bug (right, with Bob drifting with his couch in the opposite direction). Mary's trajectory in the second frame has been properly computed from the first frame using the Lorentz transform.}
\label{pic_twin}
\end{figure}

For both Bob and Mary, the advancement of their age is given by the elapsed proper time which will depend on the paths taken by them in the space-time diagram. For example, in Figure \ref{pic_twin}, left, the elapsed proper time between the event ``Mary left'' and the event ``Mary returned'' is, for Bob, $\Delta \tau_{\rm Bob}=\int_{\rm M-left}^{\rm M-ret}{\rm d} t \sqrt{1-u_{\rm Bob}^2/c^2}$ integrated along Bob's world line and for Mary, $\Delta \tau_{\rm Mary}=\int_{\rm M-left}^{\rm M-ret}{\rm d} t \sqrt{1-u_{\rm Mary}^2/c^2}$ integrated along Mary's world line. But $u_{\rm Bob}=0$ in this frame, whereas it is non-zero at least some of the time for Mary. Thus, in this frame the integrand for Mary is always smaller than (or equal to) that for Bob. Thus, when Mary returns, she has aged less than Bob did.

Furthermore, because proper time is a relativistic invariant, this calculation does not depend on the frame in which it is conducted. For example, Lady Bug (frame $S'$) is flying very fast past Bob and Many and the couch and watches the Bob and Mary saga unfold in Figure \ref{pic_twin}, right. Now $u'_{\rm Bob}$ is no longer zero, and Mary's path in the Minkowski diagram has shifted as well, but the path integrals will yield the same answer in the $S'$ frame as they did in the $S$ frame. 

No matter how complex Mary's motion is, we can calculate her elapsed proper time using the path integral method as long as we are observing her from an inertial frame.

\begin{prob} Bob and Mary are both 20 years old. Mary flies off at $0.8c$, circles around every planet in the Solar system multiple times -- all at $0.8c$ -- and comes back. Bob is now 40; how old is Mary? -- {\bf Solution.} In the Earth frame (in which Bob is stationary), 20 years elapsed. Mary's integrand $\sqrt{1-u_{\rm Mary}^2/c^2}$ is equal to 0.6 at all times, despite the complexity of her trajectory. Thus, Mary's elapsed proper time is 12 years and she is now 32 years old. 

Mary cannot instantaneously accelerate from 0 to 0.8c in Bob's frame, therefore there will be some periods when she is not in fact traveling at 0.8c. But if she travels at 0.8c the vast majority of the entire time, then the contribution of those initial and final acceleration and deceleration chunks to the overall path integral is negligible, so they do not affect our conclusions. 
\end{prob}

\clearpage
\begin{figure}[!htb]
\centering
\includegraphics[scale=0.8]{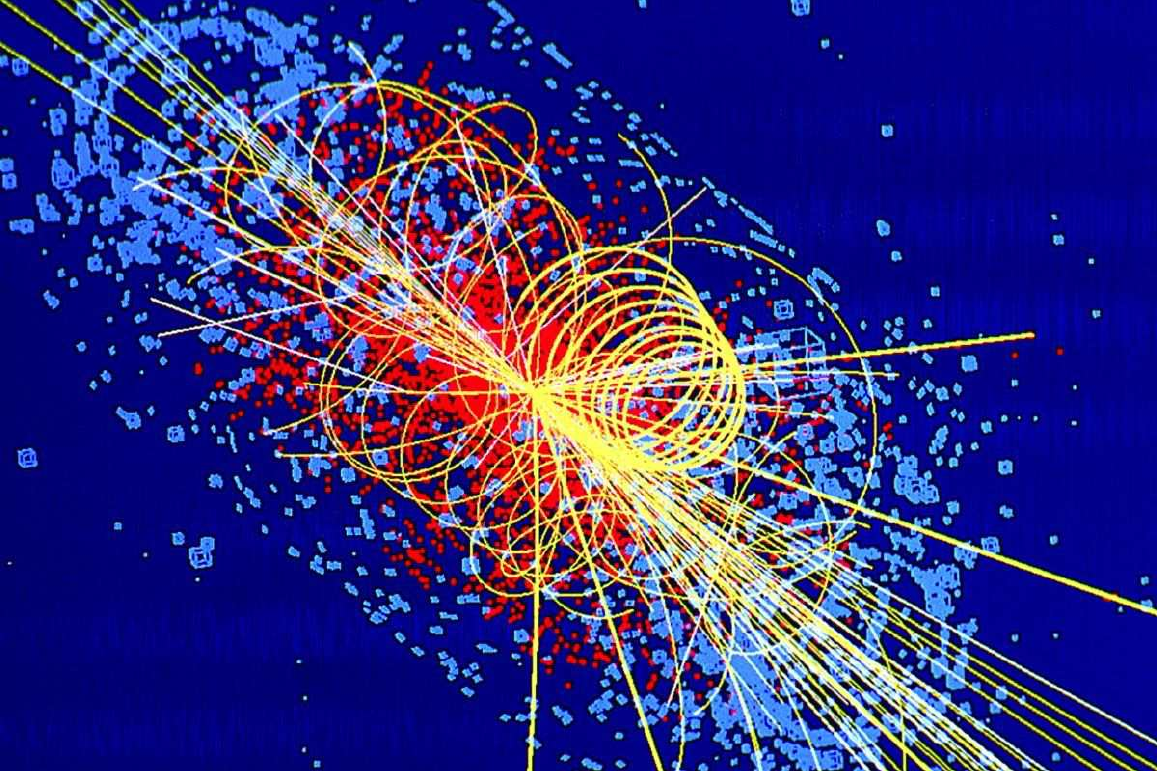}
\caption{Another LHC CMS event simulation, image by Lucas Taylor, from \url{http://cms.web.cern.ch/news/about-higgs-boson}.}
\end{figure}

\clearpage
\section{Addition of velocities}
\label{sec_velocity}

How does one measure velocities? Again, it is helpful to visualize the measurement in terms of events. Suppose we are observing the soccer game either from the tracks frame or from the train frame. There are rulers and clocks everywhere. As the ball flies through space, it breaks the clocks which stop. Thus, in the tracks frame we can take two broken clocks separated by a small physical distance $\Delta \vec{r}$ and read off their time separation $\Delta t$ and define $\vec{u}=\Delta\vec{r}/\Delta t$, just as we did in classical mechanics. Although the system of reference moving with the train is more difficult to visualize, the same thing can be done in $S'$, with $\vec{u'}=\Delta\vec{r'}/\Delta t'$.

In classical mechanics, we know how the velocity vector transforms from one frame to another: $\vec{u'}=\vec{u}-\vec{v}$. In this chapter we determine how this vector transforms in Special Relativity. 

Suppose the ball was kicked away from the flower pot $x=0, y=0, z=0$ at $t=0$. This is event number 1. Event number 2 is the ball breaking a clock a short distance away $\Delta x, \Delta y, \Delta z$ at time $\Delta t$. The definition of velocity as measured in frame $S$ is then
\begin{eqnarray}
\Delta x=u_x\Delta t;\\
\Delta y=u_y\Delta t;\\
\Delta z=u_z\Delta t.
\end{eqnarray}
In frame $S'$, the space-time coordinates of the event number 1 are (0,0,0,0). The space-time coordinates of the event number 2 can be found from the Lorentz transform (\ref{eq_lorentz_final1}): 
\begin{eqnarray}
c\Delta t'=\gamma c \Delta t - \beta \gamma \Delta x=\gamma c \Delta t - \beta \gamma u_x\Delta t;\\
\Delta x'=-\beta\gamma c\Delta t + \gamma \Delta x = -\beta \gamma c \Delta t+\gamma u_x \Delta t;\\
\Delta y'=\Delta y = u_y\Delta t;\\
\Delta z'=\Delta z = u_z\Delta t.
\end{eqnarray}
It is extremely important to remember that $\beta$ and $\gamma$ refer to the motion of the train and therefore use the velocity $v$ of the train! 

The velocities we measure in $S'$ are
\begin{eqnarray}
u_x'=c\frac{\Delta x'}{c\Delta t'}=c\frac{-\beta \gamma c \Delta t+\gamma u_x \Delta t}{\gamma c \Delta t - \beta \gamma u_x\Delta t}=\frac{-v+u_x}{1-vu_x/c^2};\label{eq_vel_tran1}\\
u_y'=c\frac{\Delta y'}{c\Delta t'}=c\frac{u_y\Delta t}{\gamma c \Delta t - \beta \gamma u_x\Delta t}=\frac{u_y}{\gamma(1-vu_x/c^2)};\label{eq_vel_tran2}\\
u_z'=c\frac{\Delta z'}{c\Delta t'}=c\frac{u_z\Delta t}{\gamma c \Delta t - \beta \gamma u_x\Delta t}=\frac{u_z}{\gamma(1-vu_x/c^2)}.\label{eq_vel_tran3}
\end{eqnarray}
If instead we throw a vase on the train with velocity $\vec{u'}$ and want to calculate its velocity in the $S$ frame, then all we need to do is flip the sign of $v$ in these equations.

We recover the Galilean transform of velocity by taking the limit $c\rightarrow\infty$: in this case in the equations above, $\gamma\rightarrow 1$ and $1-vu_x/c^2\rightarrow 1$, so we get $u_x'=u_x-v, u_y'=u_y, u_z'=u_z$. 

We consider one-dimensional motion along the $x$-axis and use velocity transformation equation $u_x=(v+u_x')/(1+vu_x'/c^2)$. Suppose we throw some object on the train, which must have $u_x'<c$. The velocity of the train is always $v<c$. Can we ever observe this object moving with velocity $>c$ in the tracks frame? We denote $u_x'=c-\delta$, where $\delta>0$ is how much smaller the velocity is than the speed of light in the train frame. We need to perform the comparison
\begin{equation}
\frac{v+u_x'}{1+vu_x'/c^2}?c.\label{eq_velocity_comparison}
\end{equation}
Multiplying both sides by the denominator and plugging in $u_x'=c-\delta$, we re-write this as
\begin{equation}
v+c-\delta \,?\, c+\frac{v}{c}(c-\delta),
\end{equation}
which is equivalent to $\frac{v}{c}\delta ? \delta$, where the sign is always $<$. Thus $u_x<c$: if an object is moving with a velocity smaller than the speed of light in one frame, it is moving with a velocity smaller than the speed of light in any other frame. 

What about photons? Plugging $u_x'=c$ into the velocity transformation formula, we find $u_x=(v+c)/(1+v c/c^2)=c$. Thus if something (most commonly a photon) is moving with a speed of light in one frame, it is also moving with the speed of light in any other frame. Of course, this is not a new finding: to formulate Special Relativity we postulated that the speed of light must be the same in all frames, and we are recovering the same result using the velocity addition formulae derived from our postulate. This calculation a `sanity check': it is helpful to reassure ourselves that our theory is internally consistent. 

What about that lighthouse spot? Suppose we have a sequence of events separated by $u_x'>c$. Although we know that physical objects with mass cannot move with such velocities, we do know that certain things (like the lighthouse spot) are allowed such velocities. We denote $u_x'=c+\delta$, where $\delta$ is positive. Then we need again to find the sign in (\ref{eq_velocity_comparison}). Multiplying both sides by the denominator, we find $v+c+\delta \, ?\, c+v+\delta v/c$, and because the velocity of the train $v$ is always $<c$, the sign is $>$. Therefore, if something (like the spot of light in the lighthouse) is moving with velocity $>c$ in one frame, it will be moving with velocity $>c$ in any other frame. 

All these examples are simply restating the same facts we learned previously about intervals: time-like intervals stay time-like in all systems, space-like intervals stay space-like in all systems, and light-like intervals stay light-like in all systems. 

\clearpage
\begin{figure}[!htb]
\centering
\includegraphics[scale=0.7]{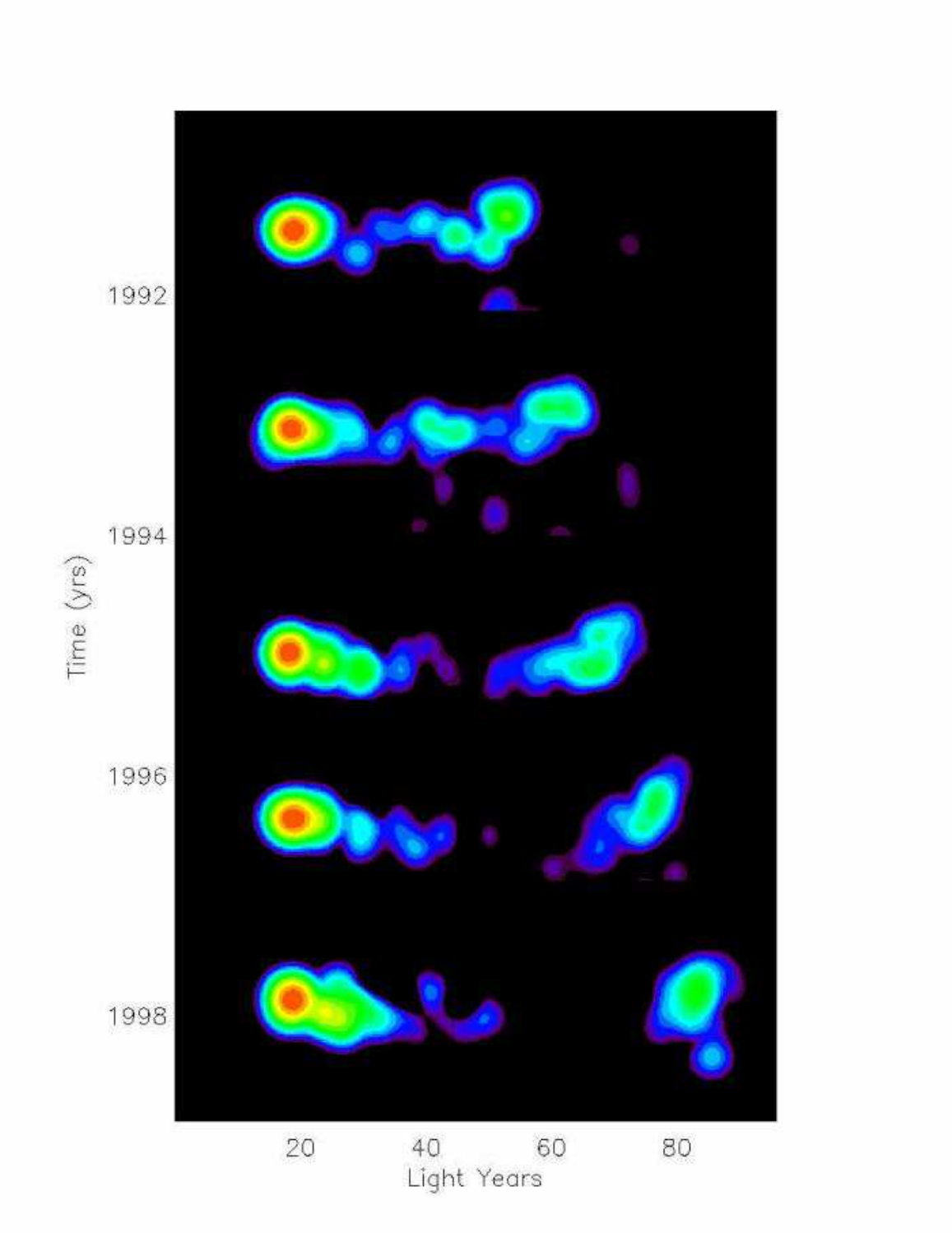}
\caption{Apparent superluminal motion in quasar 3C279 is shown as a sequence of five radio snapshots made over seven years \citep{wehr11}. The blob on the right appears to have covered a distance of $\sim 25$ light years between 1991 and 1998. Image by Glenn Piner, from \url{http://images.nrao.edu/387}.}
\label{pic_superlum}
\end{figure}

\clearpage
\section{4-vectors: velocity, energy-momentum}

\subsection{Need for new definitions of energy and momentum}
\label{sec_need}

\begin{figure}[htb]
\centering
\includegraphics[scale=0.8, clip=true, trim=0.5cm 16cm 2cm 0.5cm]{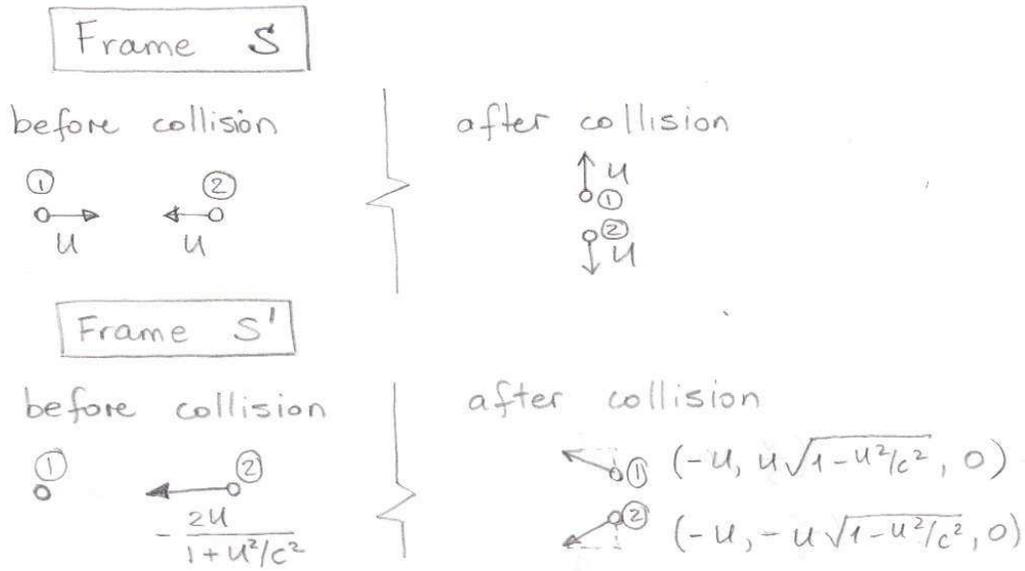}
\caption{The same elastic collision viewed from two different frames.}
\label{pic_coll}
\end{figure}

Consider an elastic collision of two identical particles shown in Figure \ref{pic_coll}, top. In classical mechanics, with $\en=\sum_i m_iu_i^2/2$ and $\vec{p}=\sum_i m_i \vec{u}_i$ this is a perfectly acceptable situation in which both the energy and the momentum of the system are conserved during the collision: $\en_{\rm before}=\en_{\rm after}$ and $\vec{p}_{\rm before}=\vec{p}_{\rm after}$. 

Now we determine if this collision is compatible with relativistic mechanics. We move into the reference frame $S'$ moving with velocity $u$ to the right -- this frame is co-moving with particle 1 before the collision (Figure \ref{pic_coll}, bottom). We denote $x$ to be the axis parallel to the initial velocity of particle 1. We introduce the $xy$ and the $x'y'$ plane to be the plane of the collision and $z$ and $z'$ to be perpendicular to this frame. Using relativistic velocity transformations from the previous section, we can calculate all the components of the particle velocities in this new frame, as marked on the plot. 

In the frame $S$ before the collision, particle 1 was moving with velocity vector $(u,0,0)$ and particle 2 with $(-u,0,0)$. In the frame $S'$ moving with $u$ along the $x$-axis, the velocity vectors become $(0,0,0)$ (particle 1 is at rest in the co-moving frame) and $(-2u/(1+u^2/c^2),0,0)$, where we used equation (\ref{eq_vel_tran1}) to calculate the components of the velocity vector in the new frame. Usually we denote the frame velocity to be $v$, but in this case we picked a special inertial frame -- that moving with $u$ -- so $u$ in velocity transforms appears both for the particle and for the frame.

After the collision, the velocity vectors are $(0,u,0)$ and $(0,-u,0)$ for the particles in frame $S$. In frame $S'$ we can calculate the new velocity components using equations (\ref{eq_vel_tran1}) and (\ref{eq_vel_tran2}). The horizontal components of the velocity in this frame are $-u$ and the vertical are $\pm u/\gamma$; therefore, we find that the velocity vectors are now $(-u,u\sqrt{1-u^2/c^2},0)$ and $(-u,-u\sqrt{1-u^2/c^2},0)$. 

For the moment, we will assume that the non-relativistic expressions for energy and momentum are still valid. In frame $S$, the energy before and after the collision is $\en_{\rm before}=\en_{\rm after}=mu^2$ (there are two particles each with energy $mu^2/2$) and the momentum before and after the collision is 0. In frame $S'$ we find $\en'_{\rm before}=2mu^2/(1+u^2/c^2)^2$ and $\en'_{\rm after}=m\left(u^2+u^2(1-u^2/c^2)\right)$. The good news is that for non-relativistic velocities $u\ll c$ the energy of the system before and after the collision is the same and equal $2 m u^2$. Thus in the non-relativistic case if the energy is conserved in frame $S$, it is also conserved in $S'$. 

The bad news is that for velocities that are not small, $\en'_{\rm before}\ne \en'_{\rm after}$ -- they disagree in the order $\sim O\left(mu^2 \frac{u^2}{c^2}\right)$. Similarly, the momentum before the collision is $p'_{x, {\rm before}}=-\frac{2mu}{1+u^2/c^2}$ and the momentum after the collision is $p'_{x, {\rm after}}=-2mu$, which again are the same in the case of small velocities $u\ll c$ but not in the case of the relativistic ones. 

Conservation of energy and momentum is one of those fundamental laws of physics that is subject to the principle of relativity: if the energy and/or momentum of some system is conserved in one frame, it would be highly desirable if they were conserved in all other frames, so that we would not have to worry about which frame of reference we are using for considering fundamental processes. However, we just saw that the definitions of energy and momentum from the classical mechanics are not compatible with relativistic velocity addition. Thus if we insist on having energy and momentum conservation in Special Relativity in all inertial frames, we need new definitions of energy and momentum. 

\subsection{Four-vectors}
\label{sec_4vec}

{\bf Four-vectors (4-vectors)} are sets of four functions of physical values that transform from $S$ to $S'$ according to the Lorentz transform. For example, the difference between space-time coordinates of two events, 
\begin{equation}
\Delta s^{\mu}\equiv\left(\begin{array}{c} c\Delta t \\ \Delta x \\ \Delta y \\ \Delta z \end{array}\right)
\end{equation}
is a 4-vector, as we saw in Section \ref{sec_lorentz}. The velocity of a particle $\vec{u}$ is not a 4-vector. To begin with, it has only 3 components, not the 4 required by the definition of the 4-vector, and furthermore the relativistic transforms of velocity do not look like Lorentz transforms. 

For normal 3D space vectors, the traditional notation which we continue using here is $\vec{a}$, and it is conventional to use Latin indices for the components of the 3D vectors ($a_i$, where $i$ can take values from 1 to 3, corresponding to the projections of the vector $\vec{a}$ on the $x$, $y$ and $z$ axes). In contrast, a 4-vector is not denoted with an arrow, but with a letter and a Greek superscript: 
\begin{equation}
A^{\mu}\equiv\left(\begin{array}{c} A^0 \\ A^1 \\ A^2 \\ A^3\end{array}\right).
\end{equation}
Thus, $A^{\mu}$ is a 4$\times$1 matrix (well, sort of) with four components, and the index $\mu$ runs through these components and can take values from 0 to 3. $A^0$ is called {\bf the time-like component}, $A^{1,2,3}$ are called {\bf space-like components}. 

We can define another matrix, with 1 row and 4 columns:
\begin{equation}
A_{\mu}\equiv \left(A_0, A_1 , A_2 , A_3 \right)\equiv \left(A^0, -A^1, -A^2, -A^3 \right). 
\end{equation}
This is called the {\bf covariant} counterpart to the {\bf contravariant} 4-vector A$^{\mu}$, and there is actual physical significance to this lowering of indices and changing of the signs of the space-like components (see problem \ref{prob_covariant}). 

In our course we will predominantly use the contravariant 4-vectors, so we mostly need to make a distinction between the two to define the {\bf 4-product} of two 4-vectors:
\begin{eqnarray}
A^{\mu}B_{\mu}=A_{\mu}B^{\mu}\equiv A^0B_0+A^1B_1+A^2B_2+A^3B_3=A_0B^0+A_1B^1+A_2B^2+A_3B^3\\
\equiv A_0B_0-A_1B_1-A_2B_2-A_3B_3= A^0B^0-A^1B^1-A^2B^2-A^3B^3.
\label{eq_def_prod}
\end{eqnarray} 
This is (and should be) a bit of a cultural shock, because in the expression $A^{\mu}B_{\mu}$ the summation over $\mu$ from 0 to 3 is assumed by default but not written down explicitly. So the 4-product is always written as $A^{\mu} B_{\mu}$ instead of $\sum_{\mu=0}^3A^{\mu}B_{\mu}$, and to signify a 4-product there should always be one upper and one lower index. It is hard to imagine how this can possibly work and not lead to constant confusion, especially if we need to make a distinction between a 4-product of 4-vectors $A^{\mu}$ and $B_{\mu}$ (which consists of four terms, as in equation \ref{eq_def_prod}) and a product of the $\mu$ components of 4-vectors $A^{\mu}$ and $B_{\mu}$, which is only one term. Indeed, such distinction does sometimes need to be made, but in this case the default assumption in the relativity and electromagnetism literature is that there is summation over repeating indices, so if this is not the case the text would say so. 

An additional problem is that if we keep thinking of 4-vectors as matrices, then according to the linear algebra rules $A_{\mu}B^{\mu}$ and $A^{\mu}B_{\mu}$ would give different results: a $1\times 4$ matrix multiplied by $4\times 1$ matrix would give a single number (as in definition \ref{eq_def_prod}), but a $4\times 1$ matrix multiplied by $1\times 4$ matrix would actually give a $4\times 4$ matrix instead. Nonetheless, according to definition (\ref{eq_def_prod}) both expressions are treated equivalently for calculating the 4-product. So 4-vectors are not really matrices in the rigorous linear algebra sense, and we will often not bother arranging the contravariant 4-vectors in columns for that reason. 

{\bf Q.} So what are they, if they are not matrices?

{\bf A.} They are 4-vectors. They are mathematical objects, with some properties that are similar to those of matrices and some that are different. 

{\bf Q.} How would we interpret the expression $A^{\mu}B_{\mu}M^{\nu}N_{\nu}$? -- {\bf A.} We have here two 4-products multiplied together, $(A^0B^0-A^1B^1-A^2B^2-A^3B^3)(M^0N^0-M^1N^1-M^2N^2-M^3N^3)$. In fact, in the original expression we can re-arrange the 4-vectors in any order, e.g., $B_{\mu}N_{\nu}A^{\mu}M^{\nu}$, and it would still mean the same thing, because summation is assumed only over repeating indices, so to interpret this expression we would first collect all 4-vectors with $\mu$ indices and all 4-vectors with $\nu$ indices. Here is an example of a product which is not properly written: $A^{\mu}B^{\mu}M_{\mu}N_{\mu}$. There should always be just one upper and one lower index of the same kind to clearly indicate which two 4-vectors are being multiplied. Here is another improperly written product: $A^{\mu}B^{\mu}M_{\nu}N_{\nu}$. The repeating indices should always be one up, one down. 

One of the great things about 4-vectors is that their {\bf 4-products are relativistic invariants}. Indeed, if $A^{\mu}$ and $B^{\mu}$ are 4-vectors, then let us demonstrate that $A^{\mu}B_{\mu}=A'^{\mu}B'_{\mu}$. To do this, we use the definition of the 4-vectors: the primed and non-primed values are related via the Lorentz transform. Therefore, 
\begin{eqnarray}
A^0B^0-A^1B^1-A^2B^2-A^3B^3=\nonumber\\
(\gamma A'^0+\beta \gamma A'^1)(\gamma B'^0+\beta \gamma B'^1)-
(\beta \gamma A'^0+ \gamma A'^1)(\beta \gamma B'^0+ \gamma B'^1)-A'^2B'^2-A'^3B'^3=\nonumber\\
A'^0B'^0(\gamma^2-\beta^2\gamma^2)+A'^1B'^0(\beta \gamma^2-\beta\gamma^2)+A'^0B'^1(\beta \gamma^2-\beta\gamma^2)+\nonumber\\+A'^1B'^1(\beta^2\gamma^2-\gamma^2)-A'^2B'^2-A'^3B'^3=\nonumber\\
A'^0B'^0-A'^1B'^1-A'^2B'^2-A'^3B'^3.
\end{eqnarray}
 We have used the identity $\gamma^2-\beta^2\gamma^2=1$ which follows from the definitions of $\beta=v/c$ and $\gamma=1/\sqrt{1-v^2/c^2}$.

As an example of a 4-product of two 4-vectors, we take $A^{\mu}=(c\Delta t,\Delta \vec{r})$ and $B^{\mu}$ to be the same thing. Here we have used a normal vector in 3D space $\Delta\vec{r}$ to combine the three space-like components of the 4-vector $A^{\mu}$; such notation is commonly used. Their 4-product is $A^{\mu}B_{\mu}=c^2\Delta t^2-\Delta \vec{r}^2$, the interval between two events, which we already showed in Section \ref{sec_time} to be a relativistic invariant. When we first introduced the interval, we denoted it $\Delta s^2$, which is somewhat confusing because it can be positive, negative or zero. Now we understand that the mathematically correct way to denote the interval is $\Delta s^{\mu}\Delta s_{\mu}$, though $\Delta s^2$ is also standard notation. 

We can see that the time-like coordinate does not enter into the 4-product the same way that the space coordinates do: the signs are $+,-,-,-$. This set of signs is called {\bf Minkowski metric signature}. Time and space are intertwined in Special Relativity, and time-like and space-like components of 4-vectors do transform into one another when we apply the Lorentz transform. However, time and space are not equivalent, and as we saw in Section \ref{sec_time}, there are such things as absolute future and absolute past which cannot be switched by adjusting the reference frame. It turns out that the non-equivalence of the time and space coordinates is expressed mathematically by the different signs associated with these components in the Minkowski signature. 

One can perform many standard arithmetical operations with 4-vectors. For example, we can multiply a 4-vector by a relativistic invariant and get another 4-vector. Indeed, suppose $A^{\mu}$ is a 4-vector and $\alpha$ is an invariant. To demonstrate that $\alpha A^{\mu}$ is also a 4-vector, we need to show that it transforms according to the Lorentz transform: $\alpha' A'^{\mu}=\Lambda (\alpha A^{\mu})$. But this is true, because $\alpha'=\alpha$ and $A^{\mu}$ itself transforms according to the Lorentz transform. 

We look at the similarities and differences between 4-vectors and normal spatial vectors in the 3D space.
\begin{itemize}
\item 4-vectors have a 4-product defined as $A^{\mu}B_{\mu}=A^0B^0-A^1B^1-A^2B^2-A^3B^3$. Normal vectors have a scalar product defined as $\vec{a}\cdot\vec{b}=a_xb_x+a_yb_y+a_zb_z=\sum_{i=1}^3a_ib_i$. Thus, in both cases the vector product is defined as a linear combination of the products of the components, but these terms are combined with different signs for 4-vectors and for normal vectors in 3D space.
\item 4-products are invariant with respect to Lorentz transforms. Scalar products of 3D vectors are invariant with respect to the rotations of the coordinate system (problem \ref{prob_rot}). Using this analogy, it is often said that Lorentz transforms represent ``rotations'' in the 4D space-time. 
\end{itemize}

In the matrix notation, the 4-product $A^{\mu}B_{\mu}$ can be written as 
\begin{equation}
\left(\begin{array}{cccc} A^0 & A^1 & A^2 & A^3\end{array}\right)\left(\begin{array}{cccc} 1 & 0 & 0 & 0 \\ 0 & -1 & 0 & 0 \\ 0 & 0 & -1 & 0 \\ 0 & 0 & 0 & -1 \end{array}\right)\left(\begin{array}{c} B^0 \\ B^1 \\ B^2 \\ B^3 \end{array}\right).
\end{equation}
The diagonal matrix with the Minkowski signature on the diagonal is called {\bf Minkowski metric tensor}. The word ``tensor'' means a $4\times 4$ matrix with some special properties (roughly speaking, tensor is to matrix like a 4-vector to any set of 4 variables or like a relativistic invariant to an arbitrary physical value). The metric tensor is what one needs to know in order to calculate the length of the 4-vectors in the 4D space-time. It takes on a diagonal form in the flat empty space that we are considering in the Special Relativity course. In General Relativity, this matrix will be much more complex and will contain information about the local geometry of space (which is determined by the distribution of mass and other forms of energy).  

\begin{prob} The frame $S'$ is obtained from the frame $S$ by rotating around the $z$ axis by an angle $\phi$, although they are not moving relative to one another. Write down the transformations of the coordinates of a vector $\vec{a}=(a_x,a_y,a_z)$. Show explicitly that a scalar product of two vectors $\vec{a}\cdot\vec{b}=a_xb_x+a_yb_y+a_zb_z$ does not change as a result of the frame rotation, even though the components of both vectors do. (We know that a scalar product of two vectors is invariant under this coordinate transformation because it is equal to $ab\cos\theta$, where $\theta$ is the angle between them.) 
\label{prob_rot}
\end{prob}

\begin{prob} As we have seen, $(ct, x, y, z)$ transforms according to the Lorentz transform. Let us take a (relativistically invariant) function $f$ of these variables. How do the components of the gradient of this function in 4D space transform from one frame to the other? In other words, if you know $\left(\partial f/c\partial t, \partial f/\partial x, \partial f/\partial y, \partial f/\partial z\right)$ in the tracks frame, what are these values when measured in the train frame? As an example, consider a particular case for the function $f$ -- the interval between $(ct, x, y, z)$ and $(0,0,0,0)$. How does the 4D gradient of this function transform? This problem is for those who are interested in the physical meaning of the difference between contravariant and covariant 4-vectors, and we will not raise this subject again in these notes. 
\label{prob_covariant}
\end{prob}

\subsection{Velocity 4-vector; energy and momentum 4-vector}
\label{sec_momenergy}

We consider the motion of the ball that the kids are playing both in the track frame and in the train frame. The ball plows through space with velocity $\vec{u}$ as measured in the track frame and breaks two closely separated clocks, producing two physical events. We take a 4-vector of the difference between the two events in the limit of very small elapsed time, ${\rm d} s^{\mu}=(c{\rm d} t, {\rm d}\vec{r})$, and the elapsed proper time ${\rm d}\tau={\rm d} t\sqrt{1-u^2/c^2}$ -- a relativistic invariant. We divide the former by the latter, use the definition of the velocity $\vec{u}={\rm d}\vec{r}/{\rm d}t$, and obtain a new 4-vector:
\begin{equation}
U^{\mu}={\rm d} s^{\mu}/{\rm d}\tau=\left(\frac{c}{\sqrt{1-u^2/c^2}},\frac{\vec{u}}{\sqrt{1-u^2/c^2}}\right).
\end{equation}
This is called {\bf the 4-vector of velocity} and as such its components in the train frame can be calculated using the Lorentz transform. It can be demonstrated (Problem \ref{prob_vel}) that the Lorentz transform of the 4-vector of velocity leads to exactly the same transformation of the spatial components of the velocity that we already obtained in Section \ref{sec_velocity}. 

The 4-square of the velocity vector is $U^{\mu}U_{\mu}=\frac{c^2}{1-u^2/c^2}-\frac{u^2}{1-u^2/c^2}\equiv c^2$. Thus it is clearly a relativistic invariant, just like any 4-product of any two 4-vectors. 

The mass of the particle $m$ as measured in the proper (co-moving) frame -- also known as {\bf rest mass} in the older literature -- is a fundamental property of the particle and a relativistic invariant. Indeed, the mass of the ball at rest near the tracks as measured in the track frame does not care about how fast the train is going. Thus if we further multiply the 4-vector of velocity by $m$ we will obtain another 4-vector:
\begin{equation}
P^{\mu}=mU^{\mu}=\left(\frac{mc}{\sqrt{1-u^2/c^2}},\frac{m\vec{u}}{\sqrt{1-u^2/c^2}}\right).
\end{equation}

We look at the non-relativistic limit of the components of this 4-vector. When $u\ll c$, the spatial components are simply $m\vec{u}$ -- the non-relativistic momentum. The time-like component can be re-written using the Taylor expansion in the powers of a small parameter $u/c$:
\begin{equation}
\frac{mc}{\sqrt{1-u^2/c^2}}= mc\left(1+\frac{u^2}{2 c^2}+O\left(\frac{u^4}{c^4}\right)\right)\simeq \frac{1}{c}\left(mc^2+\frac{mu^2}{2}\right).
\end{equation}
The first term is an additive constant that does not depend on the particle's velocity and depends only on the particle's mass. In collision problems in classical mechanics (e.g., between billiard balls) the mass of the system before and after the collision is usually the same, and thus this term would be conserved  during the collision, which is why it is typically irrelevant for classical mechanics. The second term is the kinetic energy of the particle in the non-relativistic case, divided by $c$. 

Since in the non-relativistic limit the components of this new 4-vector turn into energy and momentum, we will assume for the moment that those are in fact the correct relativistic definitions of energy and momentum and see how they would work in relativistic collision problems. Thus, we have {\bf the 4-vector of energy and momentum}:
\begin{equation}
P^{\mu}=\left(\en/c,\vec{p}\right),\label{eq_momenergy}
\end{equation}
where
\begin{equation}
\en=\frac{mc^2}{\sqrt{1-u^2/c^2}}\mbox{ and }\vec{p}=\frac{m\vec{u}}{\sqrt{1-u^2/c^2}}\label{eq_momen_def}
\end{equation}
are the relativistic expressions for energy and momentum. They are often written using the Lorentz factor of the particle itself, $\gamma_u=1/\sqrt{1-u^2/c^2}$:
\begin{equation}
\en=\gamma_u mc^2\mbox{ and }\vec{p}=\gamma_u m\vec{u}.\label{eq_en}
\end{equation}
Previously we used $\beta$ and $\gamma$ to denote the motion of the $S'$ frame relative to the $S$ frame, but in equations (\ref{eq_en}) $\gamma_u$ refers to the motion of the particle as measured in some frame.  In these notes I try to use $\vec{u}$ and $\gamma_u$ for the particles and $\vec{v}, \beta$ and $\gamma$ for the frames, but many problems will not be so accommodating; you need to figure out from the context what the given values refer to (problem \ref{prob_lorentz}). 

The 4-square of the 4-vector of energy and momentum should be a relativistic invariant:
\begin{equation}
P^{\mu}P_{\mu}=\frac{\en^2}{c^2}-p^2=\frac{m^2c^2}{1-u^2/c^2}-\frac{m^2u^2}{1-u^2/c^2}=m^2 c^2, \label{eq_disp}
\end{equation}
which includes only the proper (co-moving, or rest) mass measurement and the speed of light and thus is a constant independent of the reference frame. Equation (\ref{eq_disp}) is the relativistic analog of the non-relativistic relationship between kinetic energy and momentum: $\en_{\rm kin}=p^2/2m$. For solving dynamics problems in Special Relativity, you will find it useful to remember all the permutations of equation (\ref{eq_disp}): 
\begin{equation}
\en=\sqrt{p^2c^2+m^2c^4}; \mbox{ and } p=\sqrt{\en^2/c^2-m^2 c^2}; \mbox{ and } mc^2=\sqrt{\en^2-p^2 c^2}.\label{eq_permute}
\end{equation}
In classical mechanics, the analogous expressions are
\begin{equation}
\en_{\rm kin}=p^2/(2m); \mbox{ and } p=\sqrt{2m\en_{\rm kin}} \mbox{ and } m=p^2/(2\en_{\rm kin}).
\end{equation}
They can be obtained as limiting cases of (\ref{eq_permute}) when $c\rightarrow \infty$, but remembering that the kinetic energy in classical mechanics corresponds to the difference between the total energy and the rest energy: $\en_{\rm kin}=\en-mc^2$. 

In the 4-vector (\ref{eq_momenergy}), all four components have the dimensions of momentum. Another commonly used version of the energy-momentum 4-vector is $(\en, \vec{p}c)$, which is different from (\ref{eq_momenergy}) by the invariant factor $c$ and in which all four components have the dimensions of energy. We will use both of these 4-vectors interchangeably. 

We have verified that the new definitions of energy and momentum revert to the old classical ones in the non-relativistic case $u\ll c$. We will now verify that the new definitions of energy and momentum abide by the principle of relativity. Suppose we have a collision of particles in which energy and momentum (as defined by the relativistic expressions \ref{eq_momen_def}) are conserved in some frame $S$: $\en_{\rm before}=\en_{\rm after}$ and $\vec{p}_{\rm before}=\vec{p}_{\rm after}$. Are the energy and momentum conserved in other inertial frames $S'$? Because $\en$ and $\vec{p}$ are components of the 4-vector of energy and momentum, they transform according to the Lorentz transform, and we now know how to calculate these values in $S'$:
\begin{equation}
\frac{\en'_{\rm before}}{c}=\gamma \frac{\en_{\rm before}}{c}-\beta \gamma p_{x, {\rm before}}.
\end{equation}
But because of the energy and momentum conservation in frame $S$, this is equal to $\gamma \en_{\rm after}/c-\beta\gamma p_{x, {\rm after}}$, which in turn is $\en'_{\rm after}/c$. Thus, energy is conserved in $S'$. Similarly it can be demonstrated that all the components of momentum are conserved in $S'$ as well. Therefore, unlike the definitions of energy and momentum from classical mechanics, the new definitions satisfy Einstein's Relativity Principle. 

\begin{prob} Section \ref{sec_velocity} resulted in complicated-looking equations for velocity transforms that look nothing like the Galilean transform, but not much like the Lorentz transform, either. So where is that famed mathematical beauty of the Special Relativity theory? Discover the beauty by applying the Lorentz transform to the 4-vector of velocity. Using one of the transformation equations, eliminate $1-u^2/c^2$ terms from the others and derive the velocity transform equations which should be the same as those we obtained in Section \ref{sec_velocity}.  
\label{prob_vel}
\end{prob}

\begin{prob} The Lorentz factor of a particle is $\gamma_u$. What Lorentz factor would be measured by an observer moving with velocity $0.6 c$ perpendicular to the particle? 

{\bf Solution.} We need to distinguish between the Lorentz factor of the particle and the Lorentz factor of the moving observer; we denote the latter with subscript `obs'. The frame where the particle moves with $\gamma_u$ is the frame $S$ and the frame co-moving with the observer is $S'$. Our derivation of the Lorentz transform and velocity transform equations assumed that $S'$ moves along the $x$-axis; therefore, we point the $x$-axis in the direction of motion of the observer. We know that the particle moves perpendicular to the observer in the $S$ frame, so we can point the $y$-axis in the direction of the particle. 

Method 1 (without the 4-vectors) involves the following steps. First, we convert $\gamma_u$ to velocity $u$. Second, we notice that this velocity is directed along the $y$-axis in the $S$ frame, so $\vec{u}=(0,u,0)$. Third, we apply the velocity transform from $S$ to $S'$ using equations (\ref{eq_vel_tran1})-(\ref{eq_vel_tran3}). Fourth, we calculate $\gamma_u'$ in the $S'$ frame from $\vec{u'}$. After one page of algebra, we get an answer.

Method 2 (with 4-vectors) involves exactly one line of calculations. $\gamma_u c$ is the time-like component of the velocity 4-vector of the particle $U^0$; thus, to solve the problem it is sufficient to calculate $U'^0=\gamma_u' c$, where $\gamma_u'$ is exactly what the problem is asking. The Lorentz transform for the time-like component of the velocity 4-vector is $U'^0=\gamma_{\rm obs}U^0-\beta_{\rm obs}\gamma_{\rm obs}U^1$, where $U^1=\gamma_u u_x$ and $u_x$ is the projection of the particle's velocity vector on the direction of the velocity of the observer (frame $S'$). But we are told that in the frame $S$, the observer and the particle are moving perpendicular to one another; thus, $u_x=0$. We obtain $\gamma_u'=\gamma_{\rm obs}\gamma_u=\frac{1}{\sqrt{1-(0.6)^2}}\gamma_u=5\gamma_u/4$.
\label{prob_lorentz}
\end{prob}

\begin{prob} 1 eV (electron-Volt) is defined as the energy an electron gains upon traversing a potential difference of 1 Volt. Calculate 1 eV in ergs and in Joules. What are the masses of the proton, neutron and electron in cgs and SI (MKS) systems (two significant digits and the power of 10)? Calculate the rest energy $mc^2$ of these particles in eV. 
\end{prob}

\begin{prob} Verify that in the collision problem we considered in Section \ref{sec_need} the relativistic energy and the relativistic momentum are conserved in both $S$ and $S'$ frames. 
\end{prob}

\begin{prob} 4-vectors are very special sets of 4 values, and we will not encounter all that many different 4-vectors in this course. One 4-vector which we will not discuss in detail is the 4-vector of charge and current density, $J^{\mu}=(c\rho, j_x, j_y, j_z)$, where $\rho$ is the electric charge density per unit volume and $\vec{j}$ is a 3D vector whose absolute value equals the amount of charge passing through a unit surface area per unit time and whose direction is along the charge flow. The three components $j_x, j_y, j_z$ are the projections of this vector onto the spatial axes. (a) Construct a quantity from a combination of values $\rho, j_x, j_y$ and $j_z$ that is Lorentz invariant. (b) Consider a charged cube in frame $S$; for example, we take a negatively charged one. On the microscopic level, this means that there is an excess of electrons relative to positive ions. We start moving relative to the cube along one of its edges (frame $S'$). Does the number of electrons change? Does the number of ions change? Does the size of the cube change? If so, how? Use this information and the definition of charge density (charge per unit volume) to derive the transformation of the charge density from $S$ to $S'$. (c) Now do the same calculation but use the 4-vector notation and the Lorentz transform. Your answer should be the same as in (b)! (d) The 4-vector $J^{\mu}$ can be obtained from the 4-vector of velocity $U^{\mu}$ by multiplying the latter by a relativistic invariant. Which one? What is its physical meaning?
\end{prob}

\begin{prob}
A proton has energy 3 GeV. What energy will be measured by an observer moving parallel to the proton at 0.8c? What is the kinetic energy in the lab frame and in the observer's frame?

{\bf Solution.} Method 1 (no 4-vectors) would be rather awful in this case. The steps involved are: (1) using proton energy $\en$, calculate its Lorentz factor $\gamma_u$; (2) from that, calculate velocity $u$; (3) from that, using the velocity transform (\ref{eq_vel_tran1}), calculate $u'$; (4) from that calculate $\gamma_u'$ and finally $\en'$. 

Method 2 (with 4-vectors) involves two lines of calculations. We know that the energy and momentum of this particle form a 4-vector, which transforms according to the Lorentz transform:
\begin{equation}
\left(\begin{array}{c} \en'/c \\ p_x' \\ p_y' \\ p_z' \end{array}\right)=\left(\begin{array}{cccc} \gamma & -\gamma\beta & 0 & 0 \\ -\gamma\beta & \gamma & 0 & 0 \\ 0 & 0 & 1 & 0 \\ 0 & 0 & 0 & 1 \end{array}\right)\left(\begin{array}{c} \en/c \\ p_x \\ p_y \\ p_z \end{array}\right)
\end{equation}
As always, we need to put the $x$ axis in the direction of the moving observer, because that was the assumption underlying our derivation of the Lorentz transform. With this coordinate system, only the $x$ component of the proton's momentum is non-zero, with $p_x=p$ (total momentum) since the observer and the proton are moving in the same direction. Therefore we find from the first equation of the Lorentz transform:
\begin{equation}
\en'=\gamma \en - \beta \gamma p c.
\end{equation}
In this equation $\beta$ and $\gamma$ belong to the observer and $\en$ belongs to the proton, but we do not know $p$. One way to do this would involve calculating $\gamma_u$ from $\en$, then the velocity of the proton $u$ and then its momentum from its relativistic definition (\ref{eq_momen_def}). The better way is to use one of the handy relationships between energy, momentum and mass (\ref{eq_permute}). We find: 
\begin{equation}
\en'=\gamma\en-\beta\gamma\sqrt{\en^2-m^2 c^4}.
\end{equation}
We plug in $\beta=0.8$, $\gamma=5/3$, $\en=3$ GeV and $mc^2=0.938$ GeV$\simeq 1$ GeV. We find $\en'=1.2$ GeV. In the frame which is trying to catch up to the proton, the proton is moving slower and its measured energy is lower. The kinetic energy of the proton in frame $S$ is $\en_{\rm kin}=\en-mc^2=2.1$ GeV and $\en'_{\rm kin}=\en'-mc^2=0.3 GeV$. 
\end{prob}

\subsection{Forms of energy}
\label{sec_forms}

Even when a particle is not moving, according to the relativistic definition of energy it possesses energy $\en_0=mc^2$. Einstein regarded the discovery of this {\bf rest energy} to be the most significant consequence of Special Relativity. The value of any theory is in its ability to produce testable predictions; there is now plenty of experimental evidence for the existence of this rest energy, so this prediction of Special Relativity theory was abundantly confirmed.  

The key experiment proving the existence of rest energy would be one in which massive particles disappear, releasing their rest energy in some other form. Conversely, one should also be able to create new massive particles from other forms of energy. Both processes have been amply observed, confirming {\bf the equivalence of energy and mass}:
\begin{itemize}
\item An electron-positron pair annihilates producing two photons. Thus something that has mass turns into something that does not have mass, but carries energy in the form of electromagnetic radiation. 
\item Particles can decay into other particles whose final mass is not the same as the initial mass because part of the energy is carried away in the form of kinetic energy. Perhaps the most famous example is the neutron decay: $n\rightarrow p+e+\bar{\nu_e}$. The mass of the neutron is 939.565378 MeV/$c^2$, the mass of the proton is 938.272046 MeV/$c^2$, and the mass of the electron is 0.5100998928 MeV/$c^2$. The mass of the anti-neutrino remains unknown. From the discovery of ``neutrino oscillations'' (i.e., the conversion of neutrinos from one type to another), the neutrino mass is known to be greater than zero, but it is constrained to be less than 0.3 eV/$c^2$ by astronomical observations. Thus the neutron is 0.78 MeV/$c^2$ heavier than the sum of the particles it decays into. The excess energy is released in the form of kinetic energy of decay products (i.e., the decay products are moving even if the neutron was at rest). 
\item Conversely, slamming two relatively light-weight particles together at high velocities to produce more massive particles is the whole point of particle accelerators. The faster the initial particles move, the greater the mass of the particles that can be produced. In this case, the kinetic energy of the initial particles is converted into the mass of the final product. For example, the Higgs boson discovered at the LHC in 2013 with a mass of 125 GeV/$c^2$ is produced in collisions of protons whose mass is only 0.938 GeV/$c^2$. 
\item More exotic examples include conversion of energy associated with a static magnetic field into massive particles which has been observed in astrophysical objects called magnetars. These are very compact highly magnetized stars, where the magnetic field can reach and exceed $10^{14}$ Gauss$=10^{10}$ Tesla. Because of quantum processes, this magnetic field can spontaneously decay into electron-positron pairs. In this case, the rest energy of the electrons and positrons $m_e c^2$ is created out of the energy density associated with the magnetic field $B^2/8\pi$ (in cgs units). The result is that the magnetic field of the star diminishes with time as its energy is transferred into the particles.
\item Another type of energy that needs to be taken into account in the energy conservation laws is {\bf binding energy}. For example, the nucleus of the helium atom (also called an $\alpha-$particle) is made of two neutrons and two protons, but its mass $m_{\alpha}=3727.379240$ MeV/$c^2$ is smaller than the sum of the constituent masses by 28.2924 MeV/$c^2$. The difference is due to the binding energy which can be thought of as the potential energy of the interactions among the constituents and which is usually negative, because a positive energy is required to break things apart. 
\item In the previous example, the binding energy is about $\sim$1\% of the total energy budget. Binding energy is important to take into account in nuclear physics and in particle physics. In chemistry, however, the binding energy of interaction between electrons and nuclei is usually much less significant. For example, the mass of the hydrogen atom is slightly smaller than the sum of the masses of the proton and the electron. A photon with energy 13.6 eV is required to rip apart the hydrogen atom (this value is so important in atomic physics that it has its own name and is used as an energy unit -- 1 Rydberg). This energy constitutes only a few$\times 10^{-8}$ of the rest energy of the hydrogen atom; such a small fractional difference in mass is difficult to detect even in modern experiments. Thus, in chemical reactions to a high degree of accuracy we recover the mass conservation law first proposed by Lomonosov in 1748 and Lavoisier in 1774. 
\end{itemize}

Modern literature and older literature differ somewhat on the usage of the word ``mass''. In older literature the term ``relativistic mass'' is sometimes encountered, meaning $\gamma_um$, but in modern literature ``mass'' by default means rest (proper) mass. Even more frequently, mass is given in energy units, in which case it means the rest energy of the particle divided by $c^2$. An example of such usage would be ``In LHC, protons with mass 1 GeV are moving with energy 7 TeV''. In this course we will try to remember the proper dimensions and keep usage such as ``proton mass is approximately 1 GeV/$c^2$''.

\begin{prob} Two identical particles $P$, one with energy $\en_1$ and another with $\en_2$, collide head-on and the result is one new particle $X$. Calculate the energy, the mass and the Lorentz factor of the new particle. 

{\bf Solution.} Just like in classical mechanics, this problem is solved by applying energy and momentum conservation. From energy conservation, $\en_1+\en_2=\en_X$ for the energy of the new particle. The momentum conservation $p_1-p_2=p_X$ can be re-written using the energy-momentum relationships derived above:
\begin{equation}
\sqrt{\en_1^2-m_p^2c^4}-\sqrt{\en_2^2-m_p^2c^4}=\sqrt{\en_X^2-m_X^2c^4}.
\end{equation}
We can plug in $\en_X=\en_1+\en_2$ and express $m_X$ from this relationship; then the Lorentz factor is $\gamma_X=(\en_1+\en_2)/(m_X c^2)$. 

This problem is solved relatively easily just by writing energy and momentum conservation explicitly, but in general the 4-vector notation can be quite helpful in avoiding lengthy calculations. We see how it applies in this simple case. The equality
\begin{equation}
P_1^{\mu}+P_2^{\mu}=P_X^{\mu}
\end{equation}
encompasses both energy and momentum conservation: if 4-vectors are equal, all their components must be equal. We take a 4-square of this expression: 
\begin{equation}
P_1^{\mu}P_{1,\mu}+2P_1^{\mu}P_{2,\mu}+P_2^{\mu}P_{2,\mu}=P_X^{\mu}P_{X,\mu}.
\end{equation}
For any massive particle, the 4-square of its energy-momentum 4-vector is related to its mass, and we can re-write:
\begin{equation}
m_p^2 c^2+2(\en_1\en_2/c^2-\vec{p_1}\vec{p_2})+m_p^2c^2=m_X^2c^2.\label{eq_simple_prob}
\end{equation}
 The momenta are pointing in opposite directions; thus, 
\begin{equation}
-\vec{p_1}\vec{p_2}=p_1p_2=\sqrt{\en_1^2/c^2-m_p^2 c^2}\sqrt{\en_2^2/c^2-m_p^2 c^2}
\end{equation} 
and we get the answer for $m_X$ from equation (\ref{eq_simple_prob}). 

Both the `classical mechanics' method and the 4-vector method should give the same answer. It is useful to try out both methods to see what sort of problems benefit most from the 4-vector approach.
\end{prob}

\begin{prob} In the largest linear accelerator in the world at SLAC (Stanford), electrons and positrons can be accelerated to 50 GeV. 

(a) If the electrons attain max energy, what is their Lorentz factor? What is their velocity?

(b) We are in the frame of the SLAC building, at rest relative to Earth. An electron moving with energy 50 GeV slams into a positron at rest. What is the energy (in eV; measured in the Earth frame) of the heaviest particle that can in principle be created in this process? 

(c) Same question as in (b), but if the target positron is moving head-on toward the electron at 50 GeV; if the target positron is moving at a right angle relative to the electron at 50 GeV.

(d) Now calculate the masses of the created particles (in eV) for the same three collisions: with positron at rest, with positron moving head-on, and with positron moving at a right angle. (Note: we cannot really arrange for a 50 GeV positron to move at a right angle in a linear accelerator, but we stretch our imagination here a bit.)

{\bf Solution.} 

(a) $\gamma=\en_{\rm el}/m_{\rm el}c^2=50\times 10^9/0.511\times 10^6\simeq 98000$. Velocity is very close to $c$; the fractional difference is $1-\beta\simeq 1/(2\gamma^2)=5\times 10^{-11}$, very small indeed (see eq. \ref{eq_gamma_ultrarel} and subsequent discussion).

(b) If the heaviest particle is created, then the entire product of the collision (denoted `X') is moving in one piece. By energy conservation, $\en_{\rm electron}+\en_{\rm positron}=\en_X$. Here $\en_{\rm el}=50$ GeV, $\en_{\rm pos}=0.511$ MeV, so the second term is negligibly small compared to the first one. Thus $\en_X\simeq 50$ GeV. 

(c) Energy conservation still applies, but now $\en_{\rm pos}=50$ GeV, so $\en_X=100$ GeV in both cases.

(d) To find the mass of the particle, we need to solve both the energy conservation and the momentum conservation together. This can be done explicitly, using the relationship between energy and momentum for relativistic particles, $\en=\sqrt{p^2c^2+m^2c^4}$. However, the most elegant way is via 4-vectors. 

In 4-vector notation, energy and momentum conservation: $P_{\rm el}^{\mu}+P_{\rm pos}^{\mu}=P_X^{\mu}$. Now we take a 4-square of this: $P_{\rm el}^{\mu}P_{\rm el,\mu}+P_{\rm pos}^{\mu}P_{\rm pos,\mu}+2P_{\rm el}^{\mu}P_{\rm pos,\mu}=P_X^{\mu}P_{X,\mu}$. The 4-square of the energy-momentum 4-vector of any particle is $m^2c^2$, with $m$ being the mass of this particle. We also remember that positrons and electrons have the same mass. Thus.
\begin{equation}
2 m_{\rm el}^2 c^2+2P_{\rm el}^{\mu}P_{\rm pos,\mu}=m_X^2 c^2. \label{eq_part}
\end{equation}

For the three scenarios in question, we calculate $P_{\rm el}^{\mu}P_{\rm pos,\mu}$:
\begin{itemize}
\item For the target positron at rest, the two 4-vectors of the colliding particles are $(\en_{\rm el}/c,p_{\rm el},0,0)$ and $(m_{\rm pos}c,0,0,0)$, so that their 4-product is $\en_{\rm el} m_{\rm el}$.
\item For the opposing beams, the 4-vectors are $(\en_{\rm el}/c, p_{\rm el},0,0)$ and $(\en_{\rm el}/c, -p_{\rm el}, 0,0)$. For particles with large Lorentz factors (called `ultra-relativistic particles'), $p\simeq \en/c$, and thus the 4-product is $\simeq 2\en_{\rm el}^2/c^2$.
\item For the perpendicular beams, the 4-vectors are $(\en_{\rm el}/c, p_{\rm el},0,0)$ and $(\en_{\rm el}/c, 0, p_{\rm el},0)$, with a 4-product of $\en_{\rm el}^2/c^2$. 
\end{itemize}

Plugging this into equation (\ref{eq_part}) and neglecting the rest energy of the electron by comparison to $\en_{\rm el}=50$ GeV, we find for the three scenarios, respectively, $m_X c^2=\sqrt{2\en_{\rm el} m_{\rm el} c^2}=226$ MeV; $m_X c^2=2 \en_{\rm el}=100$ GeV; $m_X c^2=\sqrt{2}\en_{\rm el}=71$ GeV. 

It is convenient to do such calculations in energy units (and therefore useful to know the rest energies of several most important particles to a couple of significant digits!). For example, in the first scenario let us translate all energies to MeV:
\begin{equation}
m_Xc^2=\sqrt{2\en_{\rm el} m_{\rm el} c^2}=\sqrt{2\times 50,000 MeV\times 0.511 MeV}=226 MeV.
\end{equation}
\end{prob}

\subsection{Ultra-high energy cosmic rays}

The highest Lorentz factor record-holders known to humanity are so-called ultra-high energy cosmic rays (UHECRs). They are probably accelerated by shock waves propagating from exploding stars or in jets emanating from black holes, but we do not yet know for sure, and it remains an area of active research. A typical energy of an UHECR is $\en=3\times 10^{20}$ eV. Assuming that it is a proton (we do not know that for sure, either; it could be a heavy nucleus, for example), we find $\gamma_u=\en/(mc^2)=3\times 10^{20}$eV/$10^9$eV$=3\times 10^{11}$. 

The kinetic energy of a baseball ($m=142$ grams) traveling at 100 km/hour is $\en_{\rm kin}=mu^2/2=5.5\times 10^8$ erg$=55$ Joules$=3.4\times 10^{20}$ eV. Thus one elementary particle packs as much energy as a regular baseball, which contains $\sim 10^{26}$ such particles! 

If the baseball moving with this kinetic energy were to hit you on the head, it would be rather noticeable. So why do we not hear alarming media reports about cosmic rays knocking people down left and right? Here are a few hypotheses that students typically express:

(1) The Earth is protected from this stuff by its magnetic field which deflects these particles. -- This is a good idea, but not at this particle energy. We will see in Section \ref{sec_dynamics} that the magnetic field of the Earth is much too small to affect the trajectory of particles with such enormous Lorentz factors. The greater the Lorentz factor, the greater the ``inertia'', the more difficult it is to deflect it. 

(2) Perhaps they do hit us, but we cannot feel it and the particles go right through. -- Let us imagine that we are the cosmic ray on approach to a brave student. What do we see? In our frame the brave student is squashed to a very, very flat student-shaped pancake (by the enormous factor of $\gamma_u$ relative to the normal thickness of the brave student). The main question is whether the student-shaped pancake is opaque or transparent to us. If it is transparent, then the cosmic ray will indeed go right through. If it is opaque, then the cosmic ray will collide with the brave student.

\begin{figure}[htb]
\centering
\includegraphics[scale=0.8, clip=true, trim=1cm 10cm 0cm 0cm]{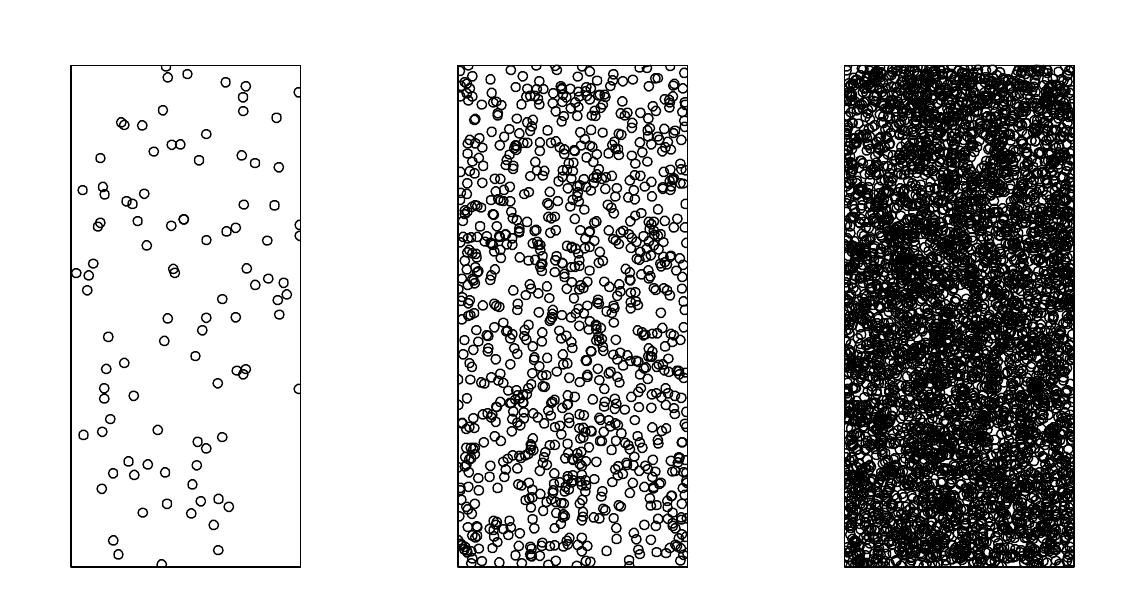}
\caption{Side views of boxes containing 100 particles, 1000 particles and 5000 particles. As the number of particles increases, the box becomes more opaque. The probability that a cosmic ray can get through can be estimated as the fraction of the area not covered by any particles.}
\label{pic_cr}
\end{figure}

The student is made of nuclei and electrons. In the classical picture the nuclei are represented by spheres with typical sizes a few$\times 10^{-13}$ cm, and they are surrounded by electrons that are $10^{-8}$ cm away. Thus an atom is quite empty, and the brave student is made mostly of empty space, with a bit of nuclei mixed in. But even though nuclei are tiny, there are lots of them. What is the probability that a cosmic ray will hit one? Looking at the brave student from the cosmic ray perspective, this probability is low if the total area of nuclei covers only a fraction of the student's area, but close to 1 if the nuclei cover most of the area or even overlap each other (Figure \ref{pic_cr}). 

We will approximate the brave student by a right rectangular prism, with height 2 m, width 50 cm, thickness 50 cm and density of water $\rho=1$ g/cm$^3$ (if you calculate the mass, you will see that I am off by a factor of a few, which is a lot in terms of the human experience, but good enough for the order-of-magnitude calculation here). Let $N$ be the total number of atomic nuclei in the student. The total mass of all nuclei is the total mass of the student: $6m_pN\simeq \rho V$, where we have assumed that the student is made of water (molecular weight $=18$, made of three nuclei, so molecular weight per nucleus $=6$). The cross-section for collision with each nucleus is not known at these energies, but we approximate it with its geometric cross-section; if the typical size of the nucleus is a few$\times 10^{-13}$ cm, then the cross-section is $\sigma \simeq 10^{-25}$ cm$^2$. This estimate is consistent with the recent LHC ATLAS measurements \citep{atla11}, though these do not quite reach the energies we are interested in. 

Then the ratio of area covered by nuclei to the area of the student is equal to $\rho T \sigma/(6 m_p)$, where $T$ is the thickness of the student. The value $\rho T/(6 m_p)$, measured in cm$^{-2}$ or m$^{-2}$ is called {\bf column density} and is a relativistic invariant. Indeed, from the standpoint of the cosmic ray, the thickness of the student is greatly contracted (by $\gamma$), but the density $\rho$ is increased by the same factor. The physical meaning of this value is the total number of particles within a column with a base 1 cm$^2$ (or 1 m$^2$ in MKS units). Furthermore, the cross-section $\sigma$ is also a relativistic invariant, since the sizes perpendicular to the motion do not change. Thus the probability of hitting the student is the same in the cosmic ray frame and in the student frame, which makes sense. Plugging in all the values, we find the ratio of area covered by nuclei to total area available to be 0.5. 

This is not comforting at all! This means that if there were such energetic particles around, they would have a decent chance of actually hitting a nucleus in our body. If they did, what would happen? When such an energetic cosmic ray hits a stationary nucleus, the energy of the collision is enormous (Problem \ref{prob_maxcoll}). Many new particles get created, also with high Lorentz factors. If they are long-lived enough, each of them will have a good chance of smashing into another nucleus, and that would in turn create another big batch of particles. We have an ever increasing number of particles at each subsequent stage; this is called a {\bf cosmic ray shower}. 

How long does it proceed? Energy conservation applies to the whole process. After the first collision, the energy of the initial particle is divided into many parts, each of those in turn is subdivided more. This proceeds until the energy of the collision is no longer sufficient to create new particles. But clearly the cosmic ray can do quite a bit of damage destroying a bunch of the nuclei in several stages of the cosmic ray shower. At the end, many less-energetic particles are produced which cannot destroy nuclei, but can do other types of damage, like ionize atoms. In other words, a cosmic ray shower is rather hazardous for health, so we would like more reassurance that this is not actually going to happen to us.

(As an aside, in late 1970s there was an incident at the particle accelerator in the Soviet Union in which a scientist leaned in front of the particle accelerator beam just as the safety mechanism failed. Although it was believed that he had received far more than a lethal dose of radiation, not only did he survive but he was able to get a Ph.D. and to continue his scientific career, though he did experience some hearing loss and epilepsy. The energies of the protons in this accelerator beam were significantly smaller than those of UHECRs, but plenty sufficient for kicking nuclei out of atoms and destroying them to produce quickly-decaying reaction products which would do further damage.) 

(3) They do not hit us because they hit the Earth very rarely. -- This is true. There are only a handful (one or two!) of the ultra-high energy cosmic rays hitting the entire planet per year. However, there is a broad range of energies for cosmic rays, and somewhat less energetic particles arrive much more frequently\footnote{\url{http://www.physics.utah.edu/~whanlon/spectrum.html}}. 

(4) The atmosphere protects us. -- Exactly. Applying the same arguments to the air in the atmosphere as we did to the pancaked student, we find that the ratio of the area covered by nuclei to available area is $\rho_{\rm atm}T_{\rm atm}\sigma/(14.5 m_p)$, where $\rho_{\rm atm}=1.2$ km/m$^3$ is the density of the atmosphere close to the ground, $T_{\rm atm}=8000$ m is the scale height of the atmosphere and 14.5 is the mean molecular weight per nucleus. Plugging in the numbers, we find this ratio to be 5. So a typical cosmic ray would encounter at least 5 nuclei on the way to the ground. 

\begin{figure}[!htb]
\centering
\includegraphics[scale=0.6]{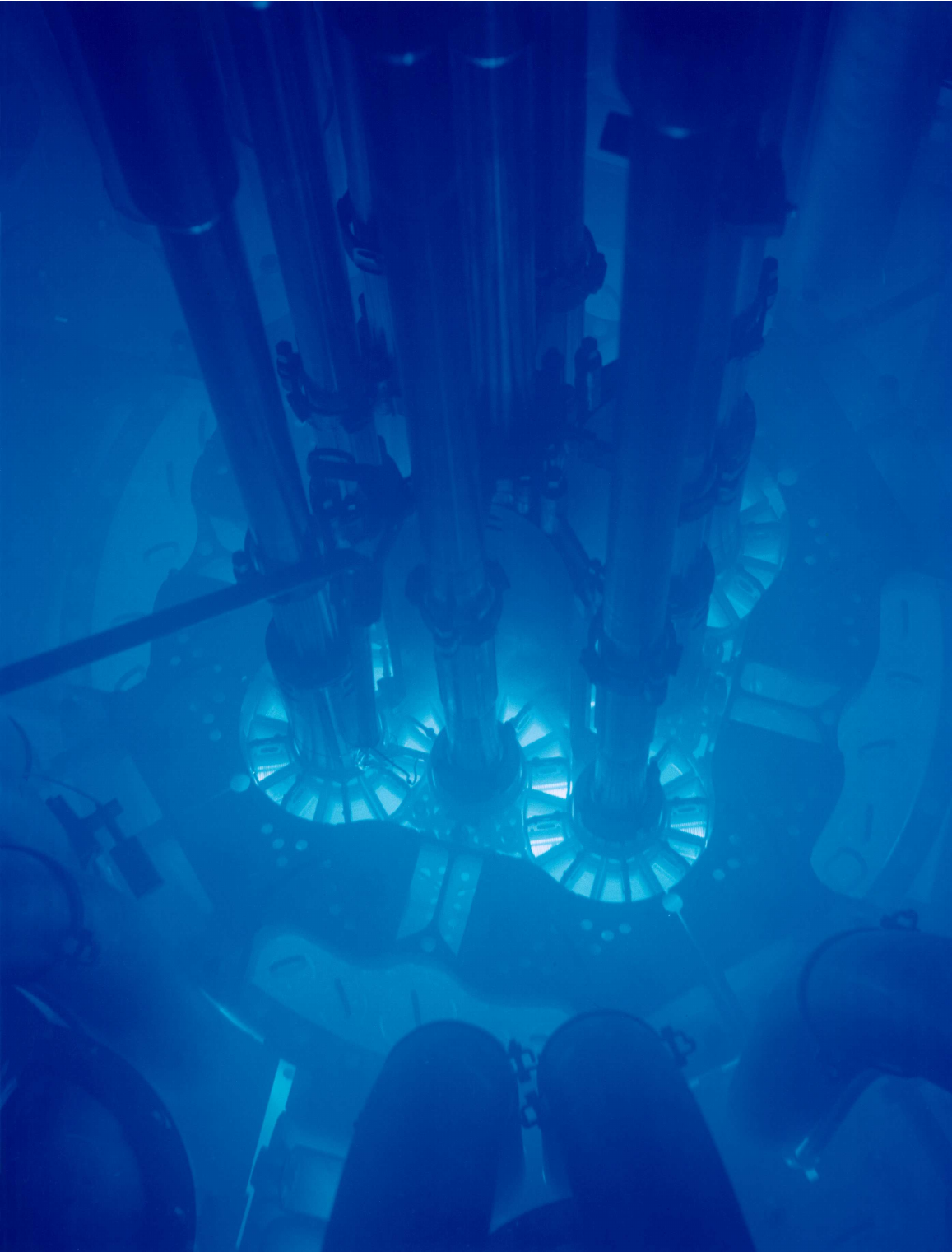}
\caption{Cherenkov radiation (produced when a charged particle moves faster than the local speed of light in a medium) glowing in the core of the Advanced Test Reactor. From \url{http://en.wikipedia.org/wiki/Cherenkov_radiation}.}
\label{pic_cher}
\end{figure}

Therefore, cosmic rays are predominantly destroyed in the upper layers of the atmosphere. They create cosmic ray showers (Figure \ref{pic_crsim}) which are detectable by various instruments on the ground, including those that use Cherenkov radiation produced by the shower particles (Figure \ref{pic_cher}). Atmosphere protects us from cosmic rays, but in space cosmic rays are bad news and do damage to the equipment over time (there are not too many students in space to hit). You can download a raw image from the Hubble Legacy Archive\footnote{\url{http://blog.galaxyzoo.org/2010/04/12/how-to-handle-hubble-images/}} and establish that it is all covered by bright streaks and spikes due to cosmic rays. 

\begin{prob}
Two particles, one with energy $\en_1$ and one with energy $\en_2$ collide, and the result of the collision is one new particle. What is the maximal mass of this particle? What is the maximal mass of the particles that can be created in a collision of a cosmic ray with the nucleus in the upper atmosphere? -- {\bf Solution}. We write energy and momentum conservation in the 4-vector notation: $P^{\mu}_1+P^{\mu}_2=P^{\mu}_X$ and take a 4-square of this equation: $P^{\mu}_1P_{1,\mu}+2P^{\mu}_1P_{2,\mu}+P^{\mu}_2P_{2,\mu}=P^{\mu}_XP_{X,\mu}$, which gives us the mass of the new particle on the right-hand side:
\begin{equation}
m_1^2c^2+2\left(\frac{\en_1\en_2}{c^2}-\vec{p}_1\vec{p}_2\right)+m_2^2c^2=m_X^2c^2.
\end{equation}
To maximize $m_X$ we need to minimize $\vec{p}_1\vec{p}_2$. The minimal value is reached when the particles collide head-on and $\vec{p}_1\vec{p}_2=-p_1p_2$, where we can plug in the values of the momentum from equations (\ref{eq_permute}). We will derive a simplified expression for the case of an ultra-relativistic collision, where $\en_i\gg m_ic^2$, in which case $\en_i\simeq p_ic$ and rest energy of the particles can be neglected. Then we obtain $m_Xc^2\simeq 2\sqrt{\en_1\en_2}$. This is a calculation appropriate for LHC, where two protons moving head-on at 7 TeV are smashed into each other. So in principle the maximal mass of the particle that can be created in such collision is 14 TeV, and it would be sitting at rest in the accelerator frame, but in practice such collisions tend to produce many lighter, but highly relativistic particles. 

For the cosmic ray, the situation is different. Only the cosmic ray is ultra-relativistic, whereas the target nucleus in the upper atmosphere is at rest. Therefore, $\vec{p}_1\vec{p}_2=0$, since $\vec{p}_2=0$. Using the approximation $\en_1\gg m_1c^2$, we find $m_Xc^2\simeq\sqrt{2\en_1m_pc^2}=800 TeV$ for an ultra-high energy cosmic ray with $\en_1=3\times 10^{20}$ eV. 

Thus the collisions of cosmic rays with the Earth's atmosphere are much more energetic than anything that can be achieved in the lab today. On the eve of the first run of the LHC, newspapers around the world were spreading panic that the LHC collisions would create a black hole that would swallow us, and there were even lawsuits trying to prevent the science operations at the LHC on these grounds\footnote{\url{https://www.nytimes.com/2008/03/29/science/29collider.html}}. There are two problems with this hypothesis: (1) If this process occurs, it has been occurring in the atmosphere for the 4.5 Gyr of the existence of our planet, producing black holes that are up to a hundred times more massive than those at LHC; (2) If such black holes are created, they interact with surrounding matter only through gravity, which is very weak. You can show that they would go right through everything (before decaying into photons due to Hawking radiation) by comparing the gravitational force exerted by the passing black hole to the electrostatic forces binding together molecules and atoms or strong forces binding together atomic nuclei. It is rather puzzling to me that with such strong arguments in favor of ``there will be no terrible consequences even if black holes are produced'', the arguments against the black hole LHC catastrophe centered instead on the fact that the black hole production at the LHC was ``unlikely''. 
\label{prob_maxcoll}
\end{prob}

\clearpage
\begin{figure}[!htb]
\centering
\includegraphics[scale=0.8]{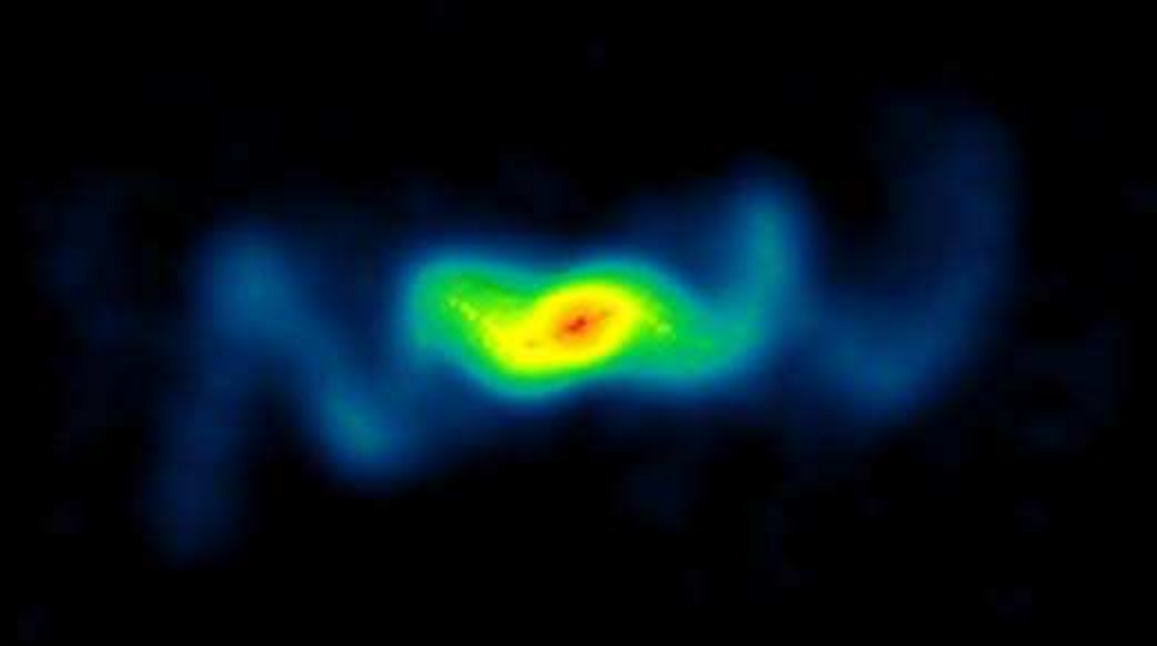}
\caption{A radio image of the cork-screw pattern left behind by mildly relativistic jets propagating from the binary star SS433 (\citealt{blun04}, downloaded from \url{http://blackholes.stardate.org/resources/article.php?p=mystery-of-ss443}). The image is made using the so-called ``false-color'' technique, in which the color is used to represent the intensity of light, not its spectral shape. In this image, the intensity declines from red to yellow to green to blue.}
\end{figure}

\clearpage
\section{Energy-momentum of photon, Doppler effect, aberration}

\subsection{Energy and momentum of massless particles}

We start with the expression $\en=mc^2/\sqrt{1-u^2/c^2}$. We will now consider the limit $m\rightarrow 0$, even though this is a bit dishonest, because in the derivation of the relativistic energy we had assumed a non-zero mass. Having said that, we see that as $u\rightarrow c$ the energy becomes infinite $\en \rightarrow \infty$, unless simultaneously we decrease the mass $m\rightarrow 0$, in which case we can arrange for a finite energy. 

Therefore Special Relativity makes two predictions (sort of):

(1) The velocity of massive particles can get arbitrarily close to $c$, but that requires arbitrarily large amounts of energy. Our physical laws and experiments suggest that infinite energy is not something we have at our disposal. Therefore, massive particles cannot move with $u=c$. The corollary is that none of the observers mentioned in these notes can move with $v=c$. It is tempting to say ``I know that observers cannot really move at $v=c$, but what would happen if they did?'' But this would be an example of a thought experiment (Section \ref{sec_timedil}) that does contradict the postulates of our theory: observers are dragging along with them a bunch of rulers and clocks and thus must have mass even in the most hypothetical of the thought experiments. Thus they cannot be accelerated to $v=c$ using a finite amount of energy. 

(2) If massless particles ($m=0$) have finite energy, they must be moving with $u=c$. This should be true for all massless particles, not just photons that lie at the foundation of Special Relativity. For these particles, the relationship between energy and momentum reads $\en=\sqrt{p^2c^2+m^2c^4}\rightarrow pc$. As you know from electricity and magnetism, this is indeed the correct relationship between the energy and momentum of an electromagnetic wave. The relationship
\begin{equation}
\en = pc \mbox{ for } m=0
\label{eq_massless}
\end{equation}
is usually the only property of massless particles needed to solve collision problems. Another useful property which follows from (\ref{eq_massless}) is that the 4-square of the 4-vector of energy and momentum of massless particles is zero:
\begin{equation}
K^{\mu}=(\en/c,\vec{p})\mbox{ and }K^{\mu}K_{\mu}=\en^2/c^2-p^2=0.
\end{equation}

\begin{prob}
Can a massive particle decay into a photon? -- {\bf Solution}. Let us assume this process is possible (illustrated in Figure \ref{pic_prob_decay}) and discuss energy and momentum conservation in this problem using three different methods. 

Method 1. Let $P^{\mu}$ be the 4-vector of energy and momentum of the initial particle and $K^{\mu}$ be the 4-vector of energy and momentum of the final photon. The laws of energy and momentum conservation during the decay require $P^{\mu}=K^{\mu}$. Let us take a 4-square of this equality: $P^{\mu}P_{\mu}=K^{\mu}K_{\mu}$. But the left-hand side of this is equal to $m^2c^2$ and the right-hand side is zero. Therefore energy and momentum cannot be conserved at the same time. Therefore, such process is impossible.

Method 2. Let us move into the frame co-moving with the particle before it decays. In this frame $S'$ before the decay the particle is at rest and its momentum $\vec{p}'$ is zero. Once the particle decays into a photon, the photon must move in some direction with velocity $c$, and therefore it will have some non-zero momentum. Therefore, momentum cannot be conserved in this frame, and such process is impossible.

\begin{figure}[htb]
\centering
\includegraphics[scale=0.8, clip=true, trim=1.0cm 6.5cm 7cm 14cm]{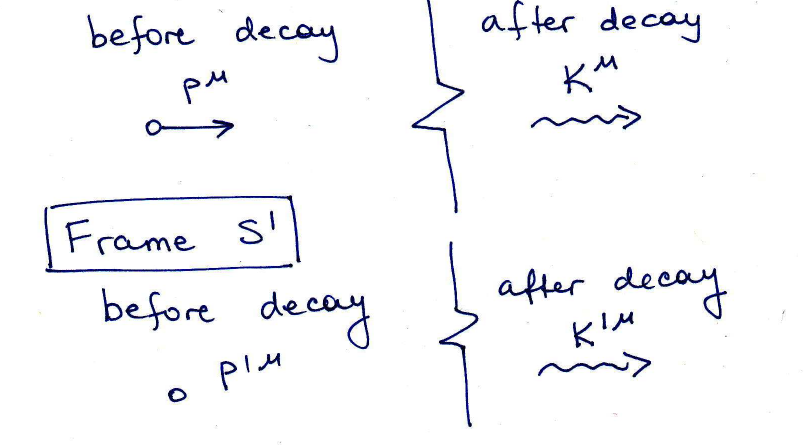}
\caption{Hypothetical decay of a particle into a photon viewed in two inertial frames}
\label{pic_prob_decay}
\end{figure}

Method 3. The two methods above are the {\bf relativistically invariant} methods of problem-solving. Another way to solve this problem is to use the methods familiar from classical mechanics and to write down energy and momentum conservation:
\begin{eqnarray}
\en_{\rm particle}=\frac{mc^2}{\sqrt{1-u^2/c^2}}=\en_{\rm photon};\\
p_{\rm particle}=\frac{mu}{\sqrt{1-u^2/c^2}}=\en_{\rm photon}/c.
\end{eqnarray}
We express $u$ from the first equation, plug it into the second equation and find that it cannot be satisfied. In this case it is possible, albeit a bit tedious, to use methods of classical mechanics, but in many problems in Special Relativity the standard methods of classical mechanics become prohibitively complex.

The conclusion is that energy and momentum conservation in Special Relativity prohibit massive particles from spontaneously decaying into photons.
\end{prob}

\begin{prob}
Can a free electron emit a photon? -- {\bf Solution.} Let us assume that such process is possible (Figure \ref{pic_prob_emit}). 

Method 1. The 4-vector of the electron before the emission of the photon is $P_b^{\mu}$. The 4-vector of the electron after the emission is $P^{\mu}_a$ and of the photon is $K_a^{\mu}$. Relativistically-invariant energy and momentum conservation is $P_b^{\mu}=P_a^{\mu}+K_a^{\mu}$. Let us take the 4-square of this expression:
\begin{equation}
P_b^{\mu}P_{b,\mu}=(P_a^{\mu}+K^{\mu}_a)(P_{a,\mu}+K_{a,\mu})=P_a^{\mu}P_{a,\mu}+2P_a^{\mu}K_{a,\mu}+K_a^{\mu}K_{a,\mu}.
\end{equation}
The left-hand side is equal to $m^2c^2$. The first term on the right-hand side is also equal to $m^2c^2$, so they cancel out. The last term on the right-hand side is zero. Therefore, energy and momentum conservation can be satisfied only if $P_a^{\mu}K_{a,\mu}=0$. 

Let us spell this product in terms of the components of the 4-vectors: $P_a^{\mu}K_{a,\mu}=\en_a\en_{ph}/c^2-p_ap_{ph}\cos\theta=p_{ph}(\en_a/c-p_a\cos\theta)=0$. But the energy of the massive particle $\en_a/c=\sqrt{p_a^2+m^2c^2}>p_a$, whereas $p_a\cos\theta\le p_a$. Therefore, $P_a^{\mu}K_{a,\mu}$ can never be 0 and therefore momentum and energy equations cannot be satisfied simultaneously.

Method 2. Let us assume that this process is possible and move into the frame co-moving with the electron before it emitted the photon. In this frame the electron is initially at rest, with total energy $\en'=mc^2$. After it emits a photon (which must have a non-zero momentum), the electron recoils with some unknown velocity $u'$ so that the momentum is conserved. The final energy of the system is $\gamma_u'mc^2+\en'_{ph}$. Because $\gamma_u'>1$ and $\en'_{ph}>0$, the final energy of the system is greater than before the electron emitted the photon. Thus, this process is impossible because the energy and the momentum cannot be conserved at the same time. 

\begin{figure}[htb]
\centering
\includegraphics[scale=0.8, clip=true, trim=0.5cm 16.5cm 0.5cm 1.5cm]{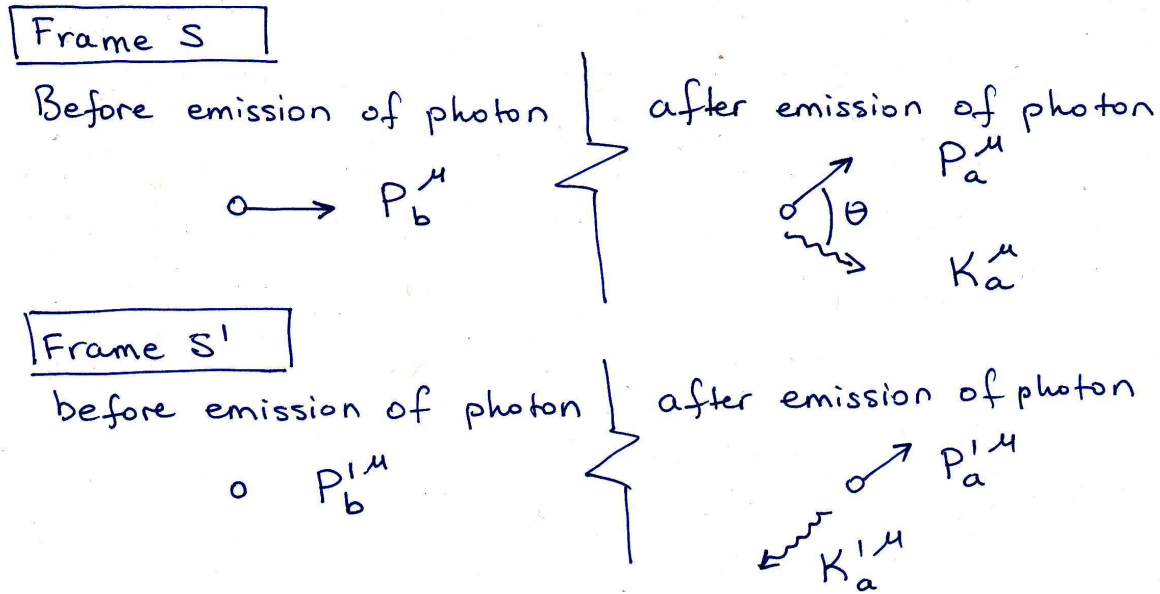}
\caption{Hypothetical emission of a photon by a free particle viewed in two inertial frames}
\label{pic_prob_emit}
\end{figure}

Method 3. Again, it is possible to write explicitly energy and momentum conservation and plow through various algebra, which is now more complicated than in the previous problem, and demonstrate that energy and momentum conservation equations are inconsistent with each other. But at this point this is too much algebra even for a simple problem, and we need to start using the methods of Special Relativity to avoid extensive computations. 

The conclusion of the problem is that a free (non-accelerating) particle cannot emit a photon. The corollary is that it is impossible ``to see'' a free particle, since ``seeing'' something involves detecting photons that it emitted. Of course charged particles can and do emit (e.g., electron moving in magnetic field produces cyclotron or synchrotron radiation), but not particles that are moving without acceleration. 
\end{prob}

\begin{prob} In the lab frame, a photon of energy $\en_1$ collides with a photon of energy $\en_2$ which is traveling in a perpendicular direction.

(a) Write down the energy-momentum 4-vectors for each photon and the total energy-momentum 4-vector of the entire system as measured in the lab frame.

(b) Compute the total energy of the photon pair in the center-of-momentum frame (i.e., the frame in which the total momentum is zero).

(c) If $\en_1=1$ keV, what is the minimum energy $\en_2$ needed to create an electron-positron pair via the process $\gamma+\gamma\rightarrow e^-+e^+$? (In particle reaction equations, $\gamma$ often denotes a photon.) 

{\bf Solution.} 

(a) Taking the $x$-axis to be along the direction of the first photon and the $y$-axis to be in the direction of the second photon, the 4-momenta are $(\en_1/c,\en_1/c, 0, 0)$ for photon 1 and $(\en_2/c, 0,\en_2/c, 0)$ for photon 2. The total 4-vector of energy-momentum is just the sum $((\en_1+\en_2)/c, \en_1/c,\en_2/c, 0)$.

(b) In any frame, the 4-square of the 4-vector of the energy-momentum of the system is the same $\en^2/c^2 - p^2 =\en'^2/c^2-p'^2$ and equal to its value calculated in the lab frame: $= (\en_1 + \en_2)^2/c^2 - \en_1^2/c^2-\en_2^2/c^2=2\en_1\en_2/c^2$. In the center-of-momentum frame $p'_{\rm cm} = 0$ and thus $\en'^2_{\rm cm}=2\en_1\en_2$, so $\en'_{\rm cm}=\sqrt{2\en_1\en_2}$.

(c) The process starts happening when the energy of the reaction in the center-of-momentum frame is sufficient to create an electron-positron pair at rest: $\en'_{\rm cm}=\sqrt{2\en_1\en_2}>2m_ec^2$ (the center-of-momentum frame is discussed in more detail in Section \ref{sec_threshold}). Therefore, the minimum energy of the second photon is
\begin{equation}
\en_2=(2m_ec^2)^2/2\en_1=2\times (511{\rm keV})^2/1{\rm keV}=522{\rm MeV}.
\end{equation}
\end{prob}

\subsection{Doppler effect and aberration via wave approach}
\label{sec_wphoton}

\begin{figure}[htb]
\centering
\includegraphics[scale=0.8, clip=true, trim=2cm 11.5cm 5cm 11cm]{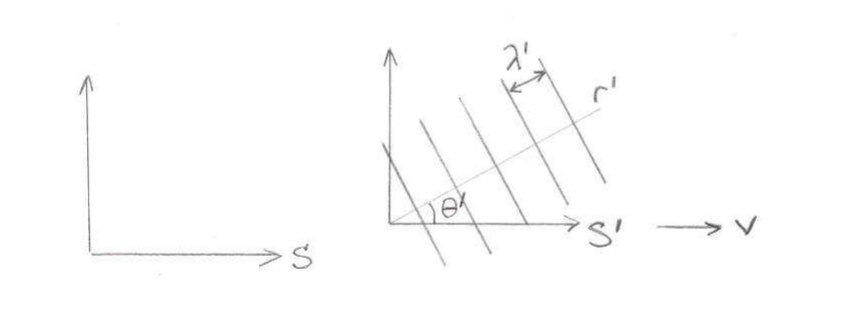}
\caption{Plane electromagnetic wave in $S'$ frame}
\label{pic_crests}
\end{figure}

The following derivation is a standard one, found in many textbooks (e.g.,\citealt{resn68}). Instead of the quantum-mechanical approach (in which we consider photon as a particle), we go back to the classical description of light as an electromagnetic wave. We consider a plane electromagnetic wave moving in the $S'$ frame at angle $\theta'$ to the $x'$ axis, and we introduce $r'$, the coordinate along the direction of the propagation of the wave. Figure \ref{pic_crests} shows the locations of the crests of the electromagnetic wave (e.g., the maximal values of field $\vec{E'}$). As a function of time and space, the amplitude of the field is $E'=E_0'\cos\left(2\pi[r'/\lambda'-\nu't']\right)$. Indeed, at $t'=0$ it is a $\cos(2\pi r'/\lambda')$ wave, so that when $r'$ changes by $\lambda'$, the argument of cosine changes by $2\pi$ and we go from one crest to the next in accordance with the definition of the wavelength. The crest that started at $r'=0$ when $t'=0$ moves forward with $r'/\lambda'-\nu't'=0$, which is satisfied when $\lambda'\nu'=c$. 

As a function of coordinates $x'$ and $y'$ in $S'$, the final expression is 
\begin{equation}
E'=E_0'\cos\left(2\pi\left[\frac{x'\cos\theta'+y'\sin\theta'}{\lambda'}-\nu't'\right]\right).
\end{equation}
We know how to transform $x', y'$ and $t'$ to $x,y$ and $t$ via Lorentz transform, so for the phase we obtain
\begin{equation}
\frac{\Phi'}{2\pi}=\frac{(\gamma x-\beta\gamma ct)\cos\theta'+y\sin\theta'}{\lambda'}-\frac{\nu'(\gamma ct-\beta\gamma x)}{c}.
\end{equation}
In frame $S$, we also have a plane wave propagating with $c$, though in a different direction and perhaps with a different frequency and wavelength:
\begin{equation}
E=E_0\cos\left(2\pi\left[\frac{x\cos\theta+y\sin\theta}{\lambda}-\nu t\right]\right)\equiv E_0\cos \Phi.
\label{eq_phase}
\end{equation}
Now comes the subtle part: the argument of the cosine -- i.e., the phase of the electromagnetic wave -- turns out to be a relativistic invariant, $\Phi=\Phi'$. It is not easy to see this without the 4-vector formalism, and we will demonstrate that this is true in Section \ref{sec_4vphoton}. There are a couple of qualitative arguments. (1) The number of crests passing by an observer must be a relativistic invariant; the number of crests is proportional to the phase; therefore the phase must be invariant. (2) The phase of the wave can be probed by interference experiments, and recording the minimum or maximum of an intensity can be turned into a physical event; thus, the phase should be the same in both frames to result in the same outcome of the interference experiment. (Both of these seem to me after-the-fact justifications which are non-trivial in light of the fact that the electromagnetic fields do transform from $S$ to $S'$). 

Nonetheless, if we accept that the phase must be the same, then we equate $\Phi=\Phi'$ and we must have the same multipliers in front of each of $x, y$ and $t$. As a result, we get three equations:
\begin{eqnarray}
\frac{\gamma \cos\theta'}{\lambda'}+\frac{\nu'}{c}\beta\gamma=\frac{\cos\theta}{\lambda};\label{eq_ab1}\\
\frac{\sin\theta'}{\lambda'}=\frac{\sin\theta}{\lambda};\label{eq_ab2}\\
\frac{\beta\gamma\cos\theta'}{\lambda'}+\frac{\nu'\gamma}{c}=\frac{\nu}{c}.\label{eq_ab3}
\end{eqnarray}
We divide equation (\ref{eq_ab2}) by (\ref{eq_ab1}) and obtain the formula for {\bf relativistic aberration of light:}
\begin{equation}
\tan\theta=\frac{\sin\theta'}{\gamma(\cos\theta'+\beta)}.\label{eq_ab4}
\end{equation}
Equation (\ref{eq_ab3}) gives us
\begin{equation}
\nu=\gamma\nu'(1+\beta\cos\theta').\label{eq_dopp}
\end{equation}
This is the {\bf Doppler effect}, which is the change of the frequency of a photon from one frame to another.

\begin{prob}Obtain the relativistic aberration of light (\ref{eq_ab4}) by applying velocity transformations with $|\vec{u}|=|\vec{u'}|=c$. 
\label{prob_ab}
\end{prob}

\begin{prob}Get into the frame of mind of an astronomer and consider a blob at distance $D$ moving with velocity $u$ at an angle $\theta$ to the line of sight to the observer. Even though the object moves in three dimensions, we can only observe its change of position projected on the plane of the sky. We ``observe'' an object when photons from it reach our telescope. Calculate the observed proper motion of the emitter on the sky, which tells you the distance the object appears to cover on the sky per unit time. For a given $u$, what is the maximal observed proper motion? Can this value be greater than $c$? If so, under what conditions on $u$? (Hint: no Lorentz transforms are necessary in order to solve this problem; it is all done in the observer's frame.) -- {\bf Solution.} Yes, the observed projected velocity can be greater than $c$. This is called apparent superluminal motion\footnote{\url{http://en.wikipedia.org/wiki/Superluminal_motion}} and is commonly observed from mildly-relativistic astronomical objects (Figure \ref{pic_superlum}).
\label{prob_superlum}
\end{prob}

\subsection{Doppler effect and aberration via 4-vector approach}
\label{sec_4vphoton}

We now have the 4-vector of energy and momentum for the photon, $K^{\mu}=(\en/c,\vec{p})$ and the relationship between photon energy and momentum, $\en=pc$. We also have the wave description of the aberration and Doppler effect, which involves the frequency $\nu$ and the wavelength $\lambda$ of the wave. How are the two approaches related to one another? 

Furthermore, we have obtained two complicated-looking equations for electromagnetic waves, equations (\ref{eq_ab4}) and (\ref{eq_dopp}). They look nothing like Lorentz transform, nothing like anything from the classical mechanics, and not even much like the velocity transforms (though can be connected to those as in Problem \ref{prob_ab}). So again (cf. Problem \ref{prob_vel}) we must ask: where is that famed mathematical beauty of the Special Relativity?

To answer these questions, we re-derive the aberration and the Doppler formulae using the 4-vector approach. Following the same notation as in Figure \ref{pic_crests}, we assume that the photon is the $S'$ frame has the components of 4-vector $K'^{\mu}=(\en'/c, \en'\cos\theta'/c,\en'\sin\theta'/c,0)$, where we have already taken into account the relationship between the energy and the momentum of the photon $p'=\en'/c$. 4-vectors, including the 4-vector of energy and momentum of a photon, transform using the Lorentz transform:
\begin{equation}
\left(\begin{array}{c} \en/c \\ \en\cos\theta/c \\ \en\sin\theta/c \\ 0 \end{array}\right)=\left(\begin{array}{cccc} \gamma & \gamma\beta & 0 & 0 \\ \gamma\beta & \gamma & 0 & 0 \\ 0 & 0 & 1 & 0 \\ 0 & 0 & 0 & 1 \end{array}\right)\left(\begin{array}{c}  \en'/c \\ \en'\cos\theta'/c \\ \en'\sin\theta'/c \\ 0  \end{array}\right).
\end{equation}
This matrix equation consists of 4 different equations, the last one of them trivially satisfied (it reads $0=0$). If we take the 2nd and the 3rd lines and divide one by the other, we find 
\begin{equation}
\tan\theta=\frac{p_y}{p_x}=\frac{\en'\sin\theta'/c}{\gamma(\beta\en'/c+\en'\cos\theta'/c)}=\frac{\sin\theta'}{\gamma(\beta+\cos\theta')},
\end{equation}
and we recover equation (\ref{eq_ab4}). From the first equation we obtain
\begin{equation}
\en=\gamma(1+\beta\cos\theta')\en'. 
\end{equation}
This looks very similar to equation (\ref{eq_dopp}), except for the energy instead of the frequency. Therefore, if energy is proportional to frequency, then we recover Doppler effect using both the wave and the particle approach.

The coefficient of proportionality is the {\bf Planck constant}, so that 
\begin{equation}
\en=h\nu,
\end{equation}
where $h=2\pi\hbar$ and $\hbar=1.05\times 10^{-27}$ cgs units or $\hbar=1.05\times 10^{-34}$ MKS units. $\hbar$ has the units of angular momentum. Of course this relationship is not something we can obtain from classical physics of classical electromagnetism; the appearance of $\hbar$ makes it clear that this is a quantum-mechanical (or rather, quantum-electrodynamical) relationship. 

Here are some of the commonly used ways of representing the 4-vector of the energy and momentum of photons:
\begin{equation}
K^{\mu}=\left(\frac{\en}{c},\vec{p}\right)=\left(\frac{h\nu}{c},\frac{h\nu\vec{n}}{c}\right)=\left(\frac{\hbar\omega}{c},\frac{\hbar\omega\vec{n}}{c}\right),
\end{equation}
where $\omega=2\pi\nu$ is the angular frequency of the wave and $\vec{n}$ is the unit vector in the direction of propagation of the photon.

We introduce a 3D vector $\vec{k}=\omega\vec{n}/c\equiv 2\pi \vec{n}/\lambda$, which is called {\bf the wave vector} (applicable to any wave, not just the photons, with $c$ that needs to be replaced by the wave propagation speed). Its value is equal to $2\pi/\lambda$ and it is pointed in the direction of the propagation of the wave or the photon. With this notation, $\vec{p}=\hbar \vec{k}$ and
\begin{equation}
K^{\mu}=(\hbar \omega/c,\hbar \vec{k}).
\end{equation}
Of course, whichever way we present these newly defined values, $K^{\mu}K_{\mu}=0$. 

Finally, we can demonstrate that the phase of the electromagnetic wave is a relativistic invariant, a fact that we already used in Section \ref{sec_wphoton}. Using the setup described in that Section, we start with the expression for the phase (\ref{eq_phase}) and we relate the phase to the newly introduced 4-vector:
\begin{equation}
\Phi=2\pi \left(\frac{r}{\lambda}-\nu t\right)=\frac{2\pi r}{\lambda} - 2\pi\nu t=\vec{r}\vec{k}-\frac{\omega}{c}ct=-\frac{1}{\hbar}\left(\frac{\hbar\omega}{c} ct - \hbar\vec{k}\vec{r}\right)\equiv -\frac{1}{\hbar}K^{\mu}S_{\mu},
\end{equation}
where $K^{\mu}$ is the 4-vector of energy and momentum of the photon and $S^{\mu}$ is the 4-vector of space-time coordinates of the point with a given phase. Because the phase is a 4-product of 4-vectors (multiplied by a fundamental constant), it is a relativistic invariant.

\begin{prob}
A blob of gas is moving directly away from us at velocity $v$, while producing an emission line with a laboratory frequency $\nu_0$. What is the observed frequency of the emission line? -- {\bf Solution.} As usual, our first step is to direct $x$ axis parallel to the velocity vector of the moving frame, which is that of the blob of gas. The blob emits photons in all directions, as seen from the blob, but the only ones that we are receiving are those directed back at us; therefore $\cos\theta'=-1$ (Figure \ref{pic_prob_dopp}). The co-moving frequency $\nu'$ is the laboratory frequency $\nu_0$. We plug this into equation (\ref{eq_dopp}) and find
\begin{equation}
\nu=\nu_0\frac{1-\beta}{\sqrt{1-\beta^2}}=\nu_0\sqrt{\frac{1-\beta}{1+\beta}}.
\end{equation}
As the object is moving away, frequency is becoming lower, observed light moves toward redder wavelengths. 

\begin{figure}[htb]
\centering
\includegraphics[scale=0.8, clip=true, trim=2cm 6.5cm 5cm 17cm]{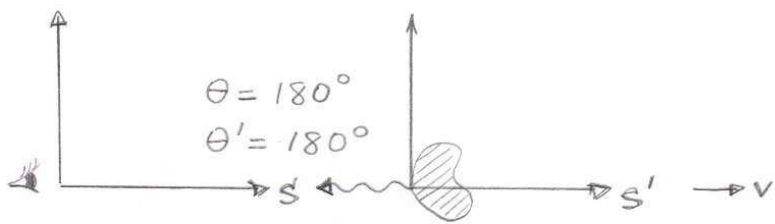}
\caption{Calculation of the Doppler effect for the emitter moving directly away from us}
\label{pic_prob_dopp}
\end{figure}

Let us Taylor-expand this in the orders of $\beta$: $\nu=\nu_0\left(1-v/c+O((v/c)^2)\right)$. This is an effect known to you from non-relativistic wave mechanics, for example in application to sound waves (when $c$ in this equation needs to be replaced by the speed of sound): as an ambulance is approaching us, $v$ is negative, frequency we hear is higher than the frequency of the ambulance siren, the pitch heard by the bystander is higher. As the ambulance passes us and moves away, $v$ is positive, frequency we hear is lower than the frequency of the siren and the pitch is lower.
\end{prob}

\begin{figure}[htb]
\centering
\includegraphics[scale=0.8, clip=true, trim=2cm 0.5cm 10cm 22cm]{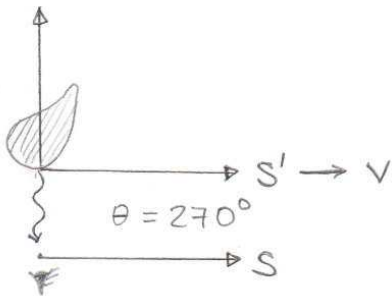}
\caption{Calculation of the Doppler effect for the emitter moving directly perpendicular to the line of sight}
\label{pic_dopp_tran}
\end{figure}

\begin{prob}
A blob of gas is moving right in front of us, perpendicular to our line of sight. What is the frequency of the observed emission? -- {\bf Solution.} Only the photons propagating with $\cos\theta=0$ will reach us (Figure \ref{pic_dopp_tran}). This is not very helpful to us in applying equation (\ref{eq_dopp}) as we did in the previous case. But from the aberration equation (\ref{eq_ab4}), we notice that $\cos\theta=0$ is equivalent to $\tan\theta\rightarrow\infty$ which happens when $\cos\theta'=-\beta$, and then with this information we now can use equation (\ref{eq_dopp}) to find
\begin{equation}
\nu=\gamma\nu_0(1-\beta^2)=\nu_0/\gamma.
\end{equation}
Unlike the case of the regular Doppler effect, there is no change in frequency to the first order of $v/c$, but light does get slightly redder in the $v^2/c^2$ terms. This is called {\bf the transverse Doppler effect.} If you were to forget about aberration of light and were to incorrectly set $\cos\theta'=0$, you would get $\nu=\gamma\nu_0$, which is exactly the opposite effect. 
\end{prob}

\subsection{Threshold for creating new particles}
\label{sec_threshold}

To finish our discussion of energy and momentum conservation in Special Relativity for particles with and without mass, here we discuss a class of problems which focus on the minimal energy requirements for creating new particles. A typical problem of this class might have a target particle at rest (A), with another particle (B) moving with high velocity and hitting target A. The problem specifies that the reaction of particles A and B may produce products of known mass (C, D, ...) and asks about the minimal energy (or momentum, or velocity) of particle B to enable this reaction. 

The most common mistake in solving `threshold' problems is to assume that the minimal incoming energy of B is achieved when the products (C, D, ...) are at rest after the collision. In this case energy conservation ($\varepsilon_{\rm B}+m_{\rm A}c^2=m_{\rm C}c^2+m_{\rm D}c^2+...$) would immediately give the requisite energy of the incoming particle. But of course this solution is incorrect because it does not conserve the momentum: in the lab frame, the momentum before the collision is non-zero, since particle B is moving, whereas if the products are at rest after the collision their momentum is zero. 

To solve threshold problems correctly, we must instead recognize that the minimum energy of the reaction is such that {\bf the products are at rest in the center-of-momentum frame} of the reaction. The center-of-momentum frame is the frame where particles A and B have a head-on collision with zero net momentum. Thus, if the products are at rest in this frame, the momentum conservation would automatically be satisfied. This means that at the threshold energy in the lab frame the products of the reaction are moving (instead of being at rest, as in the incorrect solution in the previous paragraph), but they are moving together at the same velocity.

\begin{prob}
Prove that the minimal energy $\varepsilon_{\rm B}$ to enable the reaction discussed above corresponds to the case when reaction products are at rest in the center-of-momentum frame. To do this, calculate the velocity of the center-of-momentum frame from $\varepsilon_{\rm B}$ and $m_{\rm A}$, move into this frame and use energy conservation in this frame to calculate the minimal energy $\varepsilon_{\rm B}$ to enable the reaction A+B$\rightarrow$C+D+... -- {\bf Solution:} The center-of-momentum frame $S'$ is such that $p'_{\rm A}+p'_{\rm B}=0$. Using the Lorentz transform to calculate $p'$ values from their lab values $p_{\rm A}=0$, $\varepsilon_{\rm A}=m_{\rm A}c^2$, $p_{\rm B}$ and $\varepsilon_{\rm B}$, we find that the center-of-momentum frame moves with $\beta=p_{\rm B}c/(\varepsilon_{\rm B}+m_{\rm A}c^2)$. In this frame, the energy conservation reads $\varepsilon'_{\rm A}+\varepsilon'_{\rm B}=\varepsilon'_{\rm C}+\varepsilon'_{\rm D}+...$. Again using the Lorentz transforms for the left-hand side, we find that it is equal to $\sqrt{m_{\rm B}^2 c^4+2\varepsilon_{\rm B}m_{\rm A}c^2+m_{\rm A}^2 c^4}$. The right-hand side is minimized when particles have no kinetic energy in this frame, when they are all at rest and when the total energy is equal to the sum of rest energies, which is allowed by the momentum conservation. Because the left-hand side monotonically increases with $\varepsilon_{\rm B}$, this is also the condition for minimizing $\varepsilon_{\rm B}$. 
\end{prob}

\begin{prob}
Calculate the minimum energy a pion $\pi^0$ must have in order for the process $\pi^0+p^+ \rightarrow K^++\Sigma^0$ to occur. The target proton is not moving in the lab frame, and the rest masses are 135, 494 and 1193 MeV for $\pi^0$, $K^+$ and $\Sigma^0$, respectively. -- {\bf Solution:} Minimum energy is achieved when collision products are not moving in the center-of-momentum frame, i.e., when in this frame all initial energy goes into mass creation, not kinetic energy. Labeling all particles in the reaction as it is written from 1 to 4, energy and momentum conservation in the 4-vector form are $P^{\mu}_1+P^{\mu}_2=P^{\mu}_3+P^{\mu}_4$. Taking the 4-square of the left-hand side of this equation: $P^{\mu}_1P_{\mu,1}+2P^{\mu}_1P_{\mu,2}+P^{\mu}_2P_{\mu,2}=m_1^2c^4+2\varepsilon_1m_2c^2+m_2^2c^4$. Because at the requisite threshold energy the products are moving together, the 4-square of the right-hand side of the 4-momentum equation is the square of the total mass of the products: $(m_3+m_4)^2c^4$. Solving for $\varepsilon_1$, we obtain
\begin{equation}
\varepsilon_1=\frac{(m_3+m_4)^2c^4-m_1^2c^4-m_2^2c^4}{2m_2c^2},
\end{equation}
where all masses can be plugged in the MeV/$c^2$ units to get the energy in MeV as well (proton's rest mass is 938 MeV/$c^2$), giving $\varepsilon_1=1038$ MeV. 
\end{prob}

\clearpage
\begin{figure}[!htb]
\centering
\includegraphics[scale=0.8]{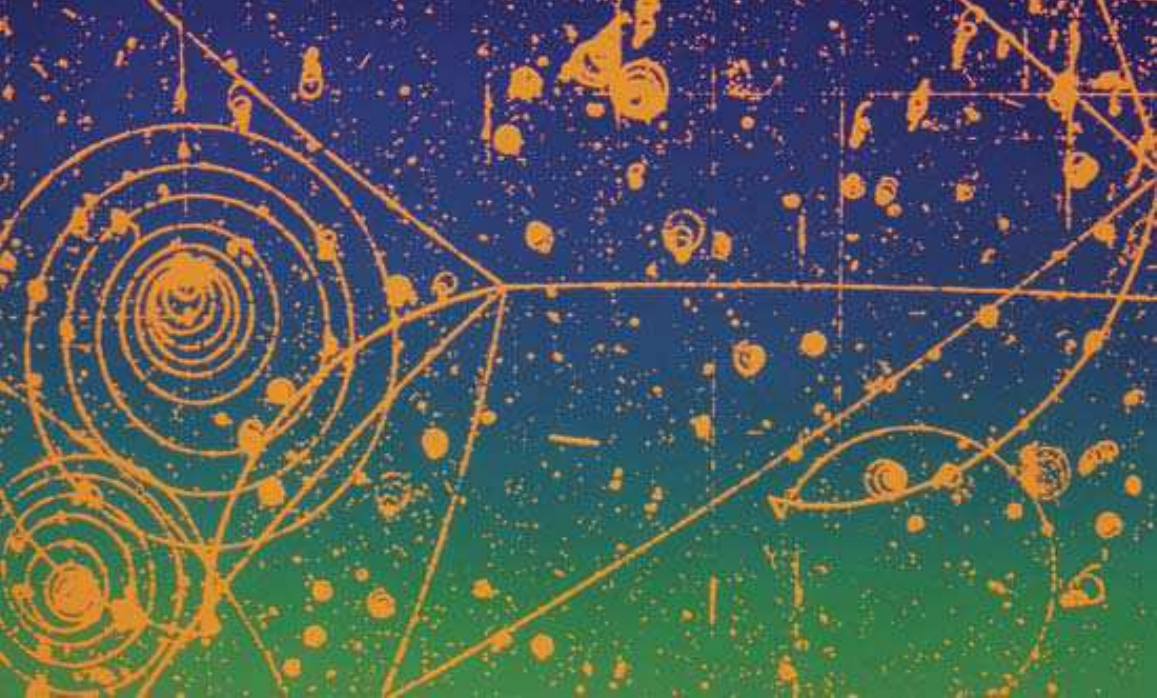}
\caption{A false-color image of particle tracks in a hydrogen bubble chamber, image by LBNL, from \url{http://kids.britannica.com/}.}
\end{figure}

\clearpage
\section{Relativistic dynamics}
\label{sec_dynamics}

\subsection{Forces and accelerations in Special Relativity}

The goal of this section is to generalize the 2nd law of Newton to the relativistic case. Because such a law is the fundamental law of nature, it should be the same in all inertial frames. Furthermore, it needs to reduce to the well-known non-relativistic case for particles moving with small velocities, $u\ll c$. 

Let us consider the classical mechanics formulation of the 2nd law of Newton using momentum instead of the usual way with acceleration ($\vec{F}=m\vec{a}$):
\begin{equation}
\frac{{\rm d}\vec{p}}{{\rm d} t}=\vec{F}.\label{eq_force}
\end{equation}
It turns out that this law is applicable in Special Relativity when we use the relativistic expression for momentum, $\vec{p}=\gamma_um\vec{u}$, although at the moment this is not saying much because we do not yet know how to calculate any forces in Special Relativity.

Furthermore, this is not a relativistically invariant (i.e., ``beautiful'' in the Special Relativity sense) way of formulating the 2nd law of Newton. The quantities on either side of this equation are neither relativistic invariants, nor 4-vectors, and we have a nasty feeling that both sides will have to be recalculated (transformed) if we consider this equation in another inertial frame (see \citealt{resn68} for the many pages of algebra on this issue). Postponing the question of how this equation transforms from one frame to another and how to ``beautify'' this equation for it to be more in line with the relativistically invariant approach (Section \ref{sec_fields}), in this section we consider the properties of this equation, discuss the best methods of solving it and develop intuition for the types of effects that are present in Special Relativity that are not present in classical mechanics. 

We start by discussing the relationship between force and acceleration in Special Relativity. We re-write:
\begin{equation}
\vec{F}=\frac{{\rm d}\vec{p}}{{\rm d}t}=\frac{\rm d}{{\rm d}t}\left(\frac{m\vec{u}}{\sqrt{1-u^2/c^2}}\right)=\frac{m({\rm d}\vec{u}/{\rm d}t)}{\sqrt{1-u^2/c^2}}+\vec{u}\frac{\rm d}{{\rm d}t}\frac{m}{\sqrt{1-u^2/c^2}}.
\end{equation}
The change in velocity per unit time is acceleration, $\vec{a}\equiv{\rm d}\vec{u}/{\rm d}t$, which can be defined in a similar way we defined the velocity for the relativistic case: the soccer ball flies through space, breaks some clocks, thus we can record its trajectory as a function of time, and therefore its velocity vector as a function of time, and therefore its acceleration. The second term of this equation can be multiplied and divided by $c^2$ to get the rate of change of the particle's energy:
\begin{equation}
\vec{F}=\frac{m\vec{a}}{\sqrt{1-u^2/c^2}}+\frac{\vec{u}}{c^2}\frac{{\rm d}\en}{{\rm d}t}.\label{eq_acc}
\end{equation}

The energy of the particle changes because of the work done on the particle: ${\rm d}\en=\vec{F}\cdot{\rm d}\vec{r}$, so the rate of change of energy is ${\rm d}\en/{\rm d}t=\vec{F}\cdot\vec{u}$. Finally we obtain
\begin{equation}
\vec{F}=\frac{m\vec{a}}{\sqrt{1-u^2/c^2}}+\frac{\vec{u}}{c^2}\left(\vec{F}\cdot\vec{u}\right).
\end{equation}

As we take the $c\rightarrow \infty$ limit, we recover the non-relativistic relationship $\vec{F}=m\vec{a}$, but the awful truth is that in general acceleration is not proportional to force in Special Relativity, and moreover they are not even necessarily parallel to one another. This makes acceleration a very non-intuitive value in Special Relativity, and we aim to avoid it whenever possible. 

\begin{prob}
Instead of relying on the analogy with the classical mechanics, derive equation ${\rm d}\en=\vec{F}\cdot{\rm d}\vec{r}$ formally from the relativistic relationship between energy and momentum and equation \ref{eq_force}. -- {\bf Solution:} $\varepsilon=\sqrt{p^2c^2+m^2c^4}$, so 
\begin{equation}
{\rm d}\varepsilon=\frac{1}{2\sqrt{p^2c^2+m^2c^4}}2c^2\vec{p}\cdot{\rm d}\vec{p}=\frac{c^2}{\varepsilon}\vec{p}\cdot{\rm d}\vec{p}=\vec{u}\cdot{\rm d}\vec{p}=\vec{u}\cdot\vec{F}{\rm d}t=\vec{F}\cdot{\rm d}\vec{r}.
\end{equation}
\end{prob}

\subsection{Equation of motion in the electromagnetic field}

Here are the equations of force acting on a charged particle in electromagnetic field in Special Relativity:
\begin{eqnarray}
\vec{F}=q\left(\vec{E}+\frac{\vec{u}}{c}\times\vec{B}\right)\mbox{ in cgs units }\label{eq_force_cgs}\\
\vec{F}=q\left(\vec{E}+\vec{u}\times\vec{B}\right)\mbox{ in MKS units }\label{eq_force_mks}.
\end{eqnarray}
It is a great relief that these equations are the same as those in the non-relativistic theory. Just like the 2nd law of Newton needs to be supplied by empirically determined forces in order to understand how a particle reacts to a force, the same way in Special Relativity these expressions for the force are the results of an enormous body of experimental work. 

We need to understand how these expressions can be different in the two different unit systems. Let us place an electron into static electric and magnetic fields, each 1 cgs unit strong (the cgs unit of the magnetic field is widely known as Gauss). Plugging in all values in cgs units into the force equation in cgs units, we should find force in cgs units of force, which is dyne:
\begin{equation}
\vec{F}=4.8\times 10^{-10}\left(\vec{1}\mbox{ cgs unit}+\frac{\vec{u}}{c}\times\vec{1}\mbox{ Gauss }\right),\mbox{ in dynes}.
\end{equation}

Now we consider exactly the same physical situation (i.e., the same strengths of the electric and magnetic fields), but use MKS units for everything; therefore, first we need to translate the values of the fields into MKS units\footnote{\url{http://en.wikipedia.org/wiki/Centimetre-gram-second_system_of_units}}. For example, 1 cgs unit worth of electric field is $3\times 10^4$ Volts/m and 1 Gauss$=10^{-4}$ Tesla. Applying the MKS equation for force (\ref{eq_force_mks}), we find
\begin{equation}
\vec{F}=1.6\times 10^{-19}\left(\vec{1}\times 3\times 10^4+\frac{\vec{u}}{c}\times\vec{1}\times 3\times 10^8 \times 10^{-4}\right),\mbox{ in Newtons},
\end{equation}
where in the last term we multiplied by $c$ and divided by $c$. Pulling out the common factor of $3\times 10^4$ out of the parentheses, we obtain
\begin{equation}
\vec{F}=4.8\times 10^{-15}\left(\vec{1}+\frac{\vec{u}}{c}\times\vec{1}\right),\mbox{ in Newtons}.
\end{equation}
Because Newtons are $10^5$ times stronger than dynes, this is indeed a completely identical force as the one we calculated in cgs units. Thus the same physical situation viewed in two different unit systems results in the same force, as it should. 

The first lesson of relativistic dynamics is that acceleration is a terrible value to use in Special Relativity. (Look at equation \ref{eq_acc}. Just look at it.) When confronting a dynamics problems, always look for a way to avoid velocity and acceleration and instead use energy and momentum. For particles moving in the electric and magnetic field, by far the most useful formulations of the equations are
\begin{eqnarray}
\frac{{\rm d}\vec{p}}{{\rm d}t}=\vec{F}=q\left(\vec{E}+\frac{\vec{u}}{(c)}\times \vec{B}\right),\label{eq_force_mom}\\
\frac{{\rm d}\en}{{\rm d}t}=\vec{u}\cdot\vec{F}=q\vec{u}\cdot\vec{E}.\label{eq_force_en}
\end{eqnarray}
Here $c$ in parentheses is present or absent depending on whether we are using cgs or MKS units. In the second equation, we have taken into account that the magnetic force is always perpendicular to the particle's velocity $\vec{u}$, and therefore it does not do work on the particle. These equations are often usefully supplemented by the relativistic relationships between energy and momentum (\ref{eq_permute}). 

\begin{prob}Calculate the velocity acquired by an electron traversing a potential drop of $10^6$ Volts, starting from rest. -- {\bf Solution.} Using energy conservation, we find $\en_{\rm final}-\en_{\rm initial}=q\Delta V=1$ MeV. The final energy is $\en_{\rm final}=\gamma_{\rm final}mc^2$, whereas the initial energy is $\en_{\rm initial}=mc^2$, which for an electron is 0.511 MeV. We obtain $\gamma_{\rm final}=2.957$ and $u=0.94c$.

To see the differences between the relativistic behavior of particles and the non-relativistic behavior of particles, we also calculate the velocity we would have obtained by (incorrectly) applying the non-relativistic expressions: $mu^2/2=q\Delta V$, so $u=\sqrt{2q\Delta V/m}$. This equation is valid in both cgs and MKS units, so let us make sure we get consistent results from both. Using MKS units for everything
\begin{equation}
u=\sqrt{\frac{2\times 1.6\times 10^{-19}\times 10^6}{9.1\times 10^{-31}}}=6\times 10^8 \mbox{m/s},
\end{equation} 
whereas in cgs units the same voltage is equal to $10^6/300$ cgs units, so
\begin{equation}
u=\sqrt{\frac{2\times 4.8\times 10^{-10}\times 10^6/300}{9.1\times 10^{-28}}}=6\times 10^{10} \mbox{cm/s}.
\end{equation}
The two systems agree (as they should) that the velocity would have been $2c$ in this calculation. Thus the second lesson of relativistic dynamics (the first being that acceleration is awful) is that in comparison with the non-relativistic case, relativistic particles are having a harder time getting accelerated to higher velocity. A relativistic particle has a high ``effective inertia'', which is clear from Figure \ref{pic_gamma}: once we get to velocities not too far away from $c$, all the work done on the particle goes into increasing its Lorentz factor $\gamma_u$, which barely changes velocity, since $\gamma_u$ is rising so steeply with $u$ in this regime.
\end{prob}

\begin{prob}An electron moving with total energy 1 MeV is entering a region of static electric field directed perpendicular to its initial velocity (Figure \ref{pic_prob_efield}). Calculate the component of velocity that the electron will acquire after it has traversed a potential drop of $\Delta V=10^6$ Volts. -- {\bf Solution.} We introduce coordinate $x$ to be parallel to the field and $y$ to be along the initial motion of the particle. In classical mechanics and non-relativistic electromagnetism, we would write the force equation along the $x$ axis, calculate the acceleration, integrate it and obtain $u_x$, which is what the problem is asking. The issue with using this method in relativistic dynamics is that projecting (\ref{eq_acc}) really does not help much: we will obtain
\begin{equation}
a_x=\frac{{\rm d}u_x}{{\rm d} t}=\frac{\sqrt{1-u^2/c^2}}{m}\left(qE-u_x\frac{q\vec{E}\cdot\vec{u}}{c^2}\right).
\end{equation}
Even though the left-hand side depends just on the value we need to find ($u_x$), the right-hand side has a complex dependency on both $u_x$ and $u_y$. We would need to solve two coupled, non-linear differential equations for $u_x$ and $u_y$, which is not for the faint-hearted. 

\begin{figure}[htb]
\centering
\includegraphics[scale=0.8, clip=true, trim=1cm 20cm 13cm 1.5cm]{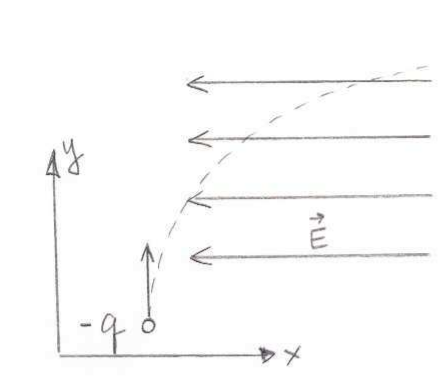}
\caption{Electron entering a region of perpendicular electric field}
\label{pic_prob_efield}
\end{figure}

This problem is as easy as they get (one particle in static electric field), so there must be a less cumbersome way. We will use equations (\ref{eq_force_mom}-\ref{eq_force_en}) -- there is one equation for energy and two for the two projections of the momentum equation (nothing interesting is happening along the $z$ axis). Spelling them out with $\vec{E}=(-E,0,0)$ (where the minus sign is introduced so that the negatively charged electron moves along the positive $x$ direction and $q$ is now the absolute value of the charge), we find
\begin{eqnarray}
\frac{{\rm d}\en}{{\rm d}t}=qEu_x;\\
\frac{{\rm d}p_x}{{\rm d}t}=qE;\\
\frac{{\rm d}p_y}{{\rm d}t}=0.\label{eq_force_py}
\end{eqnarray}
Integrating the first equation, we find $\en_{\rm final}-\en_{\rm initial}=q\Delta V$. This enables us to calculate $\gamma_{\rm final}=3.914$ and the absolute value of the final velocity $u_{\rm final}=0.967c$. This final velocity is made up of two components, $u_{\rm final}=\sqrt{u_{x,{\rm final}}^2+u_{y,{\rm final}}^2}$. In classical mechanics we would have said that since there is no force along the $y$ direction, the $u_y$ component of the velocity should not change, which would have led us to the answer for $u_{x,{\rm final}}$, but in Special Relativity the $y$ component of the velocity does change, as can be seen from (\ref{eq_force_py}). 

Since $p_y$ is the same in the beginning and the end of the problem, we find $\gamma_{\rm final}u_{y,{\rm final}}=\gamma_{\rm initial}u_{y,{\rm initial}}$, where the initial Lorentz factor and the $u_y$ velocity can be easily calculated from $\en_{\rm initial}=1$ MeV. With a little bit of algebra to put all these pieces together, we find $u_{x,{\rm final}}=\sqrt{1-\gamma_{\rm initial}^2/\gamma_{\rm final}^2}=0.87 c$. 

Comparing this value with the answer to the previous problem, we see that despite the same electric field and voltage drop, the particle in this problem acquired slightly smaller $x$ component of the velocity. This is another manifestation of the previously suggested increased effective inertia of the relativistic particles. In classical mechanics, both particles would have acquired the same $u_x$ because the $y$ component of the velocity would have no bearing on the $x$ component of the equation of motion. 
\end{prob}

\begin{prob}What is the trajectory of a particle moving in uniform magnetic field, perpendicular to the field? -- {\bf Solution.} The equation of motion is now ${\rm d}\vec{p}/{\rm d}t=q\vec{u}\times\vec{B}/(c)$, where again the $c$ in parentheses depends on which unit system we choose for calculating the final values: in cgs the factor of $c$ is present and in MKS it is absent. Particle energy does not change, therefore in the expression $\vec{p}=\gamma_um\vec{u}$ the term $\gamma_u$ can come out of the differential. We now have $\gamma_um{\rm d}\vec{u}/{\rm d}t=q\vec{u}\times\vec{B}/(c)$, which is to within the factor of $\gamma_u$ the same equation as the one in non-relativistic electricity and magnetism. 

We know that the solution to this equation is a circle. Over one period, $\vec{u}$ changes by $2\pi u$, time changes by period $T$, and the equation of motion can be simplified to $\gamma_u m 2 \pi u/T=quB/(c)$. Replacing $2\pi/T$ with angular frequency $\omega$, we obtain
\begin{equation}
\omega=\frac{qB}{\gamma_um(c)}.
\end{equation}
Another way of looking at this is by introducing the radius of curvature, called {\bf Larmor radius}:
\begin{equation}
R=\frac{u}{\omega}=\frac{u\gamma_um(c)}{qB}.
\end{equation}
In the non-relativistic case, we would simply set $\gamma_u=1$.

In LHC, the particles are 7 TeV protons confined to a circle by 8.4 Tesla magnets. (Magnetic fields of this order are almost at the limit of available technology for static magnetic fields; much higher magnetic fields can be achieved in the labs by compression, but they last only briefly. Lab-made values pale in comparison to the static fields in magnetars mentioned in Section \ref{sec_forms}, $\sim 10^{14}$ Gauss $=10^{10}$ Tesla.) For ultra-relativistic particles the distinction between $u$ and $c$ in the numerator of this equation is rather irrelevant:
\begin{equation}
R\simeq \frac{\gamma_umc(c)}{qB}.
\end{equation}
When this equation is used in cgs units, it conveniently has particle energy in the numerator. Putting in LHC values, we obtain $R\simeq 2.8$ km, with circumference 18 km, all covered with superconducting magnets (the only kind capable of sustaining the large current necessary to produce such fields)\footnote{\url{http://lhc-machine-outreach.web.cern.ch/lhc-machine-outreach/components/magnets.htm}, \url{http://home.web.cern.ch/topics/large-hadron-collider}}. It is the minimal size of the tunnel that needs to be built, since equipment other than magnets also needs to be fit into the accelerator along the way. As you might imagine, a 20$+$ km-long underground tunnel stuffed with superconducting magnets is rather pricey; this is precisely why high energy accelerators are so expensive. 
\end{prob}

Some internet resources describing accelerator physics would on occasion say that ``magnets accelerate particles to high energies'' in an accelerator. It is true that magnetic fields provide the centripetal force to bend the trajectory of the particles (and thus do result in an acceleration), but they do not do any work on the particles and thus cannot increase the particles' energy. 

\begin{figure}[htb]
\centering
\includegraphics[scale=0.7, clip=true, trim=10cm 16cm 1cm 1.5cm]{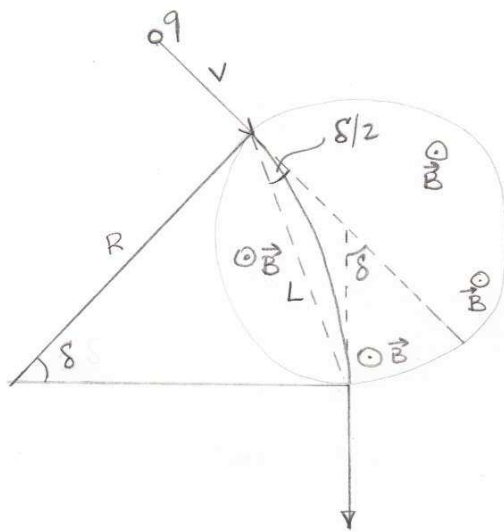}
\caption{Cosmic ray deflected by uniform magnetic field}
\label{pic_deflection}
\end{figure}

\begin{prob}
An ultra-high energy cosmic ray enters Earth's magnetosphere. Estimate by what angle it is deflected from its original trajectory. -- {\bf Solution.} This will be an order of magnitude estimate, as the actual values depend on the exact geometry of the magnetic field and the impact angle of the cosmic ray. We approximate the Earth's magnetosphere by an area of a uniform magnetic field of $B=1$ Gauss with a size $L=10^4$ km. The radius of curvature of the cosmic ray in cgs units is $R=(u\gamma m c)/(qB)\simeq \en/(qB)$. The angle of deflection (as shown in Figure \ref{pic_deflection}) is $\delta\simeq L/R$, which is an approximation applicable when the angle is very small, so we will make sure that this approximation is valid in the end. Plugging in the radius of curvature, we find $\delta \simeq (L q B)/\en=(10^9\times 4.8\times 10^{-10})/(3\times 10^{20}\times 1.6\times 10^{-12})=10^{-9}{\rm rad}\simeq 6\times 10^{-8}{\rm deg}$. The deflection angle is tiny, making it clear that the magnetic field of Earth offers no protection from such energetic particles. 
\end{prob}

\begin{prob} An unknown particle propagating in a bubble chamber decays into an electron-positron pair, whose tracks are shown in Figure \ref{pic_bubble}. (a) Which one is the electron, which one is the positron? (b) Why are the tracks spiraling inwards? (c) Estimate the mass and the energy of the unknown particle, in eV. 

\begin{figure}[htb]
\centering
\includegraphics[scale=0.3]{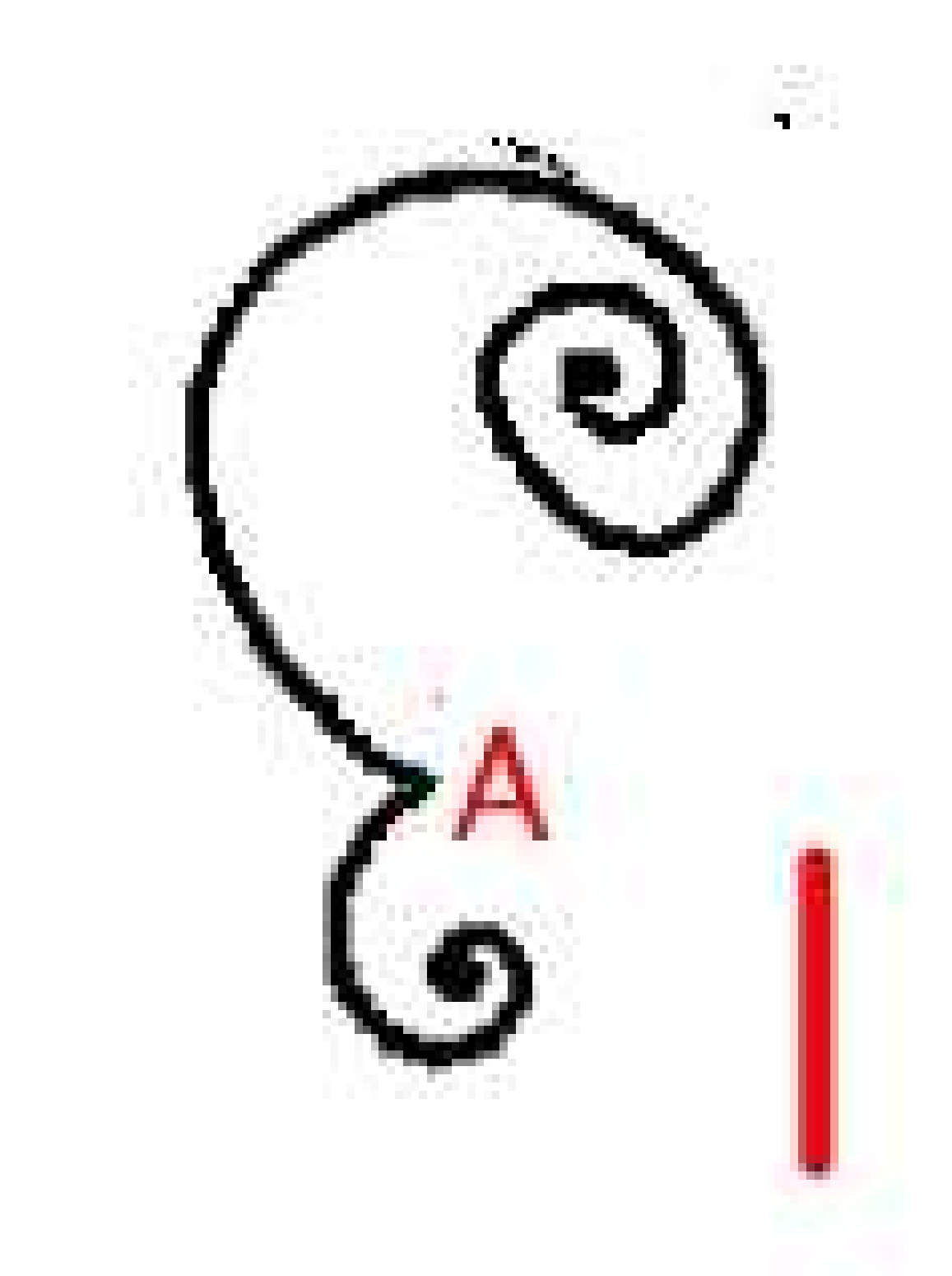}
\caption{Bubble chamber tracks of an electron-positron pair produced when an unknown particle decayed at point A. The magnetic field is 1 Tesla, directed toward the reader out of the page. The scale bar is 10 cm.}
\label{pic_bubble}
\end{figure}

{\bf Solution.} (a) The top one (the bigger curl) is positively charged, so it is a positron. (b) Because the particles are losing energy to the fluid in the bubble chamber. If they did not lose energy, their trajectories would be circles or cycloids. (c) The radius of curvature is related to the momentum via $R=pc/(eB)$. Because of energy losses, the radii of the tracks are rather difficult to calculate, but with a 30\% uncertainty or so we can estimate $R_1=13$ cm and $R_2=6$ cm for the top and the bottom tracks. Thus the momenta are $p_1 c=39$ MeV and $p_2 c=18$ MeV. The initial angle between the tracks is $\theta\simeq 70$ deg. Taking the 4-square of the 4-vector of energy and momentum conservation, we obtain
\begin{equation}
m_X^2c^4=2 m_e^2 c^4 +2 (\en_1 \en_2 - p_1 p_2 c^2 \cos\theta).
\end{equation} 
Because the Lorentz factors of the particles are rather high, $\en_i\simeq p_i c$ and the rest energy terms can be neglected (certainly these corrections are much smaller than the errors due to the crudeness of our estimates of the momenta). Thus $m_X c^2 \simeq \sqrt{2(p_1c)(p_2c)(1-\cos\theta)}=30$ MeV, whereas the energy is $\simeq p_1c+p_2c=57$ MeV.
\end{prob}

\clearpage

\clearpage
\section{Electromagnetic fields in different frames}
\label{sec_fields}

\subsection{Electric and magnetic fields are different in different inertial frames}

The good news is that the equation of motion in electromagnetic fields is the same as in the non-relativistic case, as long as we avoid using acceleration and use momentum instead:
\begin{equation}
\frac{{\rm d}\vec{p}}{{\rm d}t}=q\left(\vec{E}+\frac{\vec{u}}{(c)}\times\vec{B}\right), \label{eq_field_mot1}
\end{equation}
and its corollary
\begin{equation}
\frac{{\rm d}\en}{{\rm d}t}=q\vec{u}\vec{E}. \label{eq_field_mot2}
\end{equation}
Again, $c$ in parentheses is present in the cgs system but not in the MKS system. 

The bad news is that just about everything in this equation -- except, as it turns out, the particle charge $q$ and of course the speed of light $c$ -- is transformed when viewed from another frame. In particular, the electric and magnetic fields are not measured to be the same in all inertial frames. This is clear from a couple of thought experiments:

(a) Moving charges create both an electric and a magnetic field. But if we move into the frame co-moving with the charges, then there is only electric field, no magnetic field. Therefore, it is clear that magnetic field measurement depends on the chosen frame: it is non-zero in the first frame, but zero in the second frame. 

\begin{figure}[htb]
\centering
\includegraphics[scale=0.8, clip=true, trim=1cm 8cm 4cm 12.5cm]{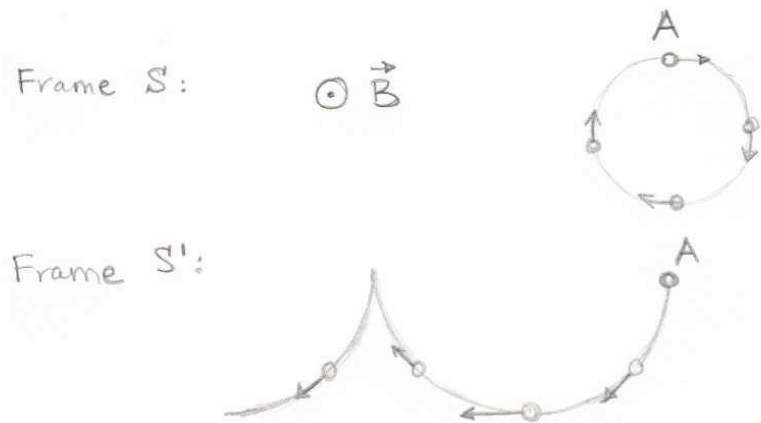}
\caption{In one frame, there is no electric field, and the particle moves on a circular orbit in uniform magnetic field. In the inertial frame which is instantaneously co-moving with the particle in point $A$, the particle accelerates from rest and therefore gains energy. Therefore there must be electric field directed downward in this frame.}
\label{pic_magn}
\end{figure}

(b) In another thought experiment, we start with a situation with no electric field, but with a uniform magnetic field in frame $S$. A particle whose velocity is perpendicular to the field moves along a circular orbit, as in Figure \ref{pic_magn}. A frame co-moving with the particle is not inertial since the particle is experiencing acceleration. But we can define {\bf an instantaneously co-moving inertial frame} $S'$: one which moves with the same velocity as does the particle at the point $A$ of its orbit. Therefore, when the particle is at point $A$, in this frame is it at rest for a moment, but then starts to accelerate along its cycloid orbit in this frame. Therefore, immediately after $A$ as seen in $S'$ the particle is gaining energy, which is not possible unless there is electric field in frame $S'$. Therefore, it is clear that the measurement of the electric field depends on the chosen frame: it is zero in $S$, but non-zero in $S'$.

\subsection{Relativistically invariant equation of motion}

For now, all we know is that equations (\ref{eq_field_mot1}-\ref{eq_field_mot2}) describe the motion of the particle in one frame and 
\begin{eqnarray}
\frac{{\rm d}\vec{p'}}{{\rm d}t'}=q\left(\vec{E'}+\frac{\vec{u'}}{(c)}\times\vec{B'}\right)\label{eq_field_pr1}\\
\frac{{\rm d}\en'}{{\rm d}t'}=q\vec{u'}\vec{E'}\label{eq_field_pr2}
\end{eqnarray}
describe the motion of the particle in another frame; we will use cgs units for now. We also know the transformation from $t$ to $t'$, from $\vec{u}$ to $\vec{u'}$, from $\vec{p}$ to $\vec{p'}$, and from $\en$ to $\en'$. We will now use these known transformations to derive transformations of $\vec{E}$ and $\vec{B}$ to $\vec{E'}$ and $\vec{B'}$. We explicitly assume that the charge $q$ does not change, for which there is abundant experimental confirmation (charge invariance, as it turns out, is a reflection of a deep symmetry of electromagnetism). 

We will start by re-writing equations (\ref{eq_field_mot1}-\ref{eq_field_mot2}) in a 4-vector form:
\begin{equation}
\frac{{\rm d}P^{\mu}}{{\rm d}\tau}=\frac{q}{c}\,\cdot \, {\rm [something]}\, \cdot U^{\nu}.\label{eq_motion}
\end{equation}
This looks like an equation from Special Relativity because it involves 4-vectors, and because it has some of the requisite elements of the modified Newton's law in place: derivative of momentum on the left-hand side and the velocity and charge on the right-hand side. Our task will now be to guess the `something' so that the 4-vector equation of motion (\ref{eq_motion}) looks exactly like (\ref{eq_field_mot1}-\ref{eq_field_mot2}) in $S$ and like (\ref{eq_field_pr1}-\ref{eq_field_pr2}) in $S'$. 

The `something' must be a function of the electric and magnetic fields. However, it cannot be a scalar function of electric and magnetic fields because then the change of momentum (related to the last three elements of the left-hand side) would be parallel to velocity, and we know that this is not right. If this is not a scalar function, what else can it be? On the left-hand side of equation (\ref{eq_motion}), we have a 4$\times 1$ matrix, and on the right-hand side $U^{\nu}$ is a $4\times 1$ matrix. So the `something' could be a $4\times 4$ matrix and the multiplication would work out from the standpoint of matrix multiplication rules.

We now spell out all the known elements of equation (\ref{eq_motion}), including the relationship between observer's time and proper time of the particle ${\rm d}\tau={\rm d}t/\gamma_u$. The unknown elements of the $4\times 4$ matrix will be named $f$ with subscripts appropriate for their place in the matrix:
\begin{equation}
\gamma_u\left(\begin{array}{c}{\rm d}(\en/c)/{\rm d}t \\{\rm d}p_x/{\rm d}t \\ {\rm d}p_y/{\rm d}t \\ {\rm d}p_z/{\rm d}t \end{array}\right)=\frac{q}{c}\left(\begin{array}{cccc} f_{00} & f_{01} & f_{02} & f_{03} \\  f_{10} & f_{11} & f_{12} & f_{13} \\  f_{20} & f_{21} & f_{22} & f_{23} \\  f_{30} & f_{31} & f_{32} & f_{33} \end{array}\right)\left(\begin{array}{c}\gamma_uc \\ \gamma_uu_x \\ \gamma_uu_y \\ \gamma_uu_z\end{array}\right).
\end{equation}
$\gamma_u$ can be cancelled out on both sides. This matrix equality contains four equations, of which the first one reads
\begin{equation}
\frac{{\rm d}(\en/c)}{{\rm d}t}=\frac{q}{c}\left(f_{00}c+f_{01}u_x+f_{02}u_y+f_{03}u_z\right).
\end{equation}
If we now compare this with equation (\ref{eq_field_mot2}), then they can be made identical if $f_{00}=0, f_{01}=E_x, f_{02}=E_y$ and $f_{03}=E_z$. 

Now let us look at the spatial components of this equation, for example
\begin{equation}
\frac{{\rm d}p_x}{{\rm d}t}=\frac{q}{c}\left(f_{10}c+f_{11}u_x+f_{12}u_y+f_{13}u_z\right).
\end{equation}
How do we make this identical to the $x$ component of the vector equation (\ref{eq_field_mot1})? Using the rules of the vector product, we can write the projection of (\ref{eq_field_mot1}) on the $x$-axis:
\begin{equation}
\frac{{\rm d}p_x}{{\rm d}t}=q\left(E_x+\frac{1}{c}\left(u_yB_z-u_zB_y\right)\right).
\end{equation}
To make these equations the same, we need to set $f_{10}=E_x, f_{11}=0, f_{12}=B_z$ and $f_{13}=-B_y$. In the same way, we can examine the $y$ and the $z$ components of the momentum equation to fill out all the unknown components of the matrix:
\begin{equation}
F^{\mu}{}_{\nu}=\left(\begin{array}{cccc} 0 & E_x & E_y & E_z \\  E_x & 0 & B_z & -B_y \\ E_y & -B_z & 0 & B_x \\ E_z & B_y & -B_x & 0 \end{array}\right),\label{eq_tensor}
\end{equation}
and the 4-vector equation of motion now reads
\begin{equation}
\frac{{\rm d}P^{\mu}}{{\rm d}\tau}=\frac{q}{c}F^{\mu}{}_{\nu}U^{\nu}.\label{eq_motion1}
\end{equation}
Here we are as usual assuming the summation over repeating indices (upper and lower $\nu$), which is equivalent to multiplying a $4\times 4$ matrix by a $4\times 1$ matrix. 

\subsection{Field transforms}

The matrix $F$ is called {\bf the tensor of electromagnetic fields}. As we already mentioned in Section \ref{sec_4vec}, a tensor is a $4\times 4$ matrix with some special properties -- roughly speaking, tensor is to matrix like a 4-vector to any set of 4 variables or like a relativistic invariant to an arbitrary physical value. An invariant does not change when we switch frames. A 4-vector transforms according to the Lorentz transform. In this section we will see how the tensor of electromagnetic fields transforms, which will allow us to calculate electric and magnetic fields in all inertial frames once we know then in one reference frame. 

We wrote the equation of motion (\ref{eq_motion1}) {\bf in a relativistically invariant form} because we know how $U^{\mu}$, $P^{\mu}$ and $\tau, q$ and $c$ transform from one frame to another: 4-vectors transform according to the Lorentz transform and invariants do not change. This is precisely what will allow us to calculate what happens to the matrix -- tensor -- $F$ as we move from one frame to another. 

So as not to drown in Greek indices, we will drop all indices for 4-vectors and tensors altogether and look at the matrix equations strictly mathematically, as we would do in linear algebra. Let $\Lambda$ be the Lorentz transform from the $S'$ to the $S$ frame, so $\Lambda$ is the matrix from equation (\ref{eq_lorentz_final2}). This means that 
\begin{equation}
U=\Lambda U',\,{\rm and}\, P=\Lambda P',
\end{equation}
where each one of these equalities is a matrix equation with a 4-vector on the left-hand side and a 4$\times$4 matrix multiplied by a 4-vector on the right-hand side. From the invariance of the proper time, we know that ${\rm d}\tau={\rm d}\tau'$. Let us now plug all these transforms into equation (\ref{eq_motion1}), to obtain:
\begin{equation}
\frac{{\rm d}\left(\Lambda P' \right)}{{\rm d}\tau'}=\frac{q}{c}F\left(\Lambda U'\right).
\end{equation}
Because we are transforming from one inertial frame into another, $\Lambda$ (which contains only the relative velocity between two frames) does not depend on $\tau$ or $\tau'$ and so can be taken out of the derivative:
\begin{equation}
\Lambda\frac{{\rm d} P'}{{\rm d}\tau'}=\frac{q}{c}F\left(\Lambda U'\right).
\end{equation}
We will now multiply the entire equation by $\Lambda^{-1}$ -- the inverse of the matrix $\Lambda$. We previously demonstrated (Problem \ref{prob_inverse}) that the inverse of $\Lambda$ can be obtained by switching the sign of the velocity, so no calculations are necessary to figure out what the elements of $\Lambda^{-1}$ should be (it is the matrix from eq. \ref{eq_lorentz_final1}). As we know, the order of multiplication does matter for matrix products; therefore we specify that we will apply $\Lambda^{-1}$ from the left-hand side:
\begin{equation}
\Lambda^{-1}\left(\Lambda\frac{{\rm d} P'}{{\rm d}\tau'}\right)=\Lambda^{-1}\left(\frac{q}{c}F\left(\Lambda U'\right)\right).
\end{equation}

While the order of multiplication of matrices is important, the multiplication is nonetheless associative, so we can rearrange parentheses: 
\begin{equation}
\left(\Lambda^{-1}\Lambda\right)\frac{{\rm d} P'}{{\rm d}\tau'}=\frac{q}{c}\left(\Lambda^{-1}F\Lambda \right) U'.
\end{equation}
We have also used the ability to carry the constant ($q/c$) between various parentheses without any harmful consequences. Now this equation can be simplified because $\Lambda^{-1}\Lambda$ in the left-hand side is a unity matrix, so they cancel each other out. We finally obtain 
\begin{equation}
\frac{{\rm d} P'}{{\rm d}\tau'}=\frac{q}{c}\left(\Lambda^{-1}F\Lambda \right)U'.\label{eq_motion2}
\end{equation}

According to the principle of relativity, all physical laws must have the same form in all inertial frames. Equation of motion is a physical law. Therefore, in the frame $S'$ it should read exactly the same as it did in frame $S$ (eq. \ref{eq_motion1}), but with all values as measured in $S'$:
\begin{equation}
\frac{{\rm d}P'}{{\rm d}\tau'}=\frac{q}{c}F' U'.\label{eq_motion3}
\end{equation}
Comparing this to our manipulated equation (\ref{eq_motion2}), we find the rule for transforming the tensor of electromagnetic radiation:
\begin{equation}
F'=\Lambda^{-1}F\Lambda.
\end{equation}
It turns out that all tensors transform this way, and this transformation rule -- multiply by $\Lambda^{-1}$ on one side and $\Lambda$ on the other -- is the definition of a tensor, analogously to how we defined 4-vectors.

$F'$ depends on the values of the electric and magnetic fields in the $S'$ frame and $F$ on the values in the $S$ frame, so this equation gives us the transforms of the field, albeit in a form which does not seem too useful:
\begin{equation}
\left(\begin{array}{cccc} 0 & E_x' & E_y' & E_z' \\  E_x' & 0 & B_z' & -B_y' \\ E_y' & -B_z' & 0 & B_x' \\ E_z' & B_y' & -B_x' & 0 \end{array}\right)=\left(\begin{array}{cccc} \gamma & -\gamma\beta & 0 & 0 \\ -\gamma\beta & \gamma & 0 & 0 \\ 0 & 0 & 1 & 0 \\ 0 & 0 & 0 & 1 \end{array}\right)\left(\begin{array}{cccc} 0 & E_x & E_y & E_z \\  E_x & 0 & B_z & -B_y \\ E_y & -B_z & 0 & B_x \\ E_z & B_y & -B_x & 0 \end{array}\right)\left(\begin{array}{cccc} \gamma & \gamma\beta & 0 & 0 \\ \gamma\beta & \gamma & 0 & 0 \\ 0 & 0 & 1 & 0 \\ 0 & 0 & 0 & 1 \end{array}\right).
\end{equation}
To obtain equations for transformations of individual components, ``all we need to do'' is multiply the three matrices on the right-hand-side and then equate the matrix elements on the left-hand side with the corresponding elements on the right-hand side. Yes, it is exactly as annoying as it sounds... So let's do it. As we remember, due to associative property we can first multiply $F\Lambda$, so our matrix equation now reads:
\begin{equation}
\left(\begin{array}{cccc} 0 & E_x' & E_y' & E_z' \\  E_x' & 0 & B_z' & -B_y' \\ E_y' & -B_z' & 0 & B_x' \\ E_z' & B_y' & -B_x' & 0 \end{array}\right)=\left(\begin{array}{cccc} \gamma & -\gamma\beta & 0 & 0 \\ -\gamma\beta & \gamma & 0 & 0 \\ 0 & 0 & 1 & 0 \\ 0 & 0 & 0 & 1 \end{array}\right)\left(\begin{array}{cccc} \beta\gamma E_x & \gamma E_x & E_y & E_z \\  \gamma E_x & \beta\gamma E_x & B_z & -B_y \\ \gamma E_y-\beta\gamma B_z & \beta\gamma E_y-\gamma B_z & 0 & B_x \\ \gamma E_z+\beta\gamma B_y & \beta \gamma E_z + \gamma B_y & -B_x & 0 \end{array}\right).
\end{equation}
One final push to multiply the two $4\times 4$ matrices on the right-hand side:
\begin{equation}
\left(\begin{array}{cccc} 0 & E_x' & E_y' & E_z' \\  E_x' & 0 & B_z' & -B_y' \\ E_y' & -B_z' & 0 & B_x' \\ E_z' & B_y' & -B_x' & 0 \end{array}\right)=\left(\begin{array}{cccc} \beta\gamma^2 E_x-\beta\gamma^2 E_x & \gamma^2 E_x-\beta^2\gamma^2 E_x & \gamma E_y-\beta\gamma B_z & \gamma E_z + \beta \gamma B_y \\ -\beta^2\gamma^2 E_x+\gamma^2 E_x & -\beta\gamma^2 E_x+\beta \gamma^2 E_x & -\beta\gamma E_y + \gamma B_z & -\beta \gamma E_z - \gamma B_y \\ \gamma E_y-\beta\gamma B_z & \beta\gamma E_y-\gamma B_z & 0 & B_x \\ \gamma E_z+\beta\gamma B_y & \beta \gamma E_z + \gamma B_y & -B_x & 0 \end{array}\right).
\end{equation}
The diagonal elements are then zero, and $f'_{01}$ and $f'_{10}$ can be simplified if we remember that $\gamma^2-\beta^2\gamma^2=1$. It is also comforting that the overall symmetry properties of the matrix are preserved as we transform between frames: we see that the elements in the first row and the first column (that involve time-like coordinates) are symmetric relative to the diagonal, whereas the spatial $3\times 3$ part of the matrix is anti-symmetric. 

We finally obtain the following relationships between electric and magnetic fields as measured in two different frames:
\begin{eqnarray}
E_x'=E_x; \, E_y'=\gamma E_y-\beta \gamma B_z; \, E_z'=\gamma E_z+\beta\gamma B_y;\label{eq_tran1}\\
B_x'=B_x; \, B_y'=\beta \gamma E_z+\gamma B_y; \, B_z'=\gamma B_z-\beta \gamma E_y.\label{eq_tran2}
\end{eqnarray}
These equations are written in cgs units which reflect well the deep connection between electric and magnetic fields, in that electric and magnetic fields have essentially the same unit of measurement. In MKS, all $B$ values need to be multiplied by $c$:
\begin{eqnarray}
E_x'=E_x; \, E_y'=\gamma E_y-\beta \gamma c B_z; \, E_z'=\gamma E_z+\beta\gamma c B_y;\\
c B_x'=c B_x; \, c B_y'=\beta \gamma E_z+\gamma c B_y; \, c B_z'=\gamma c B_z-\beta \gamma E_y.
\end{eqnarray}

\begin{prob}
Using field transform equations, demonstrate that our suspicions about the nature of the fields in frame $S'$ in Figure \ref{pic_magn} were correct. -- {\bf Solution.} In frame $S$, the electric field is zero: $\vec{E}=(0,0,0)$, whereas the magnetic field has a non-zero $z$-component: $\vec{B}=(0,0,B_0)$. The frame instantaneously co-moving with the particle as it is going through point $A$ is moving with velocity $v$ in the positive $x$-direction. According to transformation equations, in this frame we would measure electric and magnetic fields (in cgs units) to be $\vec{E'}=(0,-\beta\gamma B_0, 0)$, $\vec{B'}=(0,0,\gamma B_0)$. So indeed there is an electric force in the negative $y$-direction, which is pulling down on the particle in point $A$. 
\end{prob}

\begin{prob}
Calculate the magnetic field of a particle moving with velocity $\vec{v}$ in the non-relativistic regime, if its electric field is $\vec{E}$. -- {\bf Solution.} Let $S'$ be the frame co-moving with the particle. In this frame, there is no magnetic field, $\vec{B'}=(0,0,0)$ and the electric field is $\vec{E'}$. Let $S$ be the lab frame where we now need to calculate the values of the electric and magnetic field and relate them to one another. To the leading order in $v/c$, in all transformation equations we can set $\gamma=1$. We also need to remember to switch the signs in the velocity involved in transformation equations because we are transforming from $S'$ to $S$, not vice versa. Therefore, we obtain:
\begin{eqnarray}
E_x=E_x'; \, E_y= E_y'; \, E_z= E_z'\\
B_x=0; \, B_y=-\beta E_z'; \, B_z=\beta E_y'.
\end{eqnarray}
So in the non-relativistic case the electric field does not change (this statement is accurate to order $O(v/c)$, but not to $O(v^2/c^2)$), and we can simply say that the observed magnetic and electric field are related via $B_x=0; \, B_y=-\beta E_z; \, B_z=\beta E_y$. These equations can be grouped together into vector form:
\begin{equation}
\vec{B}=\frac{1}{c}\left(\vec{v}\times \vec{E}\right)
\end{equation}
in cgs units, and 
\begin{equation}
\vec{B}=\frac{1}{c^2}\left(\vec{v}\times \vec{E}\right)
\end{equation}
in MKS units. This is the law of Biot-Savart for a point charge, which should be familiar from the classical electromagnetism. 
\end{prob}

\begin{prob}
A region of space has an electric field $\vec{E}$ (value 1 cgs unit=30,000 Volt/meter) and a magnetic field $\vec{B}$ (value 1 Gauss$=10^{-4}$ Tesla) directed at a 45 degree angle to each other. How fast do we need to run and in what direction to make them perpendicular to one another in the moving frame? -- {\bf Solution.} This sounds like an unpleasant problem. We could write down the conversion between the fields using projections of both field vectors, and because we want the fields to be perpendicular in $S'$ we could try to find $\beta$ such that the scalar product $\vec{E'}\cdot\vec{B'}$ turns to zero. However, it turns out that this scalar product is a relativistic invariant. Which means that if the fields are perpendicular in one frame (e.g., $S'$), they will be perpendicular in all other inertial frames (including $S$), and therefore the problem has no solution! 

Let us demonstrate that $\vec{E'}\cdot\vec{B'}$ is a relativistic invariant. Indeed, using field transforms in equations (\ref{eq_tran1}-\ref{eq_tran2}):
\begin{equation}
\vec{E'}\cdot\vec{B'}=E_x'B_x'+E_y'B_y'+E_z'B_z'=E_xB_x+(\gamma E_y-\beta \gamma B_z)(\beta \gamma E_z+\gamma B_y)+(\gamma E_z+\beta\gamma B_y)(-\beta\gamma E_y+\gamma B_z).
\end{equation}
When we open all parentheses, collect all the like terms and remember that $\gamma^2-\beta^2\gamma^2=1$, we find that this is equal to $E_xB_x+E_yB_y+E_zB_z=\vec{E}\cdot\vec{B}$. Therefore, this dot product is a relativistic invariant.
\end{prob}

\clearpage
\begin{figure}[!htb]
\centering
\includegraphics[scale=0.6]{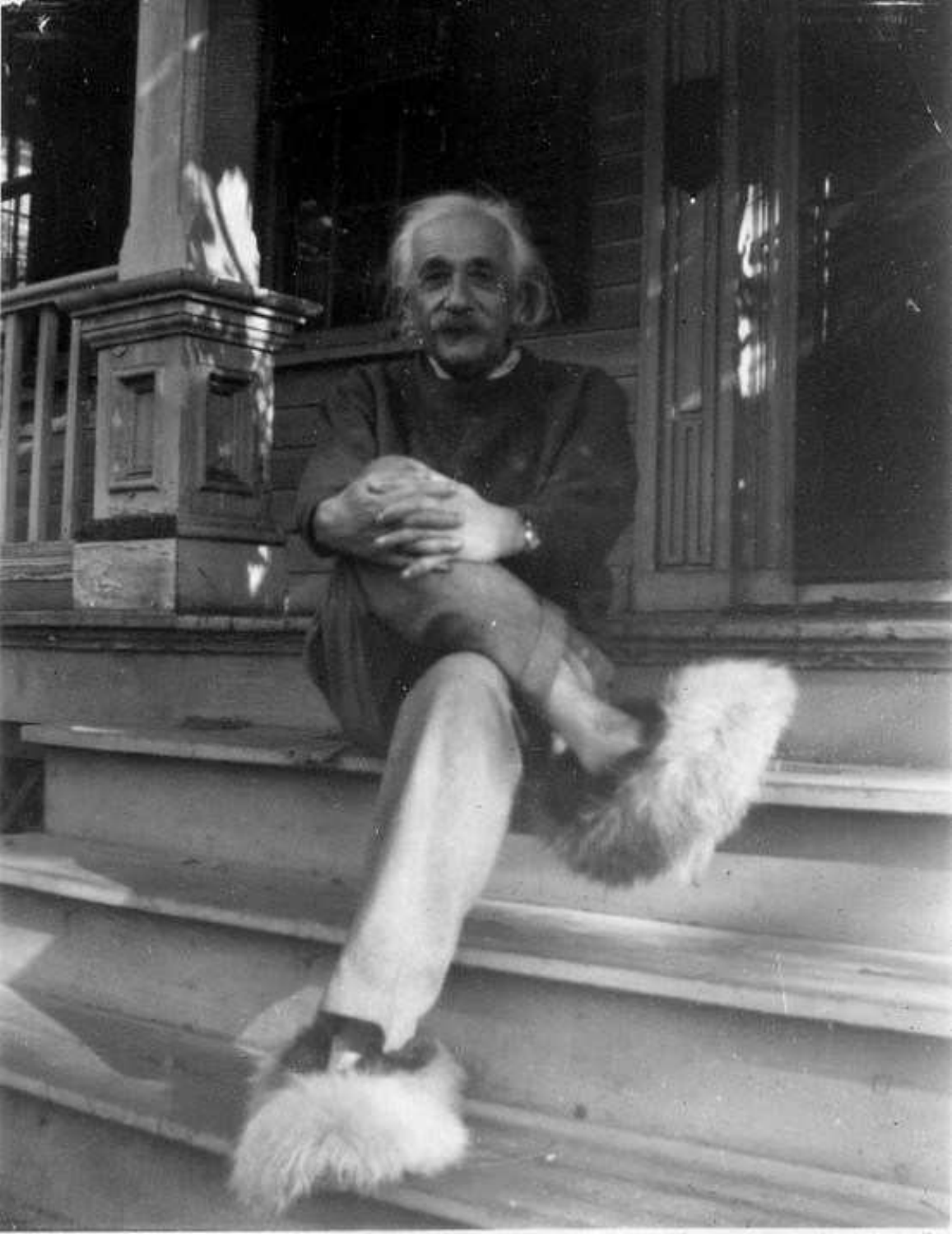}
\caption{Einstein in fuzzy slippers. Photo by Historical Society of Princeton / Gillett Griffin, from \newline \url{http://www.retronaut.com/2012/08/albert-einstein-in-fuzzy-slippers-c-1950s/}.}
\end{figure}

\clearpage
\section{Acknowledgments}

The author is grateful to all characters in the story: the train, the tracks, the flower pot, the various flags, clocks and vases that suffered for the cause, the kids and the soccer ball. Lady Bug helpfully played the role of an innocent additional inertial frame. Bob and Mary were willing to go the extra light years at enormous velocities. Bob's cat would have preferred to remain causally disconnected, but was dragged into the narrative and reluctantly participated in the lectures, though so far refused to appear in the notes. Various particles helpfully propagated at just the right energies in just the right directions. Brave students assisted with enthusiasm, even when they were pancaked by length contraction.

\clearpage
\begin{figure}[!htb]
\centering
\includegraphics[scale=0.6]{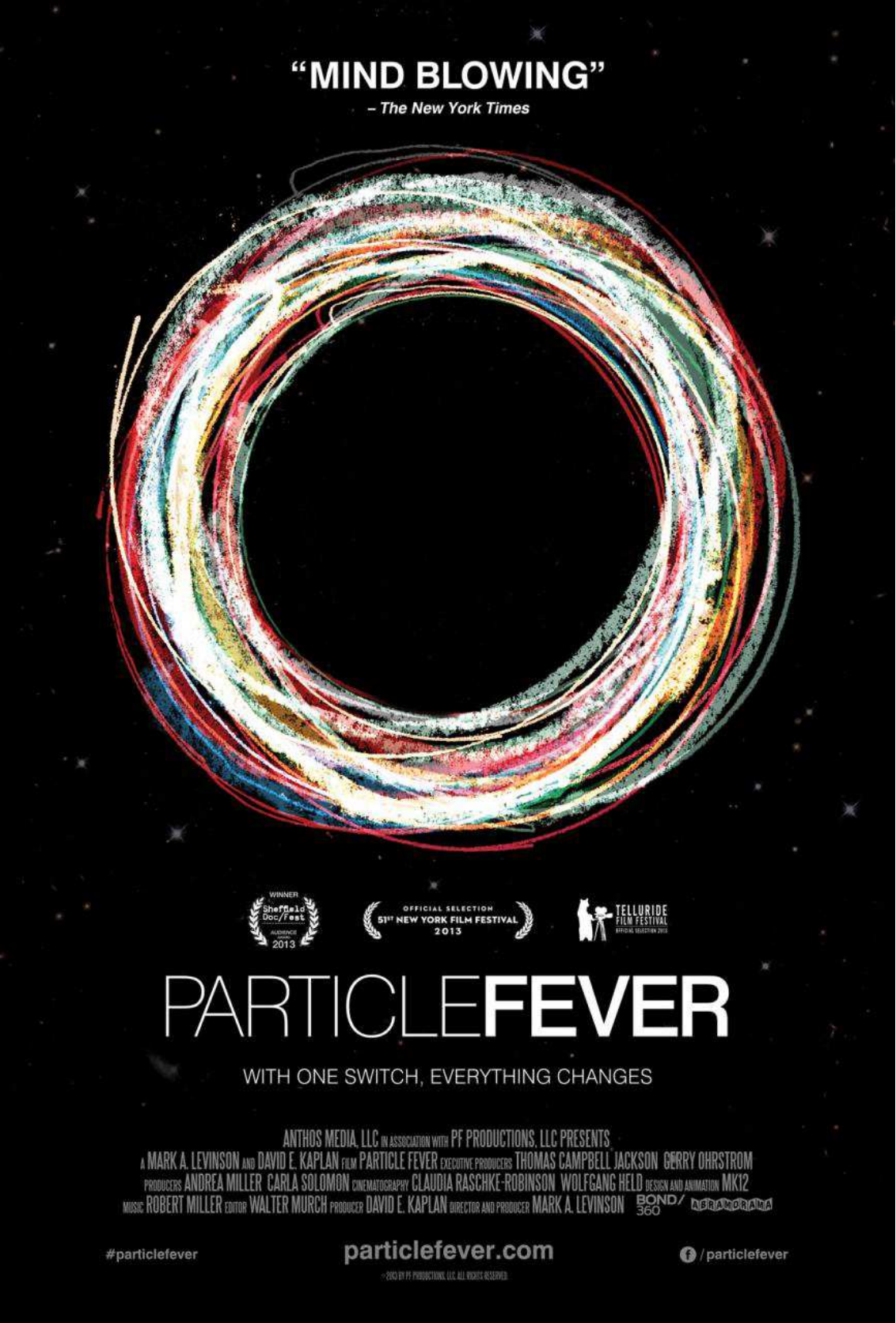}
\caption{Theatrical release poster for ``Particle Fever'', produced by M. Levinson and JHU physics professor D. Kaplan, who received the 2015 Communication Awards of the National Academies of Sciences, Engineering, and Medicine for this work. See the trailer at \url{http://particlefever.com/}}
\end{figure}

\clearpage
\begin{figure}[!htb]
\centering
\includegraphics[scale=0.8]{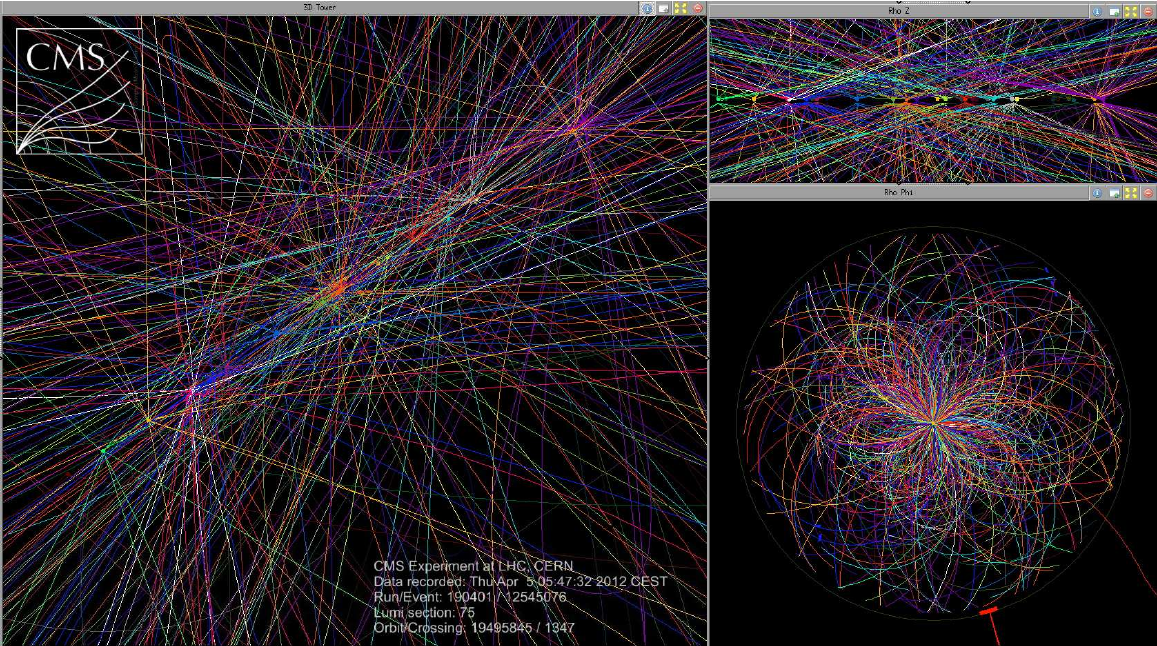}
\caption{Visualization of an actual event from LHC CMS, with 29 distinct proton-proton collisions, from \newline
\url{http://cms.web.cern.ch/news/new-world-record-first-pp-collisions-8-tev}.}
\end{figure}

\end{document}